\documentclass{aa}   
\usepackage{graphicx}
\usepackage{url}
\usepackage[breaklinks=true]{hyperref}
\hypersetup{colorlinks=true,linkcolor=blue,citecolor=blue,filecolor=blue,urlcolor=blue}
\usepackage{twoopt}
\usepackage{natbib}
\bibpunct{(}{)}{;}{a}{}{,} %% natbib format for A&A and ApJ
\usepackage{amssymb}
\usepackage{float}
\usepackage{color}
\usepackage{epstopdf}
\usepackage{ctable}
\usepackage{multirow}
\usepackage{xcolor}

\begin{document}

\title{A LOFAR-uGMRT spectral index study of distant radio halos}
%\subtitle{}

\author{
G.~Di Gennaro\inst{\ref{inst:leiden}}, %\ref{inst:hamb} \thanks{von Humboldt Fellow}
R.J.~van Weeren\inst{\ref{inst:leiden}},
R.~Cassano\inst{\ref{inst:ira}},
G.~Brunetti\inst{\ref{inst:ira}},
M.~Br\"uggen\inst{\ref{inst:hamb}},
M.~Hoeft\inst{\ref{inst:tauten}},
E.~Osinga\inst{\ref{inst:leiden}}
A.~Botteon\inst{\ref{inst:leiden}},
V.~Cuciti\inst{\ref{inst:hamb}},
F.~de Gasperin\inst{\ref{inst:hamb}},
H.J.A.~R\"ottgering\inst{\ref{inst:leiden}},
\and
C.~Tasse\inst{\ref{inst:gepi},\ref{inst:south_africa}}
}

\institute{
{Leiden Observatory, Leiden University, PO Box 9513, 2300 RA Leiden, The Netherlands}\label{inst:leiden}\\\email{digennaro@strw.leidenuniv.nl}
\and
{Istituto Nazionale di Astrofisica-Istituto di Radioastronomia, Bologna Via Gobetti 101, I40129 Bologna, Italy}\label{inst:ira}
\and
{Hamburger Sternwarte, Universit\"at Hamburg, Gojenbergsweg 112, 21029 Hamburg, Germany}\label{inst:hamb}
\and
{Th\"uringer Landessternwarte, Sternwarte 5, 07778 Tautenburg, Germany}\label{inst:tauten}
\and
{GEPI \& USN, Observatoire de Paris, CNRS, Universit\'e Paris Diderot, 5 place Jules Janssen, 92190 Meudon, France}\label{inst:gepi}
\and 
{Centre for Radio Astronomy Techniques and Technologies, Department of Physics and Electronics, Rhodes University, Grahamstown 6140, South Africa}\label{inst:south_africa}
}

\date{Received 10 June 2021; accepted 5 August 2021}

\abstract
{Radio halos are megaparsec-scale diffuse radio sources{ mostly} located at the centres of merging galaxy clusters. The common mechanism invoked to explain their origin is the re-acceleration of relativistic particles caused by large-scale turbulence.}
{Current re-acceleration models predict that a significant number of halos at high redshift should be characterised by very steep spectra ($\alpha<-1.5$) because of increasing inverse Compton energy losses. In this paper, we investigate the spectral index properties of a sample of nine clusters selected from the second Planck Sunyaev-Zel'dovich catalogue showing diffuse radio emission with the Low Frequency Array (LOFAR) in the 120--168 MHz band. This is the first time that radio halos discovered at low frequencies are followed up at higher frequencies.} 
{We analysed upgraded Giant Metrewave Radio Telescope (uGMRT) observations in Bands 3 and 4, that is, 250--500  and 550--900 MHz respectively. These observations were combined with existing LOFAR data to obtain information on the spectral properties of the diffuse radio emission.}
{We find diffuse radio emission in the uGMRT observations for five of the nine high-$z$ radio halos previously discovered with LOFAR. For those, we measure spectral indices in the range of $-1$ to $-1.4$.
For the uGMRT non-detections, we estimated that the halos should have a spectral index steeper than $-1.5$. We also confirm the presence of one candidate relic.
}
{Despite the small number of clusters, we find evidence that about half of the massive and merging clusters at high redshift host radio halos with a very steep spectrum. This is in line with theoretical predictions, although larger statistical samples are necessary to test models.}


\titlerunning{Spectral index study of high-redshift radio halos}
\authorrunning{G. Di Gennaro et al.}
\maketitle

%
%-------------------------------------------------------------------

\section{Introduction}
In the framework of the $\Lambda$CDM cosmological model, galaxy clusters form and grow via accretion of less massive systems \citep[e.g. galaxy groups or small galaxy clusters, see][]{press+schecter74,springel+06}. These events release {energy} up to $10^{64}$ erg in the intracluster medium (ICM) in a few gigayears. A fraction of this energy is dissipated by shocks and turbulence, and is used in the amplification of magnetic fields and (re)acceleration of particles, producing diffuse radio emission in the form of {radio halos} and {radio relics} \citep[see][for recent theoretical and observational reviews]{brunetti+jones14,vanweeren+19}.
Because of their intrinsic low surface brightness ($\sim$0.1--1$~\mu$Jy arcsec$^{-2}$ at 1.4 GHz), halos and relics are difficult to detect. In addition, they are characterised by steep spectra (i.e. $\alpha<-1$, with $S_\nu\propto\nu^\alpha$), and are therefore better observed at low radio frequencies (i.e. below GHz).

 \begin{table*}
\caption{Physical properties of the galaxy clusters.}
\vspace{-5mm}
\begin{center}
\resizebox{0.65\textwidth}{!}{
\begin{tabular}{lccccc}
\hline
\hline
\noalign{\smallskip}
Cluster name & $z$ & RA$_{\rm J2000}$  & Dec$_{\rm J2000}$ & $M_{\rm SZ,500}$ & kpc/$''$ \\
 & & [deg] & [deg] & [$\rm 10^{14}~M_\odot$] \\
 \noalign{\smallskip}
 \hline
 \noalign{\smallskip}
 PSZ2\,G086.93+53.18 & 0.675 & 228.50446 & +52.81074 & $5.4\pm0.5$ & 7.125\\
 PSZ2\,G089.39+69.36 & 0.680 & 208.43748 & +43.48470 & $5.7\pm0.7$ & 7.148 \\
 PSZ2\,G091.83+26.11 & 0.822 & 277.78430 & +62.24770 & $7.4\pm0.4$ & 7.676 \\
 PSZ2\,G099.86+58.45 & 0.616 & 213.6909 & +54.78029 & $6.8\pm0.5$ & 6.845 \\
 PSZ2\,G126.28+65.62 & 0.820 & 190.5975 & +51.43944 & $5.0\pm0.7$ & 7.670 \\
 PSZ2\,G141.77+14.19 & 0.830 & 70.27167 & +68.22275 & $7.7\pm0.9$ & 7.700 \\
 PLCK\,G147.3–16.6 & 0.645 & 44.105898 & +40.290140 & $6.3\pm0.4$ & 6.988 \\
 PSZ2\,G147.88+53.24 & 0.600 & 164.37923 & +57.99591 & $6.5\pm0.6$ & 6.762 \\
 PSZ2\,G160.83+81.66 & 0.888 & 186.74267 & +33.54682 & $5.7^{+0.6}_{-0.7}$ & 7.865 \\
 \noalign{\smallskip}
 \hline
\end{tabular}}
\end{center}
\vspace{-5mm}
\tablefoot{The cluster masses, $M_{\rm SZ,500}$, are taken from the Planck-SZ catalogue \citep{planckcoll16}. 
}
\label{tab:clusters}
\end{table*}

Radio halos are cluster-size structures that generally follow the distribution of the thermal cluster emission (i.e. the ICM). Their currently favoured formation scenario involves the re-acceleration of electrons via turbulence induced by  cluster merger events \citep[e.g.][]{brunetti+01, petrosian01,brunetti+lazarian07,donnert+13,brunetti+lazarian16}. In support of this scenario, radio halos are preferentially detected in dynamically disturbed clusters {\citep[e.g.][]{cassano+13,cuciti+21a}}. An additional channel may arise from  proton--proton collisions, which generate secondary electrons that then emit synchrotron radiation \citep[e.g. hadronic models,][]{blasi+colafrancesco99, dolag+ensslin00}. 
Although the non-detection of $\gamma$-ray emission from galaxy clusters rules out a dominant contribution from this channel \citep[e.g.][]{reimer+03,ackermann+10,prokhorov+14,brunetti+17,adam+21}, turbulent re-acceleration of secondary particles is still a viable model \citep{brunetti+lazarian11,pinzke+17,brunetti+17} .

Radio relics are elongated structures generally located in the cluster outskirts. It is widely believed that these are associated with propagating shock waves caused by mergers \citep[e.g.][]{rottgering+97,ensslin+98,giacintucci+08,vanweeren+10,pearce+17,hoang+18,digennaro+18}. This is also supported by the detection of strongly polarised emission at the relic position \citep[e.g.][]{vanweeren+10,digennaro+21}, which suggests amplification and compression of magnetic fields. Nonetheless, the nature of the (re)acceleration mechanism is still unclear \citep[e.g.][]{vazza+14}. Standard Fermi type-I acceleration of thermal ICM electrons  \citep[e.g.][]{drury83,ensslin+98,brunetti+jones14} sometimes requires an unrealistic shock efficiency {to justify the relic radio brightness ---in the case where electrons are accelerated} from the thermal pool--- because of the low Mach number of the shocks \citep[$\mathcal{M}\lesssim2$, e.g.][]{botteon+16,hoang+17,digennaro+19,botteon+20}. Re-acceleration of pre-existing relativistic plasma at the shock has therefore been proposed \citep[e.g.][]{markevitch+05,bonafede+14,kang+17,vanweeren+17a}. Examples that  are considered to indicate ongoing re-acceleration are still limited to a few cases \citep[e.g.][]{vanweeren+17a,digennaro+18}.

\begin{table*}[h!]
\caption{Radio observation details.}
\vspace{-5mm}
\begin{center}
\resizebox{0.95\textwidth}{!}{
\begin{tabular}{lcccccc}
\hline
\hline\noalign{\smallskip}
Cluster name & Telescope  & Project/pointing & Observation date & Observation length$^\dagger$ & Frequency coverage & Configuration \\
 & & & [dd-mm-yyyy] & [hr] & [MHz] \\
\noalign{\smallskip}\hline\noalign{\smallskip}
\multirow{3}{*}{PSZ2\,G086.93+53.18} & \multirow{2}{*}{LOFAR}  & P227+53 & 19-02-2015 & \multirow{2}{*}{8.33} & \multirow{2}{*}{120--168} & \multirow{2}{*}{HBA Dual Inner} \\
&  & P231+53 & 19-02-2015 \\
& uGRMT  & 38\_054 & 30-08-2020 & 6 & 550--900 & Band 4 \\
\noalign{\smallskip}\hline\noalign{\smallskip}
\multirow{3}{*}{PSZ2\,G089.39+69.36} & \multirow{2}{*}{LOFAR}  & P207+45 & 07-05-2015 & \multirow{2}{*}{8.33} & \multirow{2}{*}{120--168} & \multirow{2}{*}{HBA Dual Inner} \\
&  & P209+42 & 05-03-2015 \\
& uGMRT  & 38\_054 & 25-08-2020 & 6 & 250--500 & Band 3 \\
\noalign{\smallskip}\hline\noalign{\smallskip}
\multirow{4}{*}{PSZ2\,G091.83+26.11} & \multirow{2}{*}{LOFAR}  & P275+63 & 22-08-2016 & \multirow{2}{*}{8.33} & \multirow{2}{*}{120--168} & \multirow{2}{*}{HBA Dual Inner} \\
&  & P280+60 & 18-01-2019 \\
& uGMRT  & 36\_039 & 05-05-2019 & 5 & 250--500 & Band 3 \\
& uGRMT  & 36\_039 & 29-05-2019 & 5 & 550--900 & Band 4 \\
\noalign{\smallskip}\hline\noalign{\smallskip}
\multirow{4}{*}{PSZ2\,G099.86+58.45} & \multirow{3}{*}{LOFAR} & P209+55 & 30-04-2015 & \multirow{3}{*}{8.33} & \multirow{3}{*}{120--168} & \multirow{3}{*}{HBA Dual Inner} \\
& &  P214+52 & 12-05-2015 \\
& &  P214+55 & 12-05-2015 \\
& uGRMT  & 38\_054 & 30-08-2020 & 6 & 550--900 & Band 4 \\
\noalign{\smallskip}\hline\noalign{\smallskip}
\multirow{4}{*}{PSZ21\,G126.28+65.62} & \multirow{3}{*}{LOFAR} & P29Hetdex19 & 26-06-2014 & \multirow{3}{*}{8.33} & \multirow{3}{*}{120--168} & \multirow{3}{*}{HBA Dual Inner} \\
& &  P30Hetdex06 & 30-05-2014  \\
& &  P33Hetdex08 & 19-06-2014 \\
& uGRMT  & 38\_054 & 10-07-2020 & 6 & 550--900 & Band 4 \\
\noalign{\smallskip}\hline\noalign{\smallskip}
\multirow{2}{*}{PSZ2\,G141.77+14.19} & LOFAR & P068+69 & 08-09-2017 & 8.33 & 120--168 & HBA Dual Inner \\
& uGRMT  & 36\_039 & 07-05-2019 & 5 & 550--900 & Band 4 \\
\noalign{\smallskip}\hline\noalign{\smallskip}
\multirow{3}{*}{PLCK\,G147.3--16.6} & LOFAR & P044+39 & 29-11-2017 & 8.33 & 120--168 & HBA Dual Inner \\
& uGRMT  & 38\_054 & 20-06-2020 & 6 & {250--500} & Band 3 \\
& uGRMT  & 38\_054 & 16-06-2020 & 6 & 550--900 & Band 4 \\
\noalign{\smallskip}\hline\noalign{\smallskip}
\multirow{3}{*}{PSZ2\,G147.88+53.24} & \multirow{2}{*}{LOFAR}  & P165+57 & 11-05-2015 & \multirow{2}{*}{8.33} & \multirow{2}{*}{120--168} & \multirow{2}{*}{HBA Dual Inner} \\
&  & P166+60 & 13-10-2017  \\
& uGRMT  & 38\_054 & 20-06-2020 & 6 & 550--900 & Band 4 \\
\noalign{\smallskip}\hline\noalign{\smallskip}
\multirow{5}{*}{PSZ2\,G160.83+81.66} & \multirow{4}{*}{LOFAR} & P185+32 & 11-12-2019 & \multirow{4}{*}{8.33} & \multirow{4}{*}{120--168} & \multirow{4}{*}{HBA Dual Inner} \\
& &  P185+35 & 23-02-2017 \\
& &  P188+32 & 18-08-2017 \\
& &  P188+35 & 28-03-2019 \\
& uGRMT  & 38\_054 & 12-08-2020 & 6 & 550--900 & Band 4 \\
\noalign{\smallskip}\hline
\end{tabular}}
\end{center}
\vspace{-5mm}
\tablefoot{$^\dagger$The observation length includes also the time on the calibrators.}
\label{tab:obsid_info}
\end{table*}

Most of the statistical studies of diffuse radio emission are limited to the local Universe \citep[i.e. $z\sim0.1-0.4$; e.g.][]{cassano+13, kale+15, cuciti+21a,cuciti+21b}. A handful of clusters up to $z\sim0.5$ hosting diffuse radio emission have recently been reported in \cite{giovannini+20}. 
All these observations were firstly carried out at GHz frequencies, and eventually followed up at lower frequencies (i.e. $\sim100$ MHz) to investigate the spectral characteristic of the observed radio halos. This approach however misses a large fraction of (ultra-)steep spectrum halos, because radio halos with $\alpha<-1.5$ are hardly detected at GHz frequencies. 
At higher redshifts ($z\geq0.6$), only a few exceptional clusters have been studied so far (e.g. `El Gordo' at $z=0.87$, \citealt{lindner+14}, PLCK\,G147.3-16.6 at $z=0.645$, \citealt{vanweeren+14} and  PSZ2\,G099.86+58.45 at $z=0.616$, \citealt{cassano+19}). Recently, in  \cite{digennaro+20}, we  presented a statistical study of a sample of distant ($z\geq0.6$) galaxy clusters selected from the second Planck Sunyaev-Zel'dovich (SZ) catalogue \citep{planckcoll16} and observed with the LOFAR Two-Metre Sky Survey \citep[LoTSS;][]{shimwell+17,shimwell+19}. We observed that 9 out of 19 clusters host diffuse radio emission. Available X-ray observations {\citep[Chandra and/or XMM-Newton; see fig. 2 in][]{digennaro+20}} suggest that the radio halos are located in dynamically disturbed clusters. Assuming turbulent re-acceleration from the radio luminosities, we estimated magnetic field strengths similar to nearby ($z\sim0.2$) systems in the same mass range.  According to the turbulent re-acceleration scenario, a large fraction of distant radio halos should have steep integrated spectral indices ($\alpha<-1.5$) because of the strong synchrotron and inverse Compton (IC) losses. We followed up those clusters hosting diffuse radio emission in \cite{digennaro+20} with the upgraded Giant Metrewave Radio Telescope (uGMRT). This represents the first high-frequency follow up of high-$z$ halos detected at low frequencies. The observed sample is listed in  Table \ref{tab:clusters}.

Throughout the paper, we assume a standard $\Lambda$CDM cosmology, with $H_0 = 70$ km s$^{-1}$ Mpc$^{-1}$, $\Omega_m = 0.3$ and $\Omega_\Lambda = 0.7$.

\begin{table*}
\caption{Imaging parameters and image properties of the cluster sample.}
\vspace{-5mm}
\begin{center}
%\resizebox{0.85\textwidth}{!}{
\begin{tabular}{lccccc}
\hline\hline\noalign{\smallskip}
Cluster name & Central Frequency  & Resolution & $uv$ min & $uv$-taper & $\sigma_{\rm rms}$ \\
 & [MHz] & $[''\times'']$ & $[\lambda]$ & $['']$ & [$\mu$Jy beam$^{-1}$] \\
\noalign{\smallskip}\hline\noalign{\smallskip}
\multirow{4}{*}{PSZ2\,G086.93+53.18} 
& \multirow{2}{*}{144}   & $9.5\times4.5$ &  80 & -- & 91.8 \\
& & $26.0\times26.0$ & 80 & 15 & 239.4 \\
& \multirow{2}{*}{650}  & $3.8\times3.4$ &  --  & -- & 8.6 \\
& & $26.0\times26.0$ & --  & 15 & 45.5 \\
\noalign{\smallskip}\hline\noalign{\smallskip}
\multirow{4}{*}{PSZ2\,G089.39+69.36} 
& \multirow{2}{*}{144}   & $8.3\times4.8$ &  80 & -- & 64.3 \\
& & $29.0\times29.0$ &  80 & 15 & 138.2 \\
& \multirow{2}{*}{400}  & $6.4\times5.1$ & --  & -- & 57.2 \\
& & $29.0\times29.0$ & -- & 15 & 416.0 \\
\noalign{\smallskip}\hline\noalign{\smallskip}
\multirow{6}{*}{PSZ2\,G091.83+26.11} 
& \multirow{2}{*}{144}  & $6.8\times4.5$ & 80  & -- & 92.1  \\
&   & $14.0\times14.0$ & 160  & 6 & 188.4  \\
&  \multirow{2}{*}{400} & $13.8\times6.2$ &  --  & -- & 64.6 \\
&   & $14.0\times14.0$ & 160  & -- & 94.6  \\
&  \multirow{2}{*}{650} & $4.4\times2.8$ & -- & -- & 11.4 \\
&   & $14.0\times14.0$ & 160  & 6 & 38.9  \\
\noalign{\smallskip}\hline\noalign{\smallskip}
\multirow{4}{*}{PSZ2\,G099.86+58.45} 
& \multirow{2}{*}{144}  & $8.0\times4.4$ & 80 & -- & 66.3 \\
& & $18.0\times18.0$ & 180 & 10 & 153.2 \\
& \multirow{2}{*}{650}  & $5.0\times2.6$ & -- & -- & 8.2 \\
& & $18.0\times18.0$ & 180 & 10 & 33.7 \\
\noalign{\smallskip}\hline\noalign{\smallskip}
\multirow{4}{*}{PSZ2\,G126.28+65.62} 
& \multirow{2}{*}{144}  & $7.6\times4.5$  & 80 & -- & 52.3 \\
& & $25.0\times25.0$  & 80 & 15 & 117.6 \\
& \multirow{2}{*}{650}  & $7.6\times2.5$ & -- & -- & 9.8 \\
& & $25.0\times25.0$  & -- & 15 & 39.9 \\
\noalign{\smallskip}\hline\noalign{\smallskip}
\multirow{4}{*}{PSZ2\,G141.77+14.19} 
& \multirow{2}{*}{144}  & $7.6\times5.0$ &  80 & -- & 119.6 \\
& & $17.0\times17.0$ & 150 & 10 & 171.0 \\
& \multirow{2}{*}{650}  & $6.4\times3.3$ & --  & -- & 9.7 \\
& & $17.0\times17.0$ & 150 & 10 & 21.3 \\
\noalign{\smallskip}\hline\noalign{\smallskip}
\multirow{6}{*}{PLCK\,G147.3--16.6} 
& \multirow{2}{*}{144}  & $7.6\times5.9$ &  80 & -- & 152.8 \\
& & $17.0\times17.0$ & 150 & 10 & 315.6 \\
& \multirow{2}{*}{400}  & $8.0\times4.1$ &  --  & -- & 22.3 \\
& & $17.0\times17.0$ & 150 & 10 & 58.1 \\
& \multirow{2}{*}{650}  & $3.4\times3.0$ & -- & -- & 7.3 \\
& & $17.0\times17.0$ & 150 & 10 & 25.5 \\
\noalign{\smallskip}\hline\noalign{\smallskip}
\multirow{4}{*}{PSZ2\,G147.88+53.24} 
& \multirow{2}{*}{144}  & $8.6\times4.6$ & 80 & -- & 61.7 \\
& & $14.0\times14.0$ & 100 & 6 & 123.4 \\
& \multirow{2}{*}{650}  & $5.6\times2.6$ & -- & -- & 8.5 \\
& & $14.0\times14.0$ & 100 & 6 & 19.5 \\
\noalign{\smallskip}\hline\noalign{\smallskip}
\multirow{4}{*}{PSZ2\,G160.83+81.66} 
& \multirow{2}{*}{144}  & $13.1\times4.6$ & 80 & -- & 138.2 \\
& & $21.0\times21.0$ & 80 & 15 & 221.3 \\
& \multirow{2}{*}{650}  & $4.5\times2.6$ & -- & -- & 9.2 \\
& & $21.0\times21.0$ & -- & 15 & 22.7 \\
\noalign{\smallskip}\hline
\end{tabular}%}
\end{center}
\label{tab:images_parameters}
\end{table*}

\section{Observations and data reduction}

\subsection{LOFAR}
We use the same dataset presented in \cite{digennaro+20}. The sample was observed together with LoTSS \citep{shimwell+17,shimwell+19}, which consists of 8 hours of observation for each pointing (see Table \ref{tab:obsid_info}). We performed standard LOFAR data reduction, which includes direction-independent and direction-dependent calibration and imaging of the full LOFAR field of view using \texttt{prefactor} \citep{vanweeren+16,williams+16,degasperin+19}, \texttt{killMS} \citep{tasse14,smirnov+tasse15}, and \texttt{DDFacet} \citep{tasse+18,tasse+20}. Additionally, we performed extra phase and amplitude self-calibration loops using the products of the pipeline\footnote{\url{https://github.com/mhardcastle/ddf-pipeline}} and subtracting all of the sources outside of a region of $15'\times15'$ surrounding the target in order to improve the quality of the calibration, \citep{vanweeren+20}. Final imaging was done with \texttt{WSClean~v2.10} \citep{offringa+14,offringa+17} with the wideband deconvolution mode (\texttt{channelsout=6}). The images have a central frequency of 144~MHz. The systematic uncertainty due to residual amplitude errors is set to 15\% \citep{shimwell+19}.

\subsection{uGMRT}
The clusters presented in this work have been observed with the uGMRT in Band 3 (250--550 MHz) and/or Band 4 (550--900 MHz). The total length of the observation is 6 hours, except for PSZ2\,G091.83+26.11 and PSZ2\,G141.77+14.19 which have been observed for 5 hours (see Table \ref{tab:obsid_info}). Data were recorded in 2048 frequency channels with an integration time of 4 s in full Stokes mode. We used 3C286, 3C147, and 3C48 as primary calibrators, depending on the target. To process the data, we ran the Source Peeling and Atmospheric Modeling \citep[\texttt{SPAM,}][]{intema+09} on six sub-bands of 33.3 MHz bandwidth each for Band 3, and on four sub-bands of 50.0 MHz bandwidth each for Band 4. For each band, the sub-bands are then imaged together using \texttt{WSClean~v2.10} at the common frequency of 400 and 650 MHz, for Band 3 and 4, respectively. The systematic uncertainties due to residual amplitude errors are set to 8\% and 5\% for the observations in Bands 3 and 4 respectively \citep{chandra+04}.

\subsection{Imaging and integrated flux densities}\label{sec:images}
For all clusters, we produced the final deep images using \texttt{WSClean~v2.10}, with \texttt{weighting=`Briggs'} and \texttt{robust=-0.5}. For the LOFAR images, we applied an inner $uv$-cut of $80\lambda$ to remove the Galactic emission. 
The final noise levels were found in the ranges $\rm \sim50-150~\mu Jy~beam^{-1}$, $\rm \sim20-60~\mu Jy~beam^{-1}$ , and $\rm \sim7-12~\mu Jy~beam^{-1}$ for the 144, 400, and 650 MHz full-resolution images, respectively (see Table \ref{tab:images_parameters}). This means that the 650 MHz observations are about three times deeper than the LOFAR ones for detecting compact sources, that is, assuming a typical spectral index $\alpha=-0.8$. For steep-spectra sources, that is, $\alpha=-1.3,$ the sensitivities of the two arrays are similar.
The low-resolution images for both the LOFAR and uGMRT observations were produced by  applying a Gaussian taper at different resolutions to downweight the visibilities from longer baselines. We display all the images in Figures \ref{fig:images_086} to \ref{fig:images_160}. Among the nine clusters presented in this work, five of them show extended diffuse radio emission in the uGMRT observations.

To obtain the flux densities of the radio halos, we produced source-subtracted images. We applied a $uv$-cut to the data to filter out emission associated with sources of linear sizes larger than 500 kpc at the cluster redshift, and to create a clean component model of the compact sources. Given the sizes of the halo, for the LOFAR image of PSZ2\,G089.39+69.36 we employed an inner $uv$-cut of 400~kpc.
During this step, we employed multiscale deconvolution using scales of $\rm [0, 4, 8, 16] \times pixelscale$ (with the pixel size of $1.5''$, $2''$ and $1''$ for the 144, 400 and 650~MHz images, respectively) to include and subtract the diffuse emission from the radio galaxies. For the automatic deconvolution, we used a mask threshold of $1\sigma_{\rm rms}$ to subtract the faintest contaminating sources. Finally, we subtracted the compact source models from the visibilities, and tapered the $uv$-data with different Gaussian tapers (i.e. $6''$, $10''$ or $15''$). 
In case of an extended radio galaxy with a linear size $\gtrsim 500$~kpc, we cannot properly subtract the radio emission from the $uv$-data. This is the case for the candidate radio relic in PSZ2\,G091.83+26.11 and for the radio galaxy northward of  PLCK\,G147.3--16.6, which have been manually excluded from the radio halo region \citep{digennaro+20}.

We measure the radio halo flux densities for each cluster from same regions in the LOFAR and uGMRT images, encompassing the full extent of the diffuse radio emission (see Appendix \ref{apx:sub_images}).
Uncertainties on the halo flux densities are given by: 
\begin{equation}\label{eq:sb_err}
\Delta S_\nu = \sqrt{(f S_\nu)^2 + N_{\rm beam}\sigma_{\rm rms}^2 + \sigma_{\rm sub}^2} \, ,
\end{equation}
where $f$ are the systematic uncertainties due to residual amplitude errors, $\sigma_{\rm rms}$ is the map noise level, $N_{\rm beam}$ is the number of beams covering the halo region, and $\sigma_{\rm sub}$ is the uncertainty of the source subtraction in the $uv$ plane {\citep[i.e. a few percent of the residual flux from compact sources; see, e.g.,][]{cassano+13,vanweeren+20}}. The resulting flux densities are listed in Table \ref{tab:fluxes}.

\subsection{Radio halo injection and upper limits}
In those systems with no detection of diffuse emission in the uGMRT images, we derived upper limits injecting mock radio halos directly in the original visibilities \citep[e.g.][]{venturi+07,venturi+08,kale+15,cuciti+21a}. Following \cite{bonafede+17}, we modelled the radio halo brightness with a symmetric exponential profile, $I(r)=I_0\exp(-r/r_e)$, {and including power spectrum fluctuations of the type $P(\Lambda)\propto\Lambda^n$ \citep[with $\Lambda$ being the spatial scale of the fluctuations and $n=11/3$; see also][]{govoni+05,govoni+06,bonafede+09}.}
{In the exponential model,} $I_0$ is the central brightness and $r_e$ the $e$-folding radius which we assumed to be one-third of the halo radius, $R_{H}$\footnote{We define $R_{H}$ as half of the largest linear size in the LOFAR image (see Table \ref{tab:fluxes})}, \citep[][]{murgia+09}. We injected the mock halo in a region close to the cluster\footnote{The original image without compact sources is displayed in Appendix \ref{apx:sub_images}.}that is free of compact sources in order to avoid the inclusion of possible halo residuals which would bias the upper limits to lower levels \citep{bonafede+17,osinga+20,cuciti+21a}. 
We assumed different $I_0$ levels, starting from $\sigma_{\rm rms}\sqrt{N_{\rm beam}}$ and increasing this value until positive residuals were visible in the image. We define these positive residuals as an upper limit when the largest linear scale above the $2\sigma_{\rm rms}$ level is about $2r_e$ \citep{cuciti+21a}. We measured the flux density directly from the image, corresponding to $\sim80-90\%$ of the injected flux. An example of the injection procedure is displayed in Fig. \ref{fig:mock}. All the upper limits on the flux densities are reported in Table \ref{tab:fluxes}.

\section{Results}
\subsection{Spectral index maps and integrated spectral indices}\label{sec:spix}
To produce the spectral index maps of those clusters with diffuse radio emission in both LOFAR and uGMRT observations, we made images with a common inner $uv$-cut to compensate for the different interferometer $uv$-coverage. 
To emphasise the presence of the radio halo, we also applied a Gaussian taper.  
The images were then convolved to the same resolution, and re-gridded to the same pixel grid (i.e. the LOFAR image). The effective final resolutions and the noise levels of each image are listed in Table \ref{tab:images_parameters}. 
For the clusters with observations at three frequencies, we used the same procedure used in \cite{digennaro+18}, where a second-order polynomial fit was used in case of significant curvature (i.e. above the $2\sigma$ threshold, where $\sigma$ is the uncertainty associated with the second-order term). In this case, the spectral index was calculated at 400 MHz, which is the median of the total band. We blanked all the pixels below the $2\sigma_{\rm rms}$ threshold for each frequency, with $\sigma_{\rm rms}$ being the noise level reported in Table \ref{tab:images_parameters}. The spectral index uncertainty maps are obtained via 150 Monte Carlo simulations of the first- and second-order polynomial fit. We assumed the uncertainty of each flux to be given by the sum in quadrature of the noise map and the systematic flux uncertainties, that is $\Delta S_\nu=\sqrt{(f S_\nu)^2 + \sigma_{\rm rms}^2}$. For the clusters with observations at only two frequencies, we calculate the spectral index analytically, with an uncertainty of: 

\begin{equation}\label{eq:err_spix}
\Delta\alpha=\frac{1}{\ln \frac{\nu_1}{\nu_2}}\sqrt{ \left ( \frac{\Delta S_1}{S_1} \right )^2+ \left ( \frac{\Delta S_2}{S_2} \right )^2} \, .
\end{equation}

\begin{figure*}
\centering
{\includegraphics[width=0.45\textwidth]{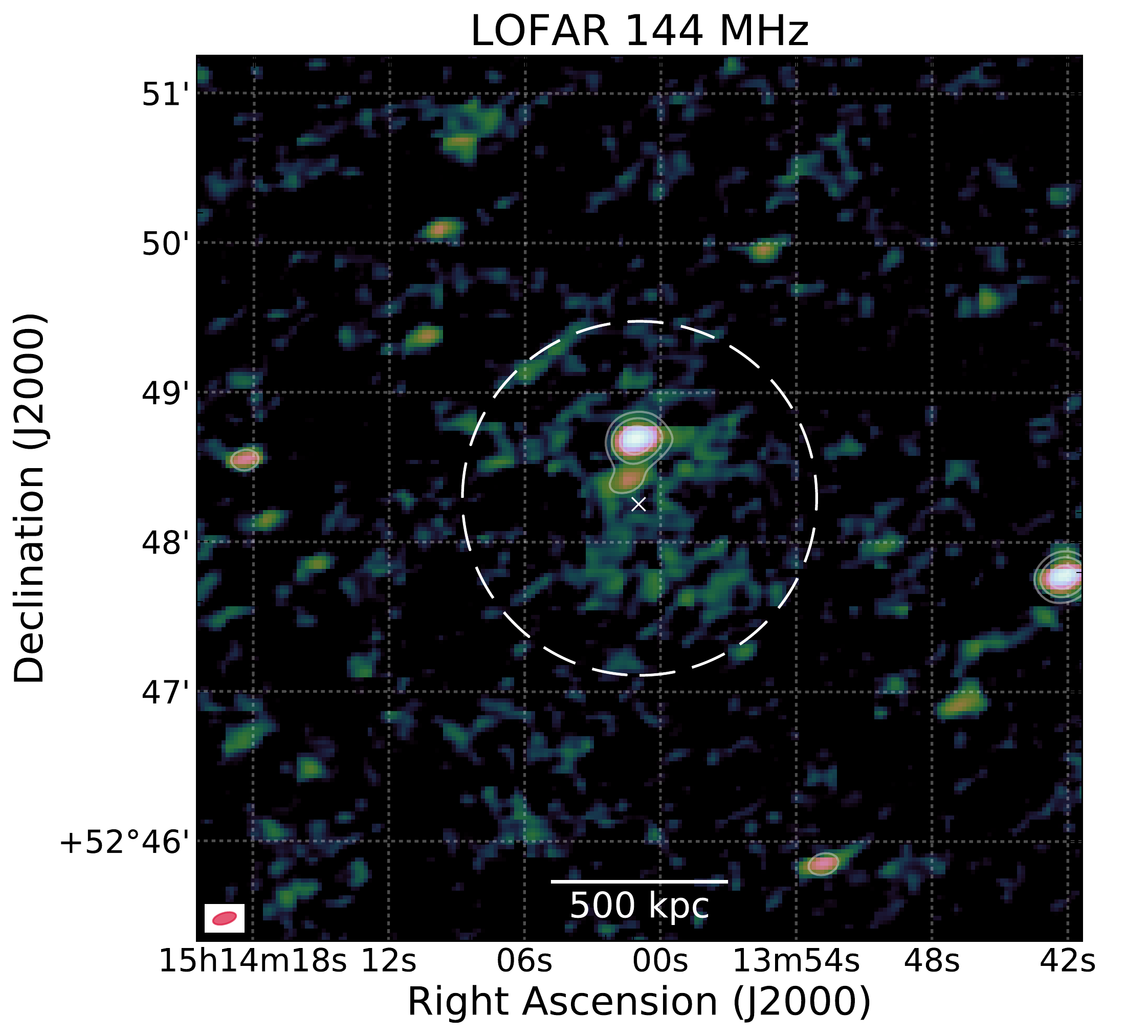}}
{\includegraphics[width=0.45\textwidth]{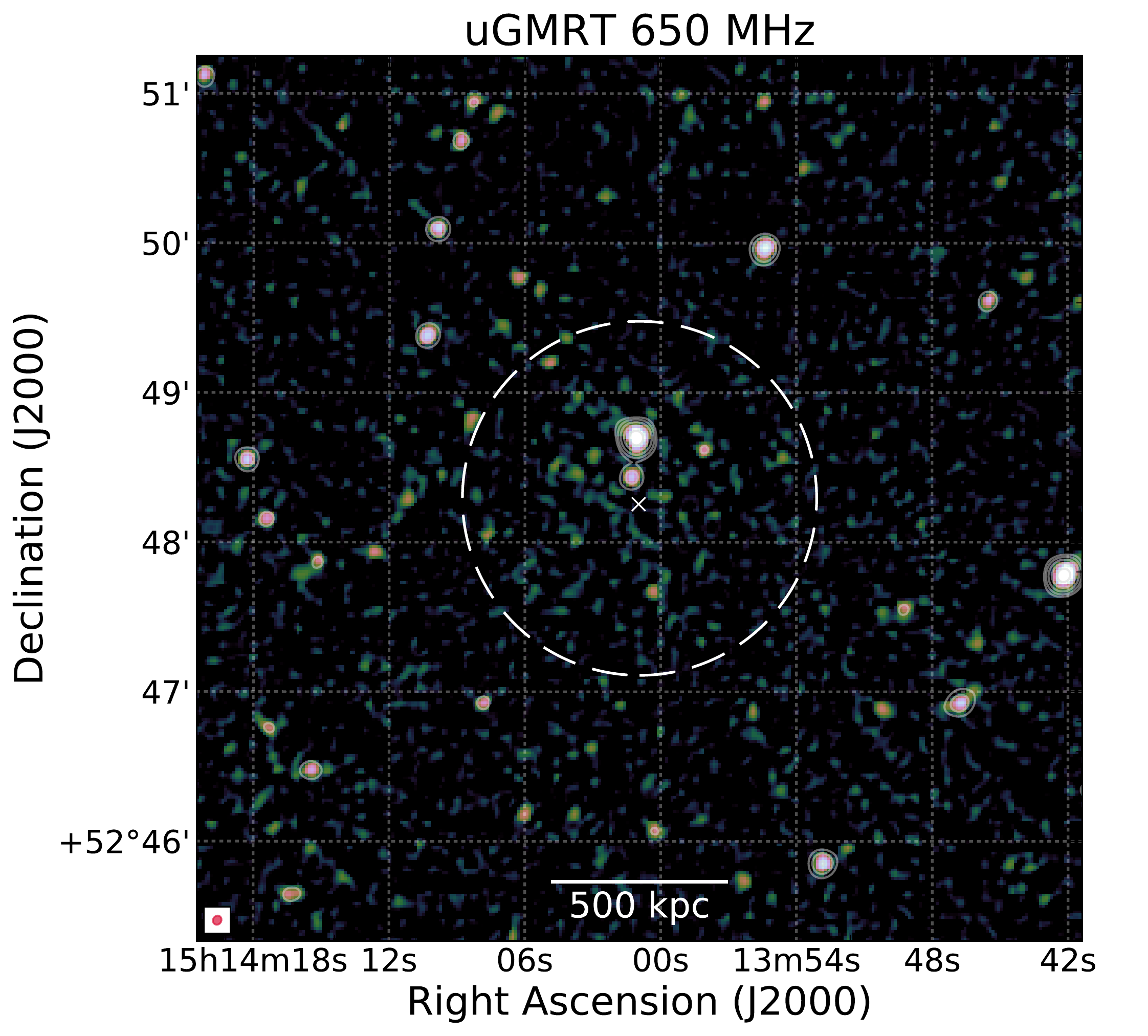}}\\
{\includegraphics[width=0.45\textwidth]{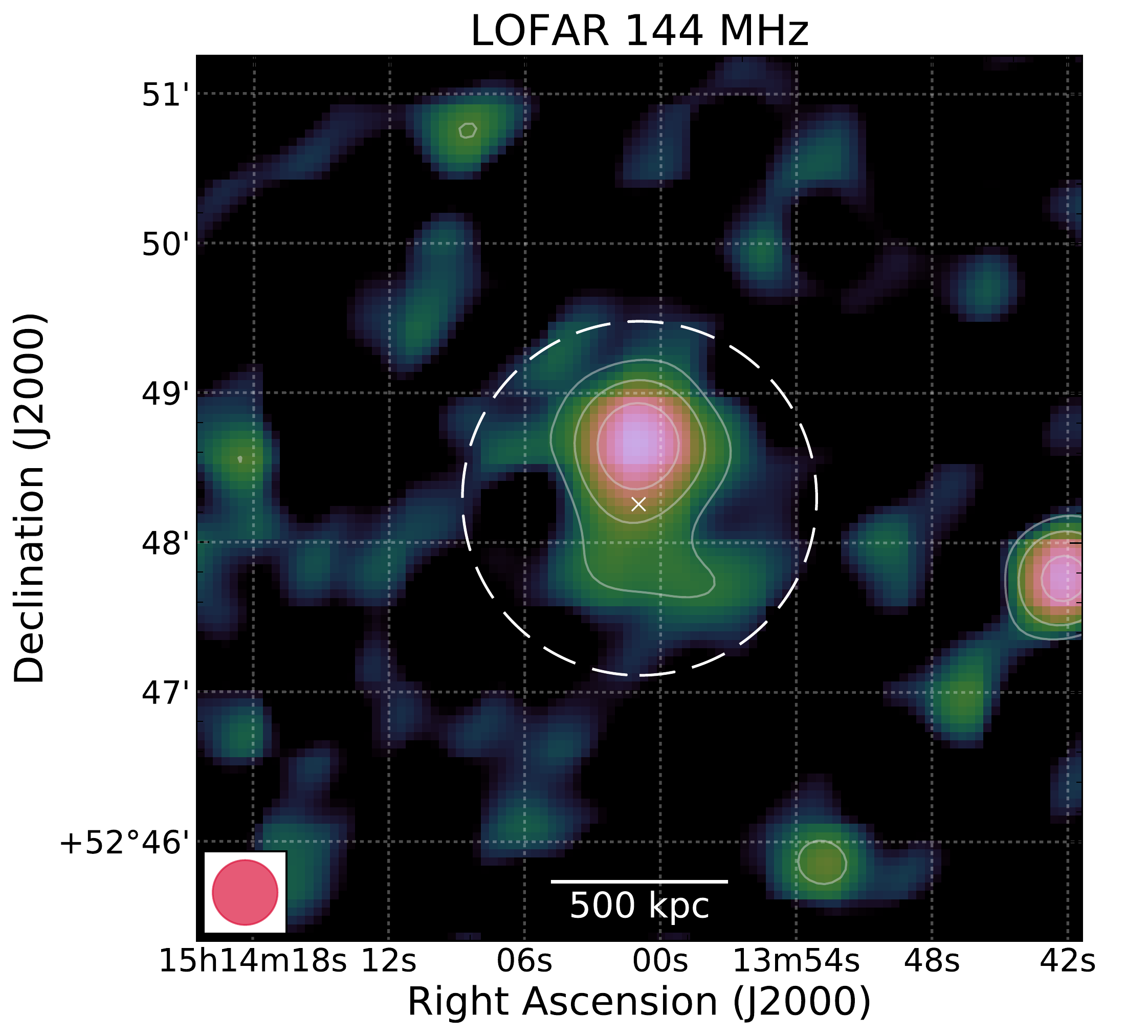}}
{\includegraphics[width=0.45\textwidth]{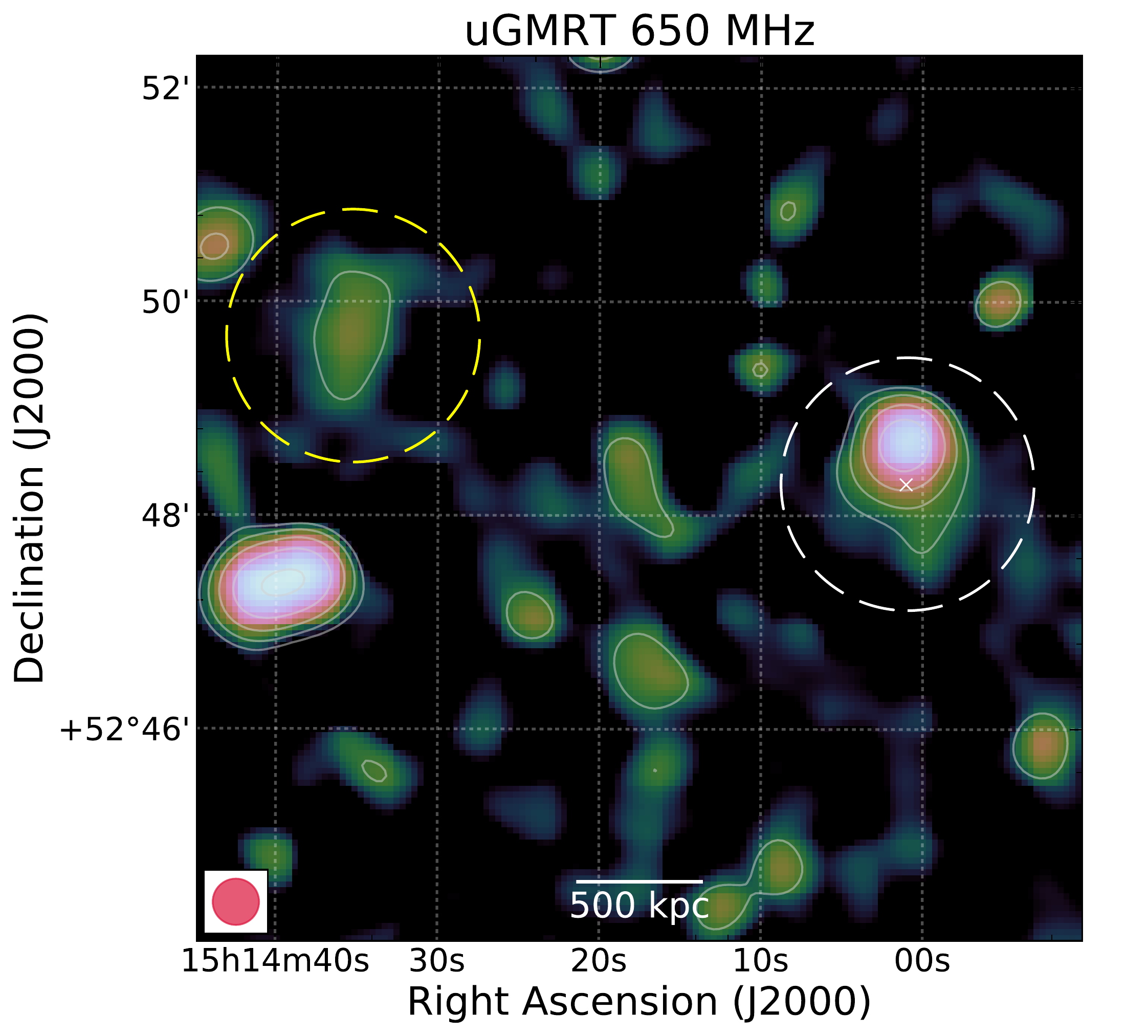}}
\caption{PSZ2\,G086.93+53.18. Top and bottom rows: Full-resolution and $26''$ images (\texttt{weighting=`Briggs'} and \texttt{robust=-0.5}) at 144 MHz (left) and 650 MHz (right). White-coloured radio contours are drawn at levels of $2.5\sigma_{\rm rms}\times[-1, 1, 2, 4, 8, 16, 32]$, with $\sigma_{\rm rms}$ being the noise level at each frequency (see Table \ref{tab:images_parameters}). The negative contour level is drawn with a dashed white line. The dashed white circle in each map shows the $R=0.5R_{\rm SZ,500}$ region obtained from $M_{\rm SZ,500}$, {with the cross showing the cluster centre}. The dashed yellow circle in the bottom right panels shows the position of the mock radio halo.}\label{fig:images_086}
\end{figure*}

\begin{figure*}
\centering
{\includegraphics[width=0.45\textwidth]{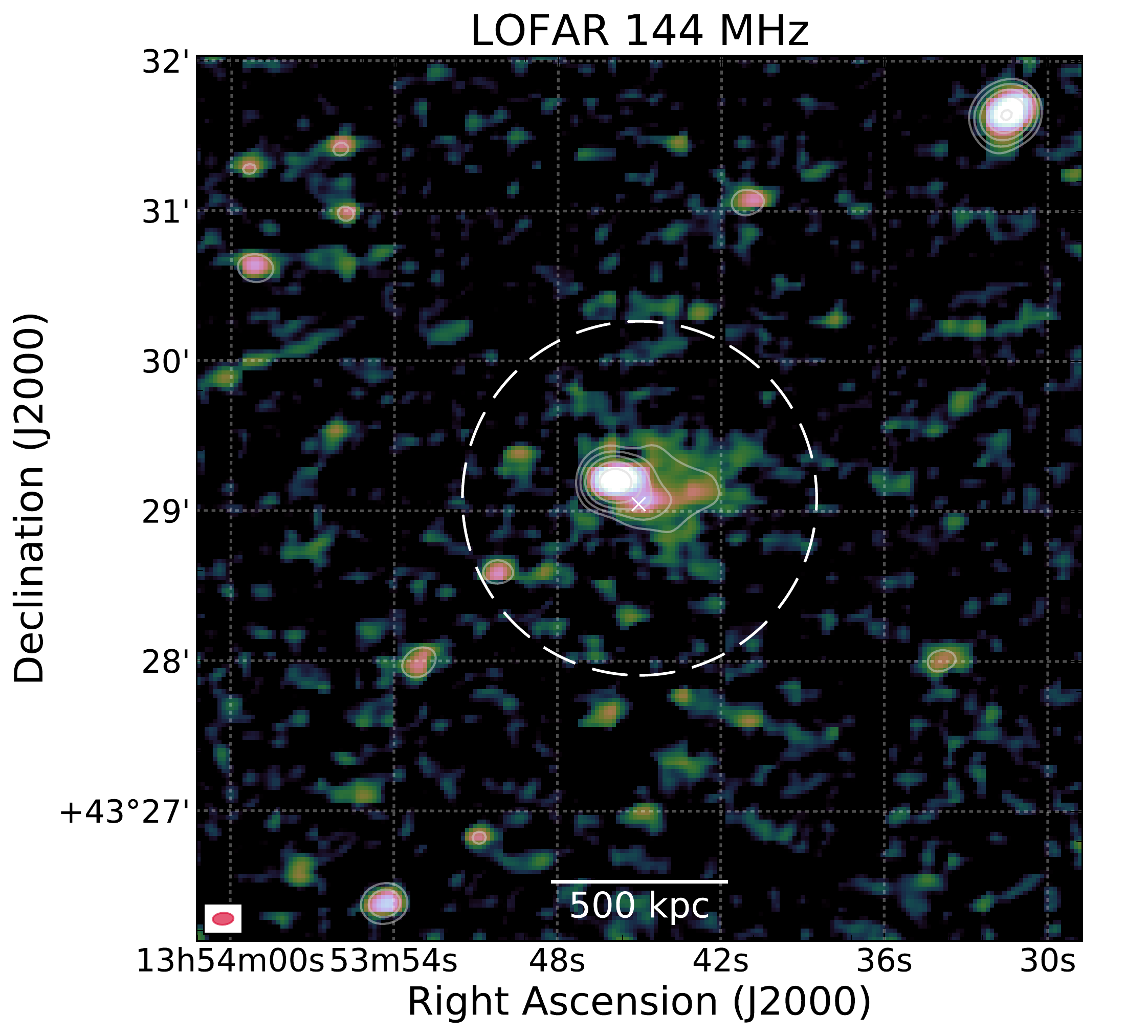}}
{\includegraphics[width=0.45\textwidth]{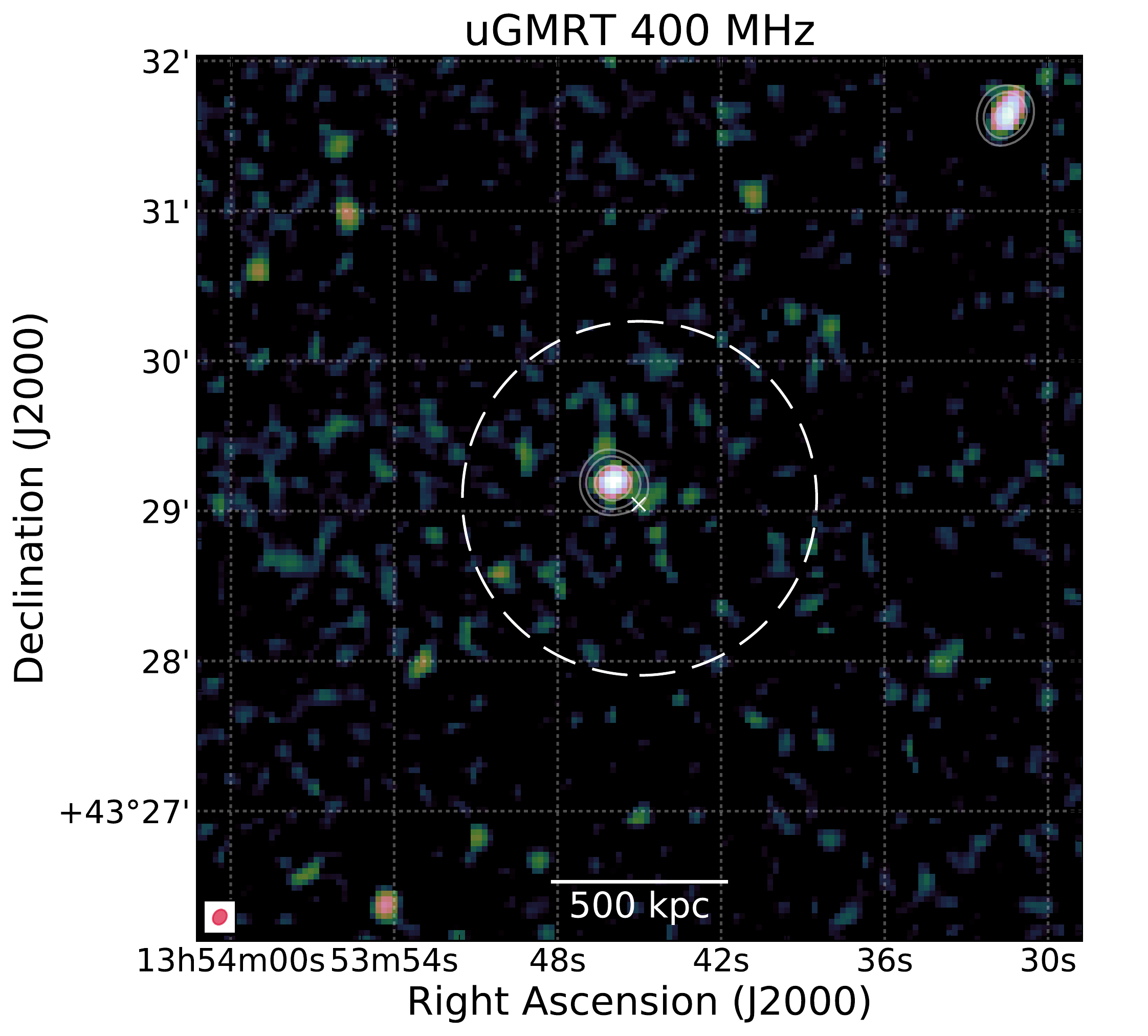}}\\
{\includegraphics[width=0.45\textwidth]{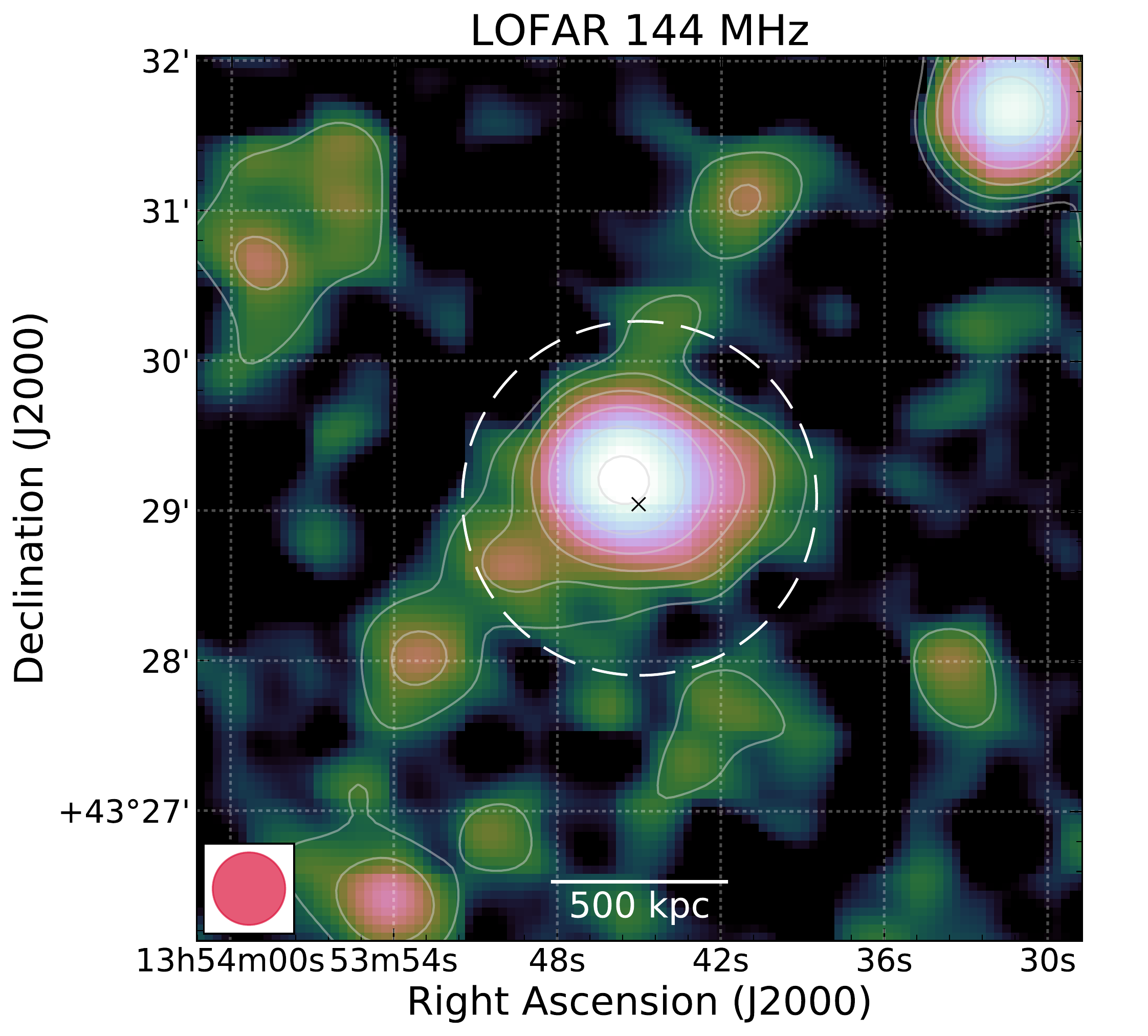}}
{\includegraphics[width=0.45\textwidth]{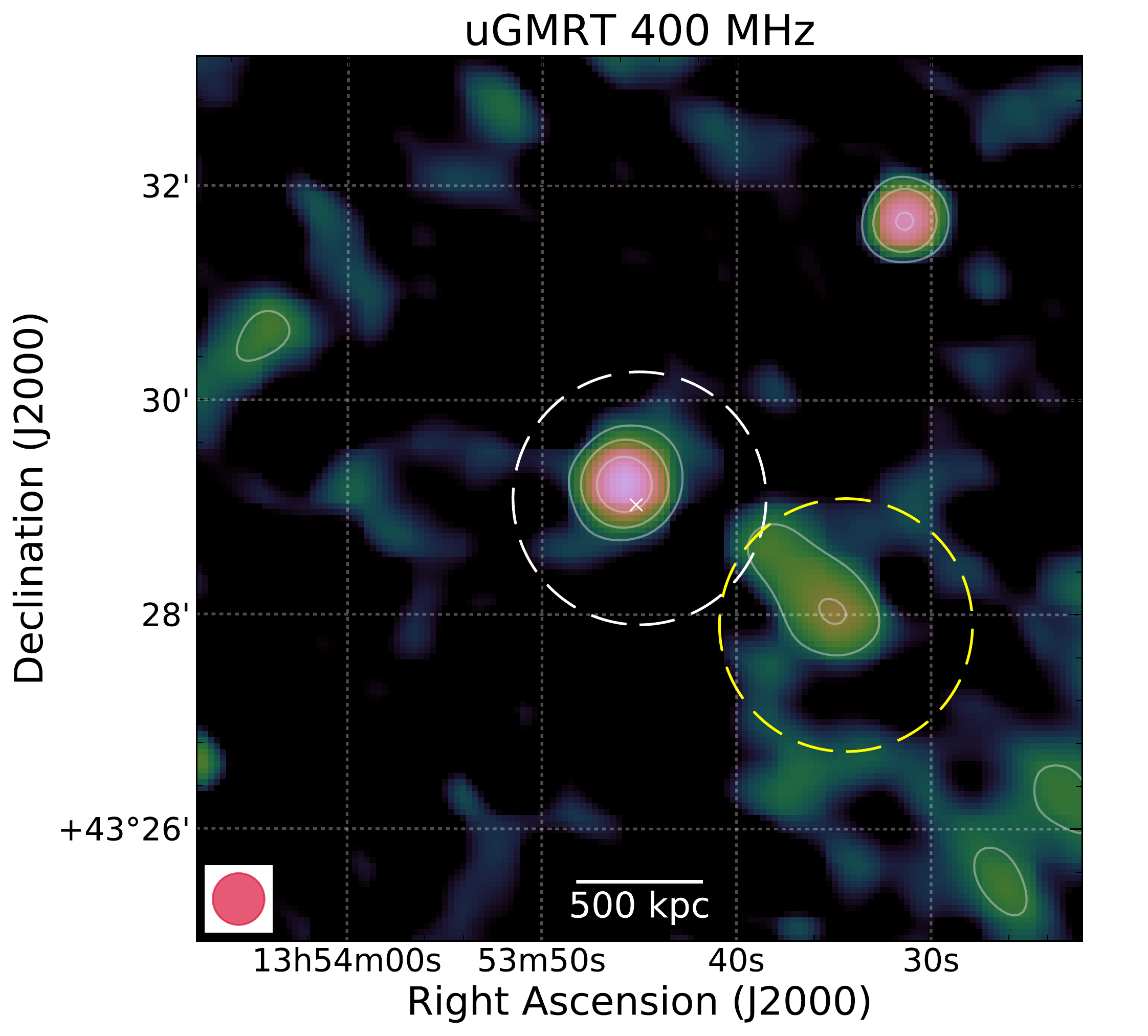}}
\caption{PSZ2\,G089.39+69.36. Top and bottom rows: Full-resolution and $29''$ images (\texttt{weighting=`Briggs'} and \texttt{robust=-0.5}) at 144 MHz (left) and 400 MHz (right). White-coloured radio contours are drawn at levels of $2.5\sigma_{\rm rms}\times[-1, 1, 2, 4, 8, 16, 32]$, with $\sigma_{\rm rms}$ being the noise level at each frequency (see Table \ref{tab:images_parameters}). The negative contour level is drawn with a dashed white line. The dashed white circle in each map shows the $R=0.5R_{\rm SZ,500}$ region obtained from $M_{\rm SZ,500}$, {with the cross showing the cluster centre}. The dashed yellow circle in the bottom right panels shows the position of the mock radio halo.}\label{fig:images_089}
\end{figure*}

\begin{figure*}
\centering
{\includegraphics[width=0.3\textwidth]{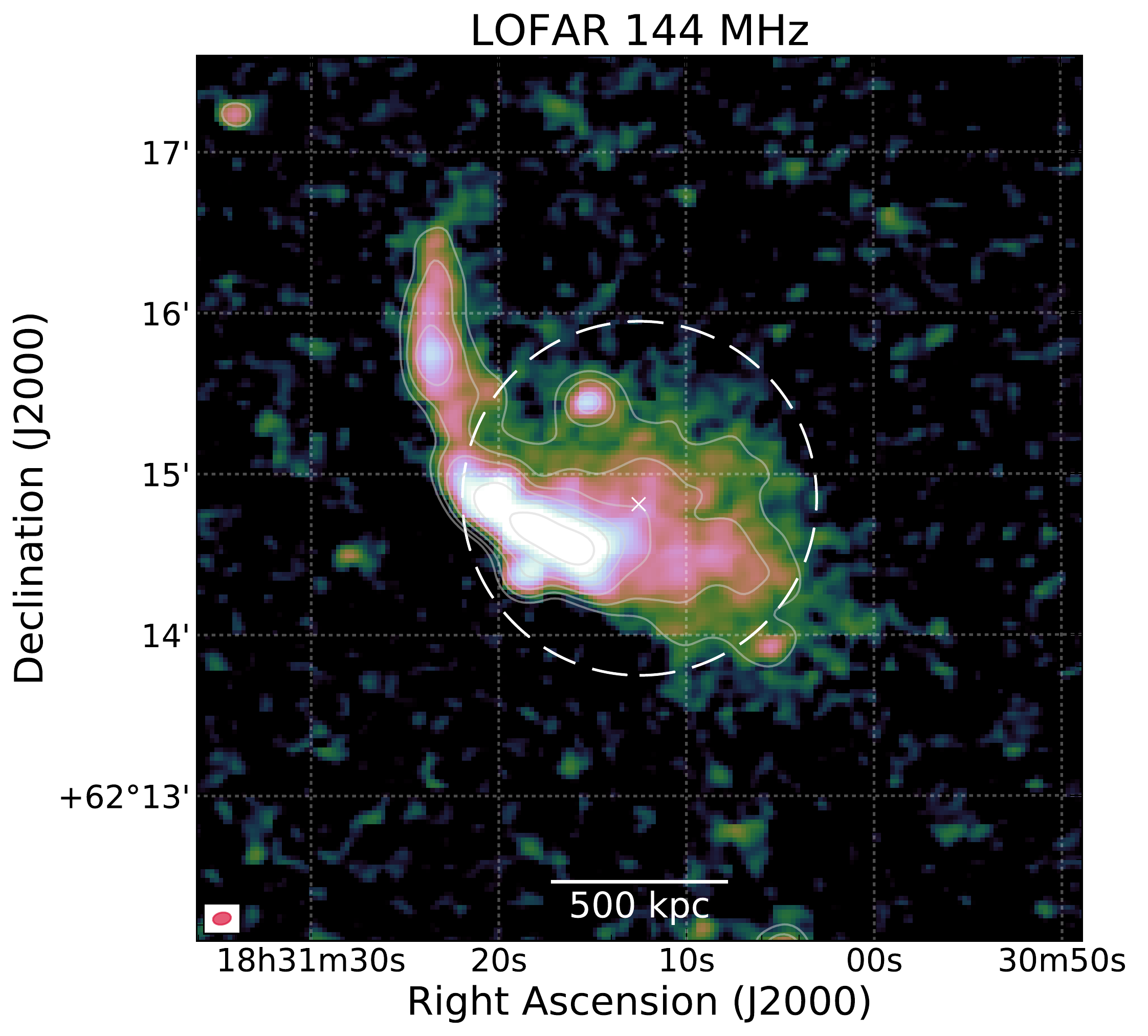}}
{\includegraphics[width=0.3\textwidth]{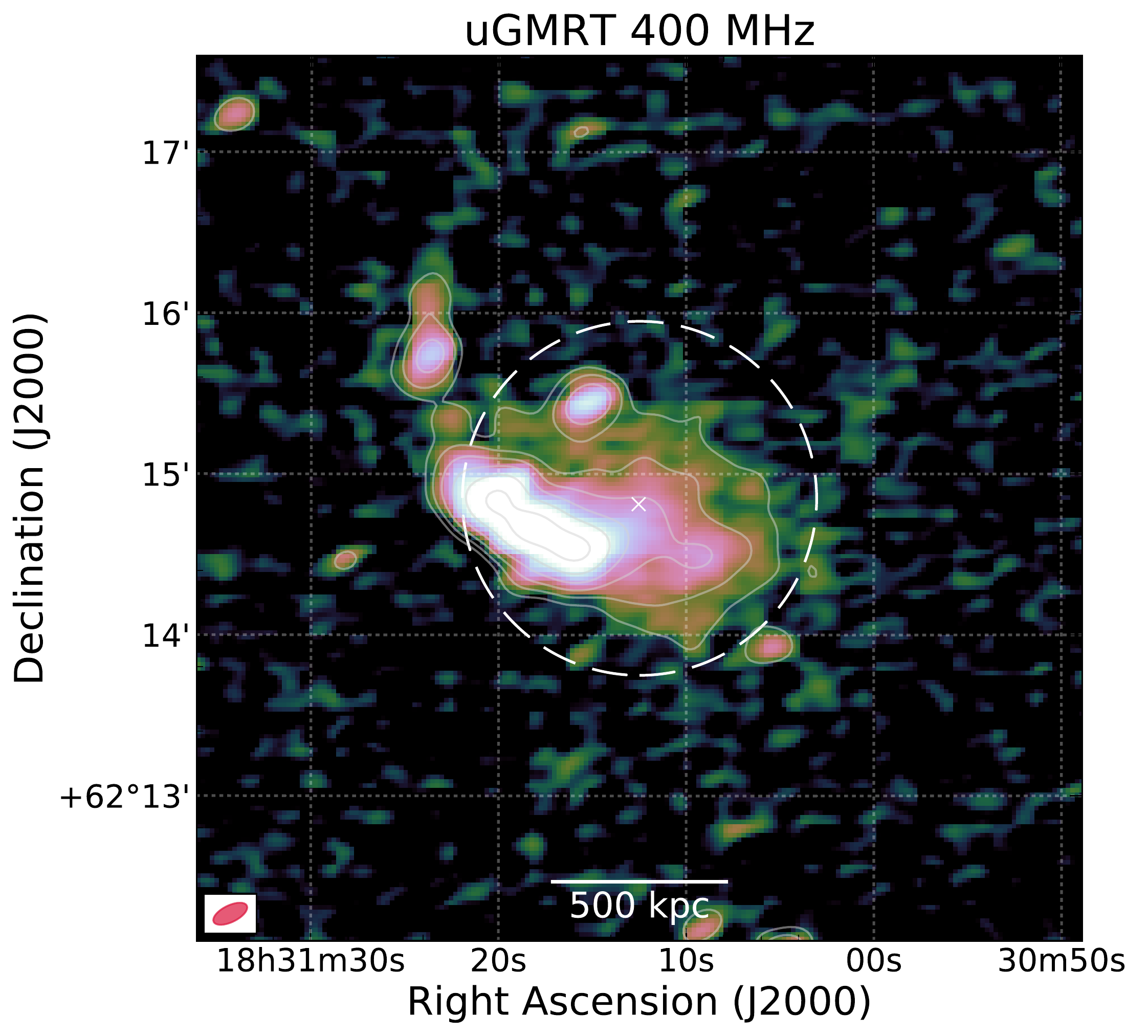}}
{\includegraphics[width=0.3\textwidth]{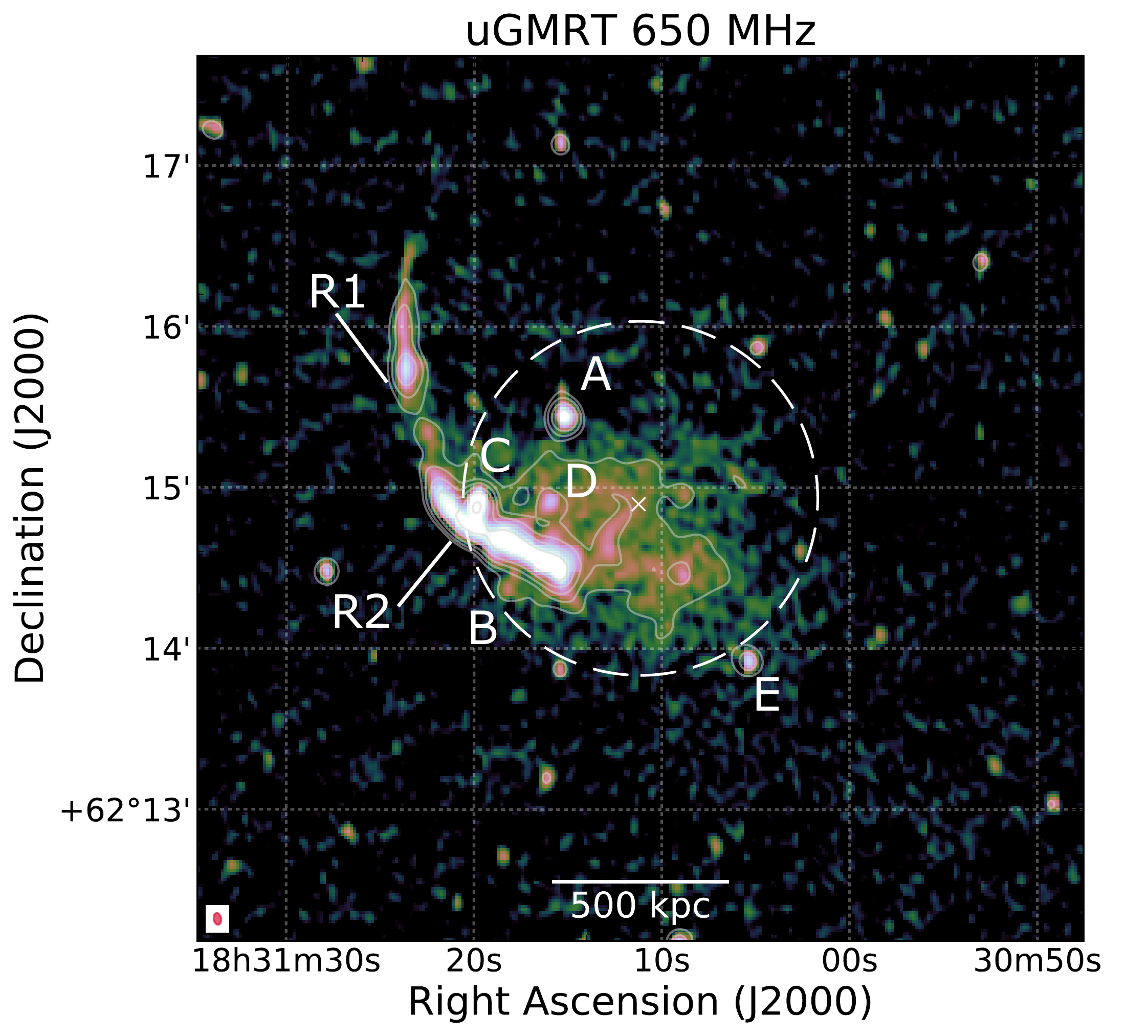}}
{\includegraphics[width=0.3\textwidth]{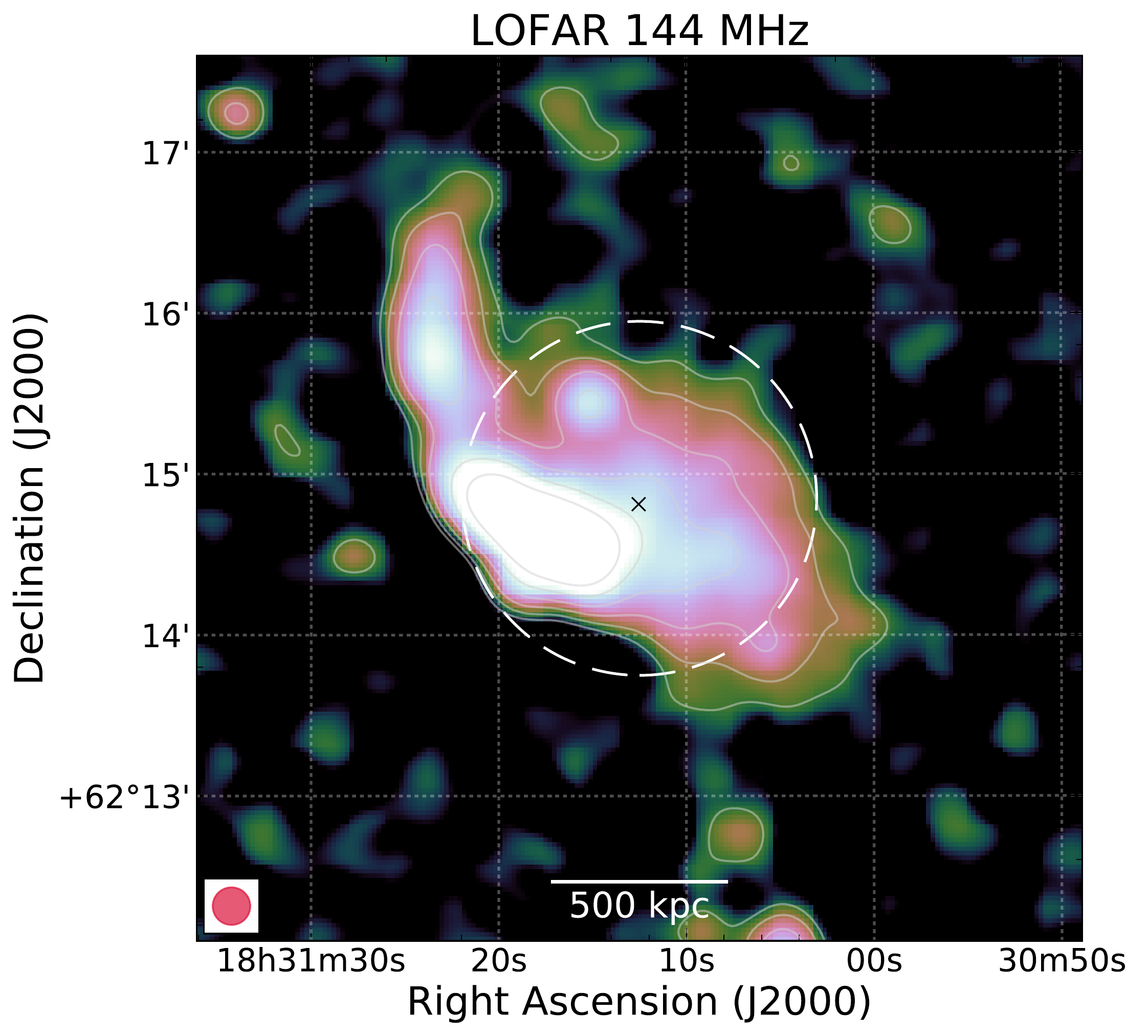}}
{\includegraphics[width=0.3\textwidth]{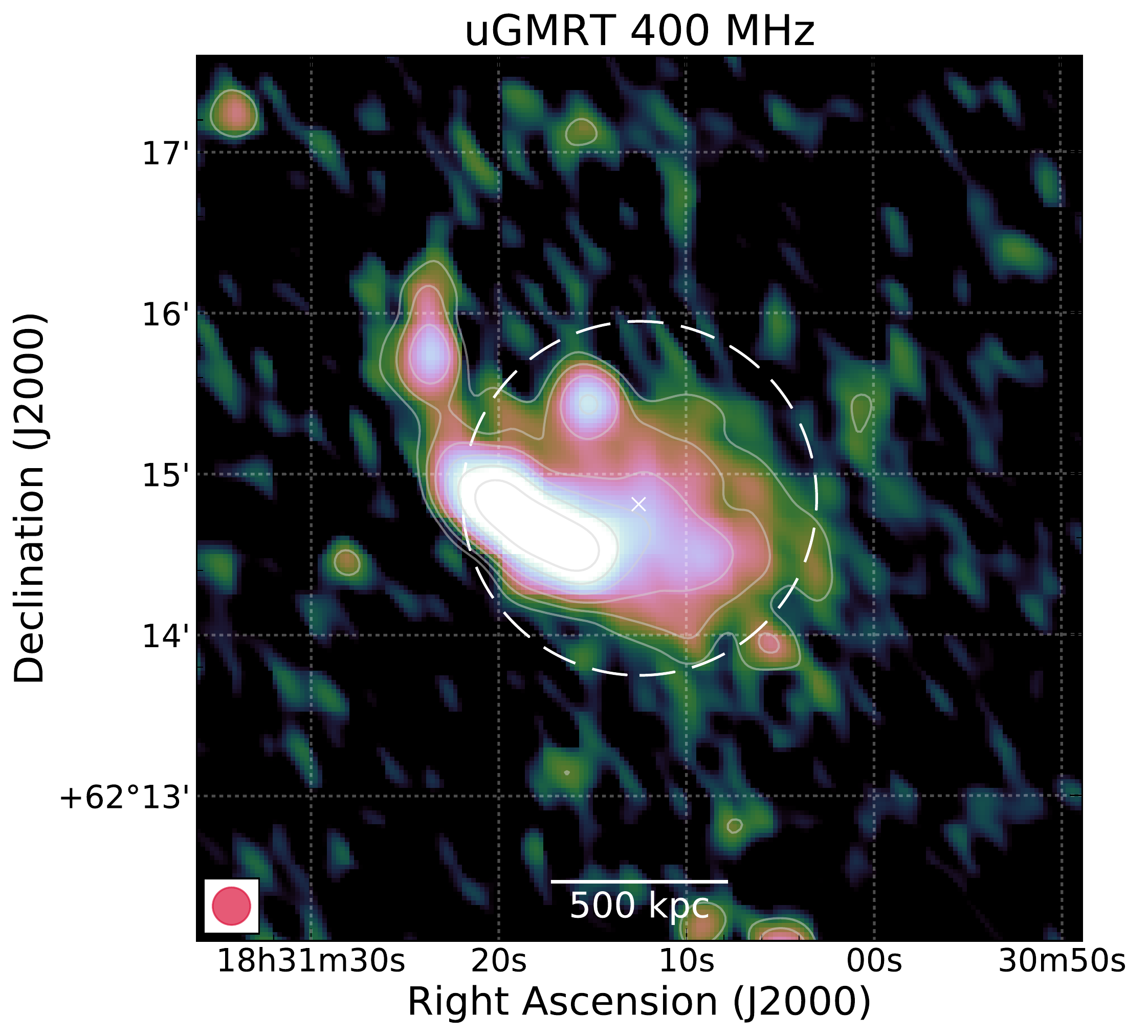}}
{\includegraphics[width=0.3\textwidth]{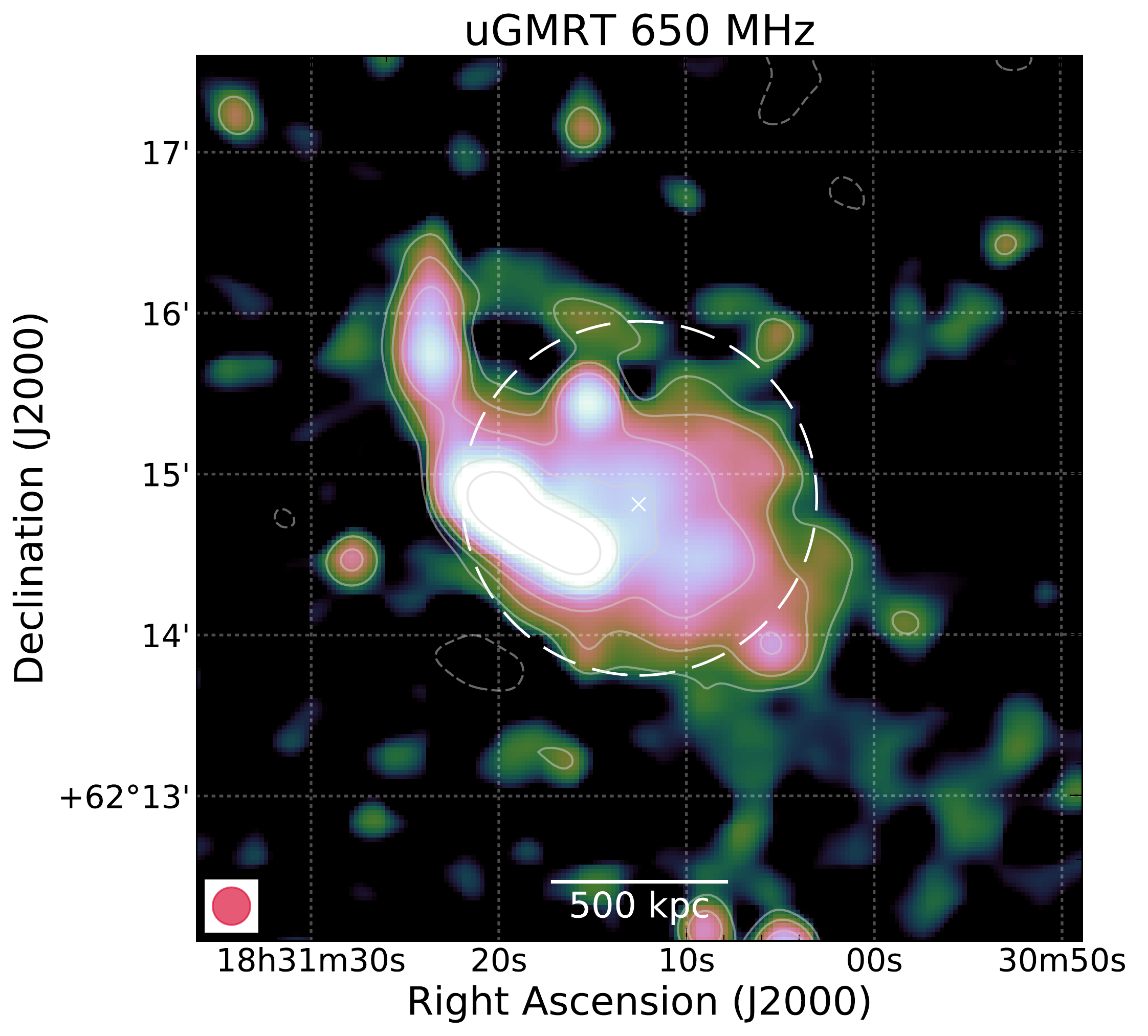}}
{\includegraphics[width=0.45\textwidth]{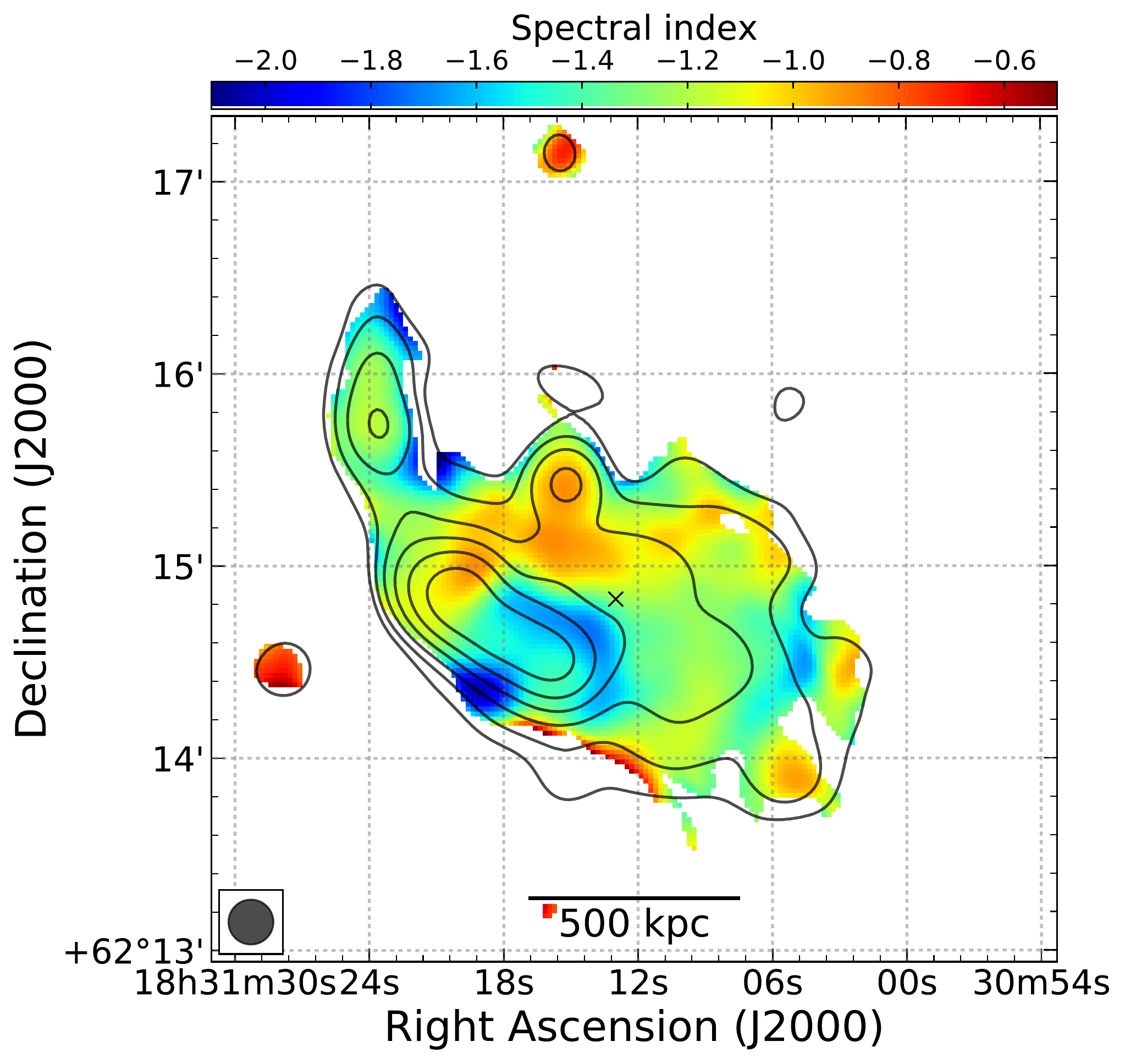}}
{\includegraphics[width=0.45\textwidth]{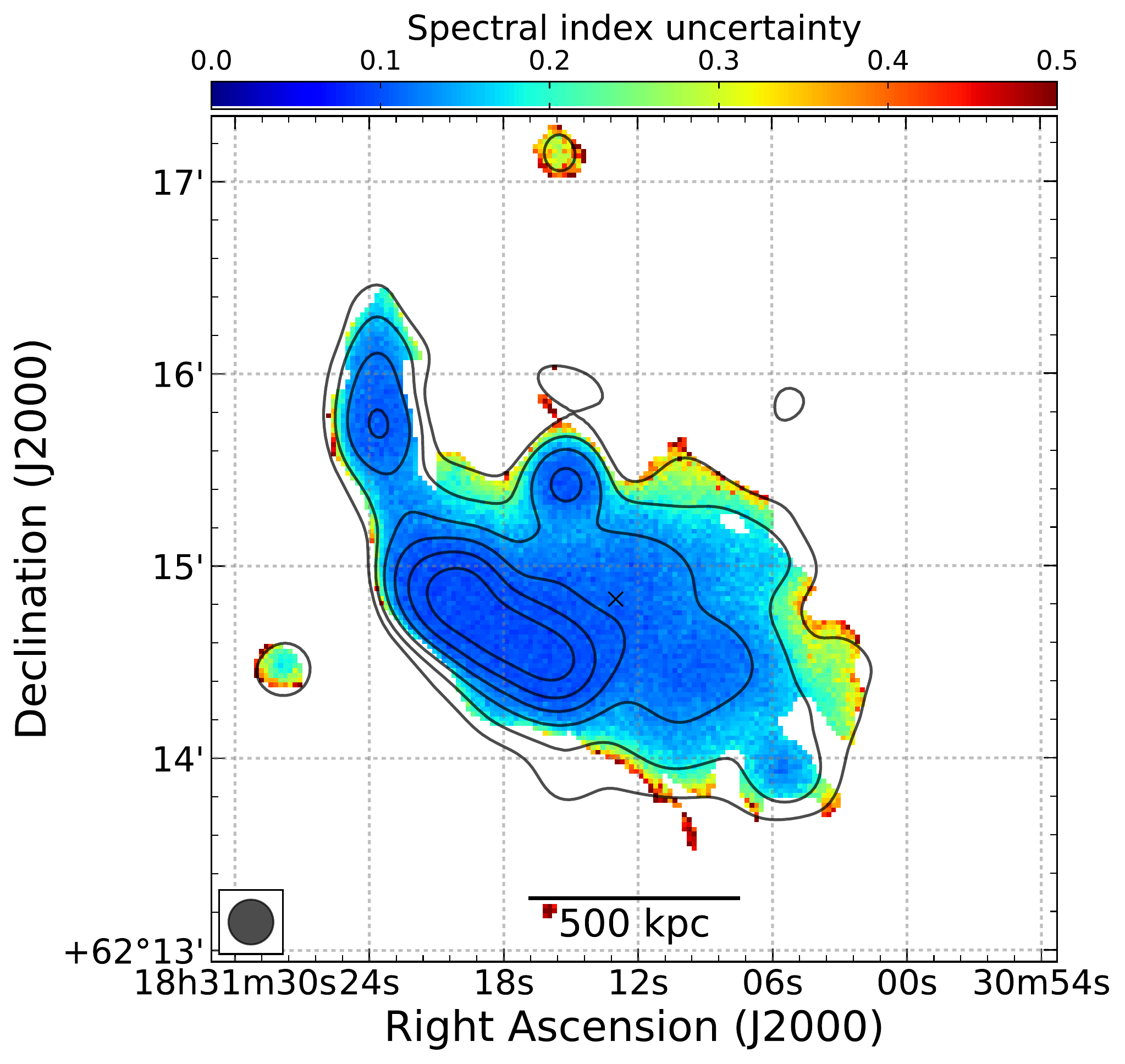}}
\caption{PSZ2\,G091.83+26.11. Top and central rows: Full-resolution and $14''$ images (\texttt{weighting=`Briggs'} and \texttt{robust=-0.5}) at 144 MHz (left), 400 MHz (middle), and 650 MHz (right). White-coloured radio contours are drawn at levels of $2.5\sigma_{\rm rms}\times[-1, 1, 2, 4, 8, 16, 32]$, with $\sigma_{\rm rms}$ the noise level at each frequency (see Table \ref{tab:images_parameters}). The negative contour level is drawn with a dashed white line. The dashed white circle in each map shows the $R=0.5R_{\rm SZ,500}$ region obtained from $M_{\rm SZ,500}$, {with the cross showing the cluster centre}. Bottom row: Spectral index map fitting at 400 MHz (see Sect. \ref{sec:spix}), at $14''$ resolution, and the corresponding uncertainty map (left and right panels, respectively). uGMRT radio contours at 650 MHz are drawn in black, at the levels $3\sigma_{\rm rms}\times[-1, 1, 2, 4, 8, 16, 32]$, with $\sigma_{\rm rms}$ being the noise level (see Table \ref{tab:images_parameters}). }\label{fig:images_091}
\end{figure*}

\begin{figure*}
\centering
{\includegraphics[width=0.43\textwidth]{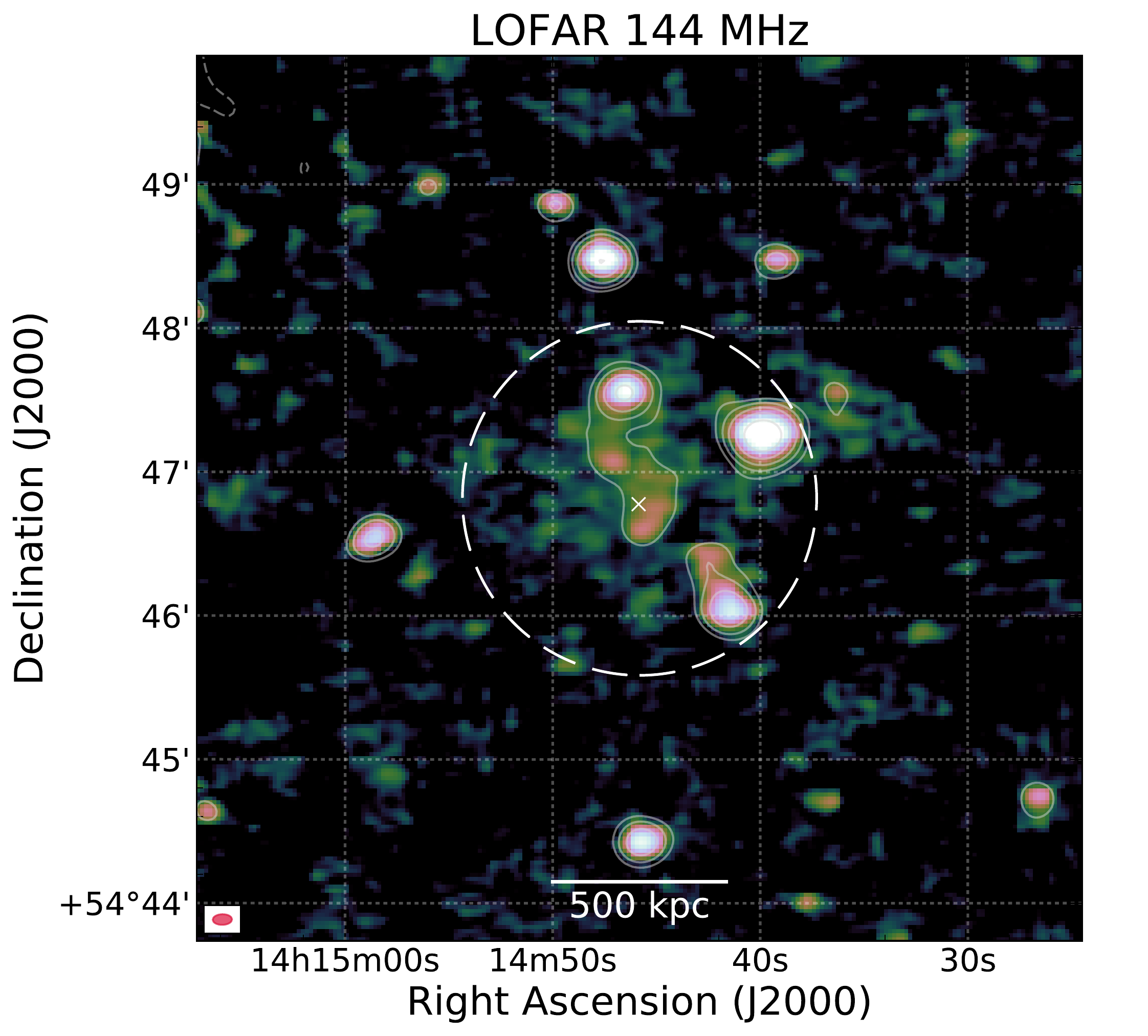}}
{\includegraphics[width=0.43\textwidth]{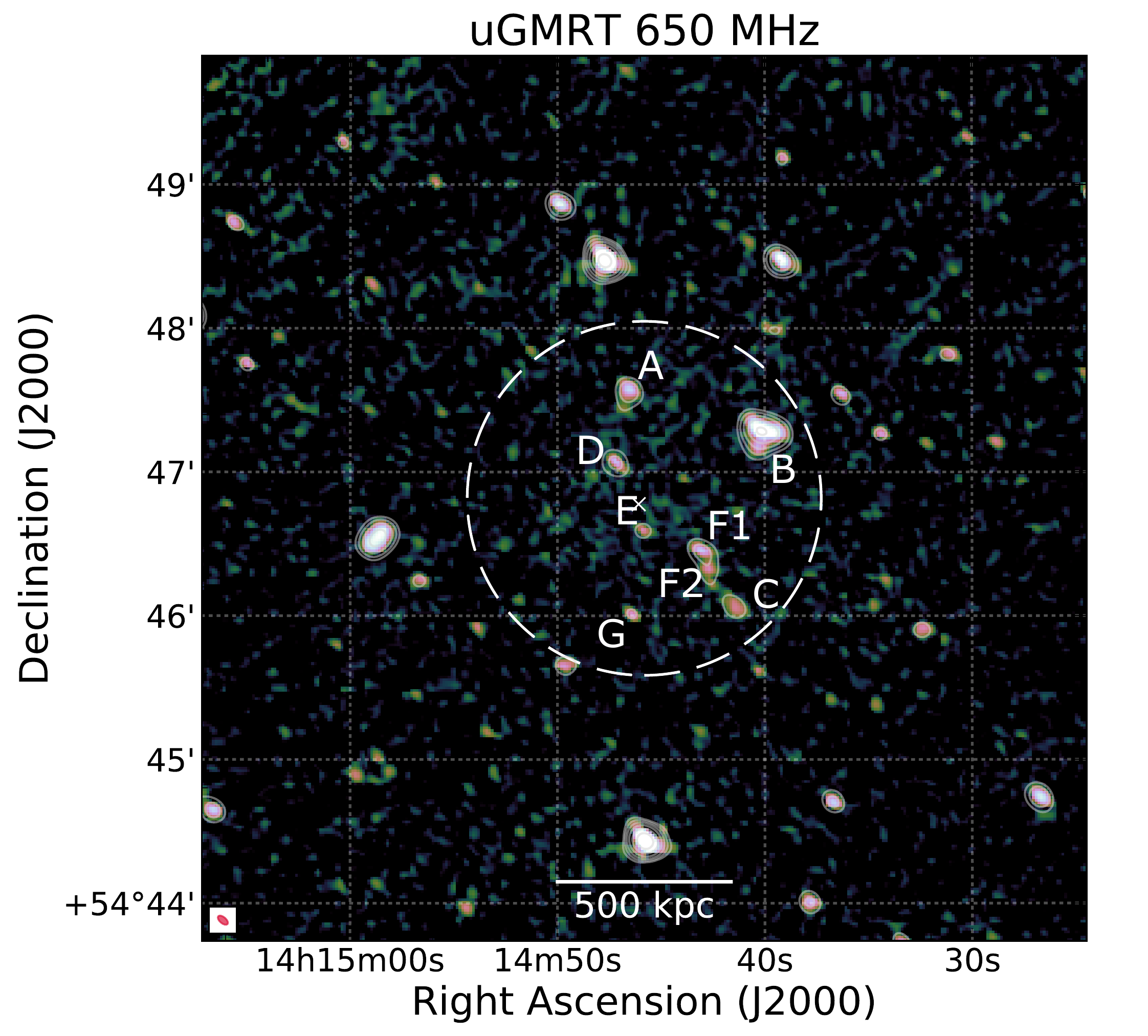}}
{\includegraphics[width=0.43\textwidth]{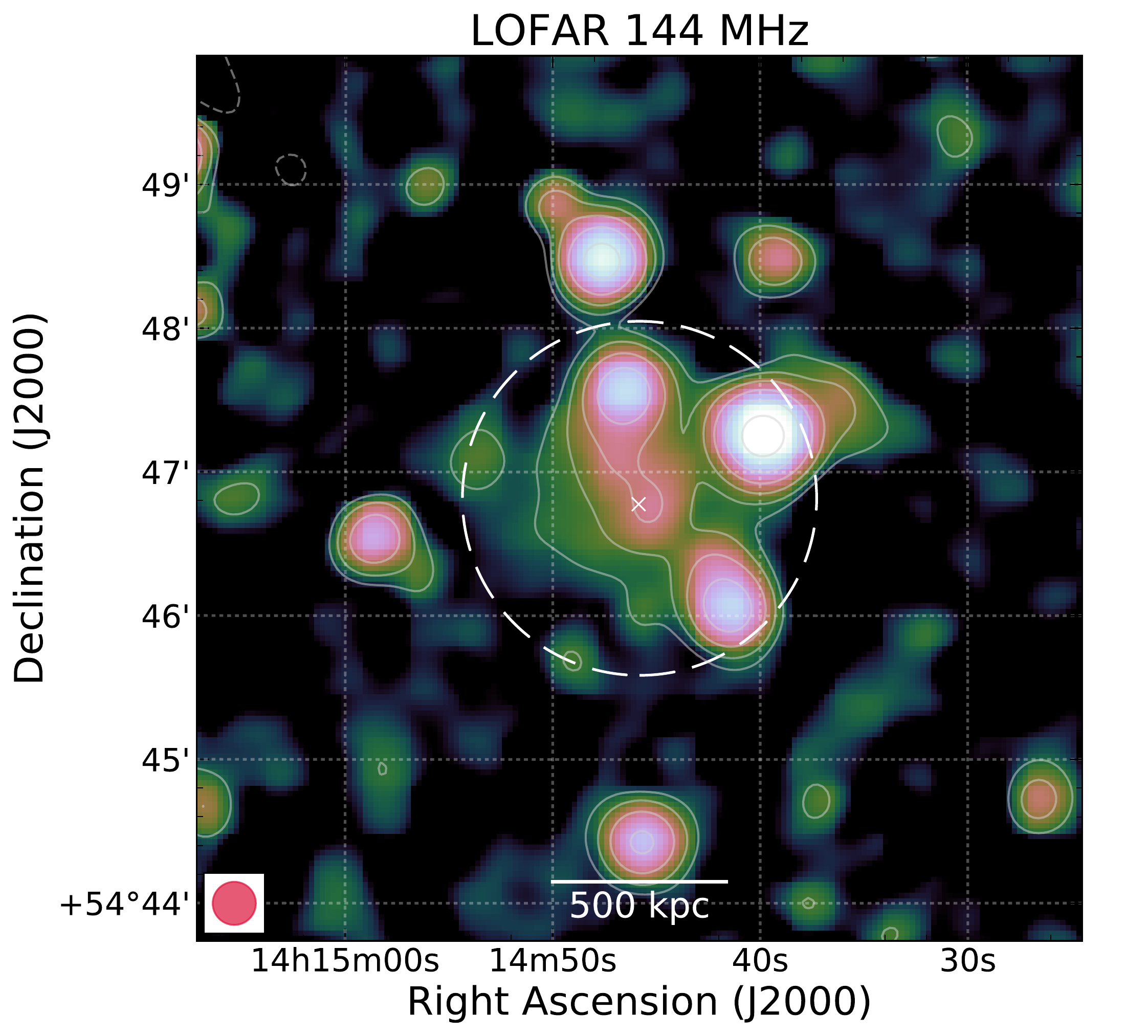}}
{\includegraphics[width=0.43\textwidth]{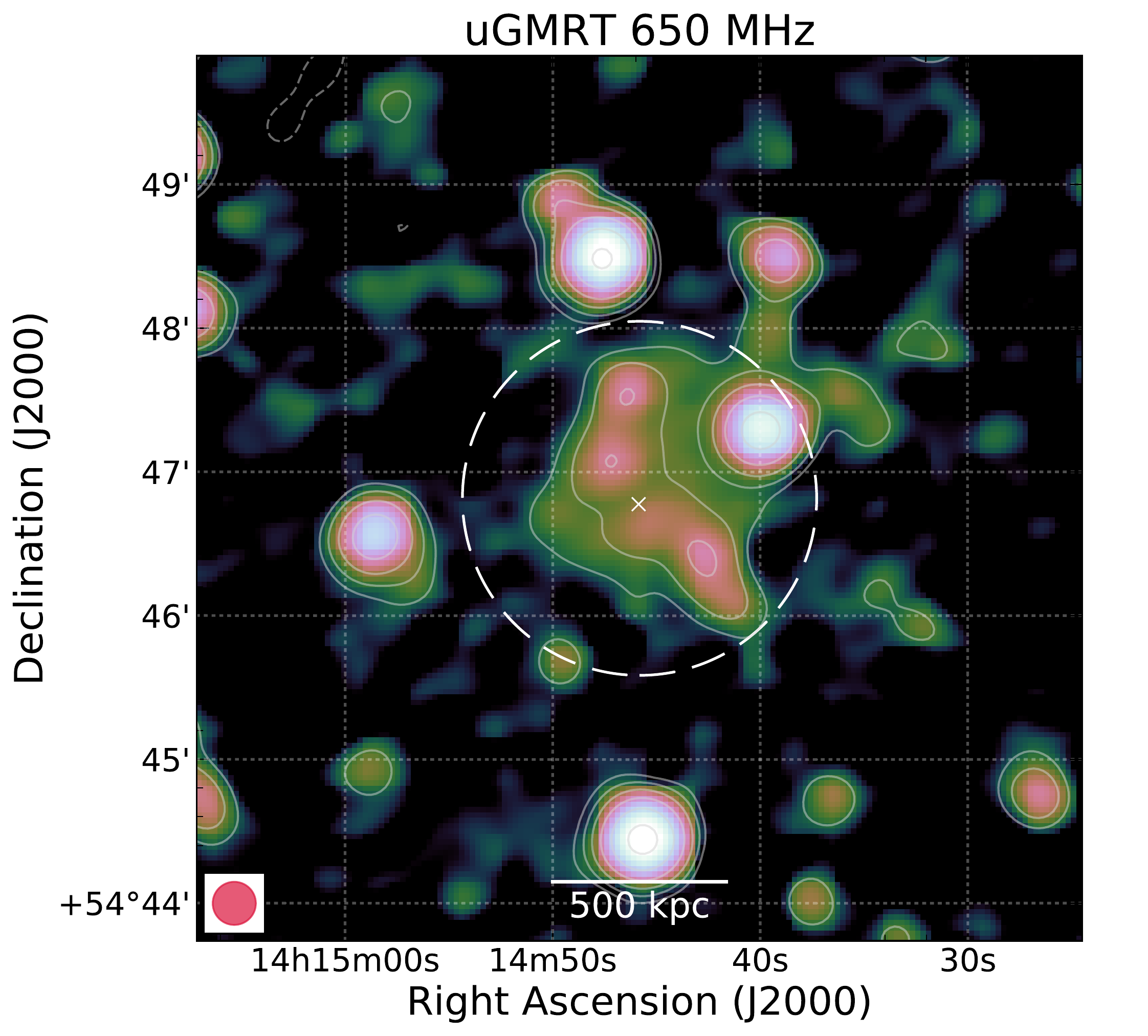}}
{\includegraphics[width=0.43\textwidth]{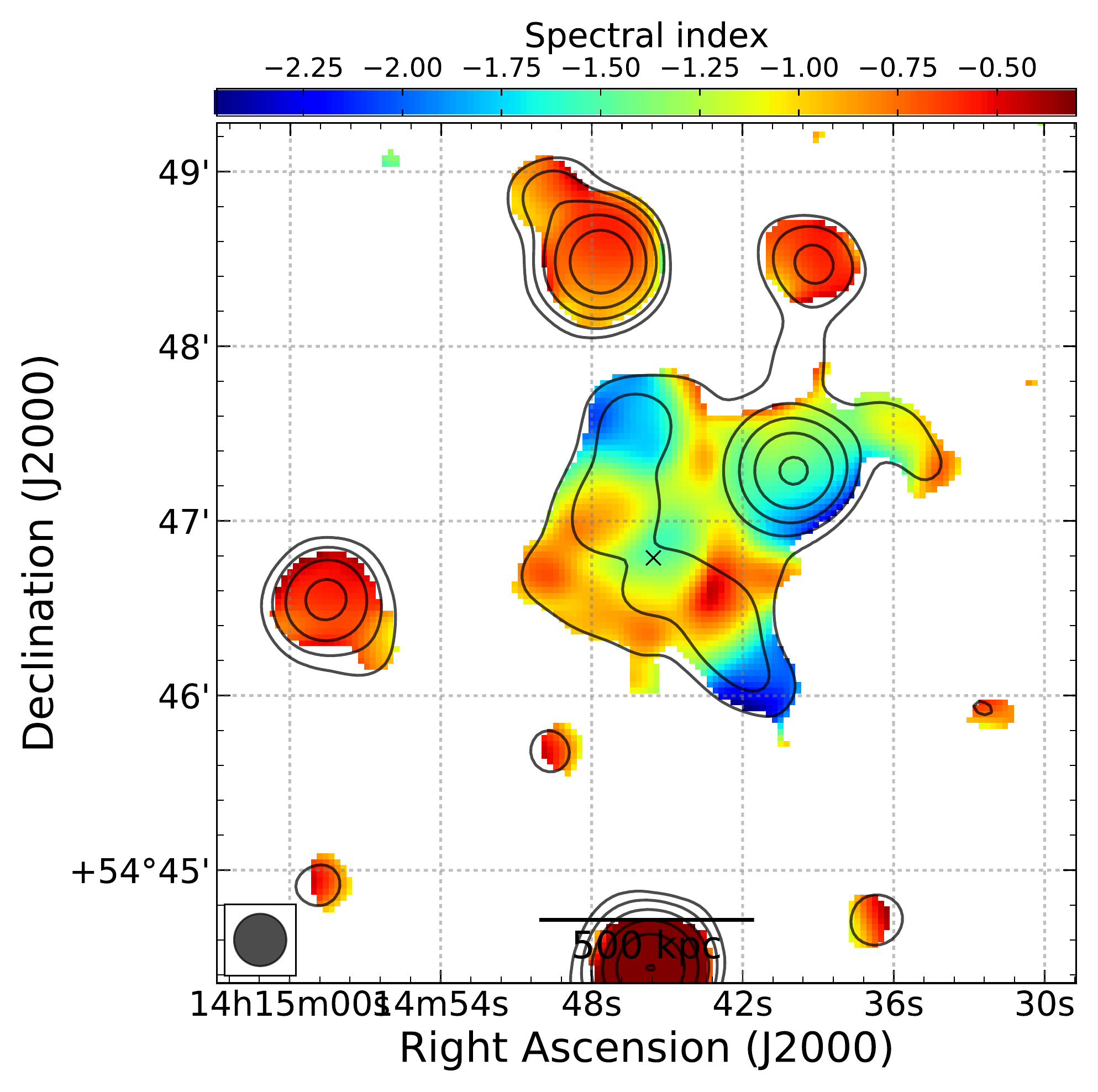}}
{\includegraphics[width=0.43\textwidth]{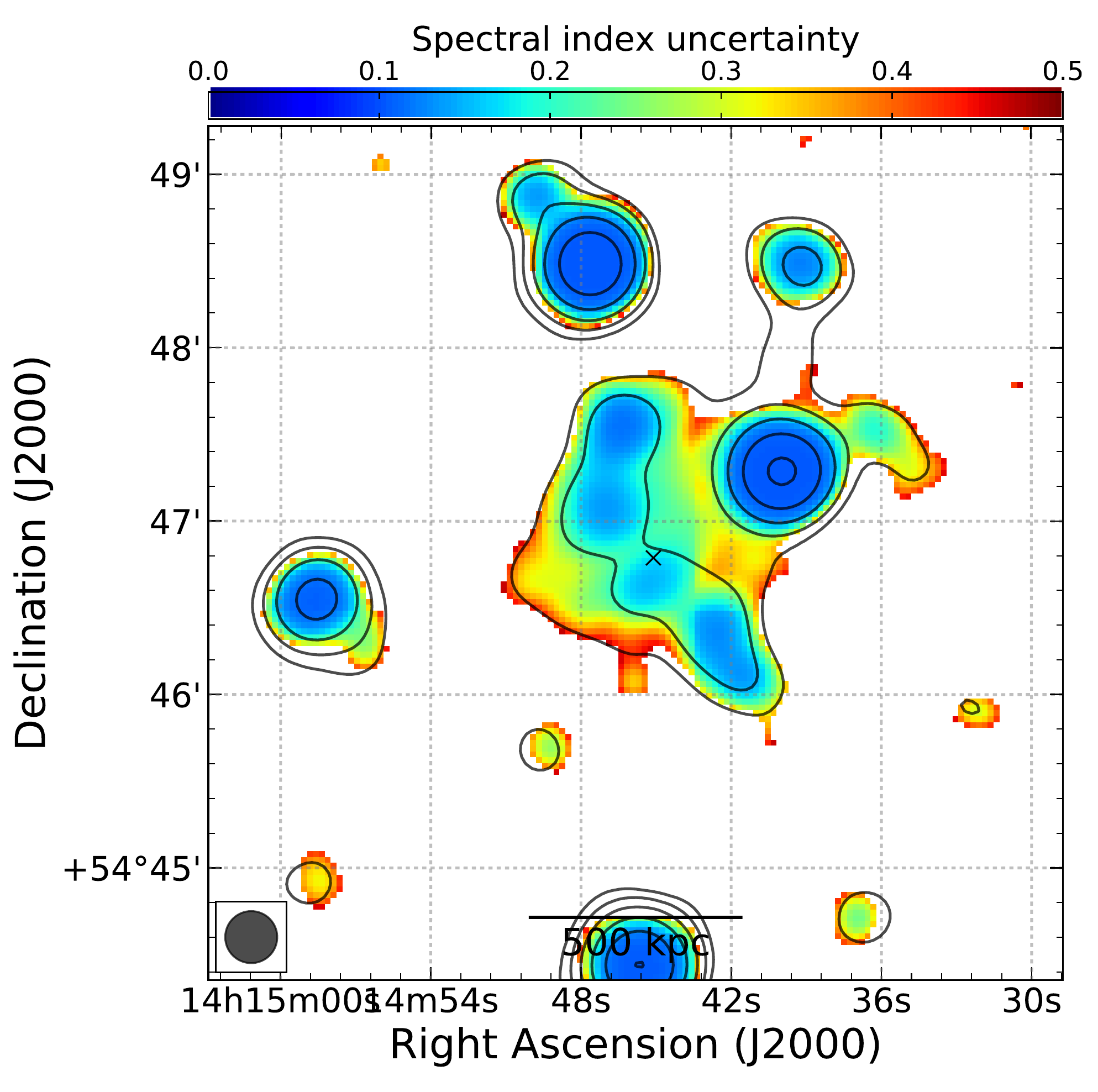}}
\caption{PSZ2\,G099.86+58.45. Top and central rows: Full-resolution and $18''$ images (\texttt{weighting=`Briggs'} and \texttt{robust=-0.5}) at 144 MHz (left) 
and 650 MHz (right). White-coloured radio contours are drawn at levels of $2.5\sigma_{\rm rms}\times[-1, 1, 2, 4, 8, 16, 32]$, with $\sigma_{\rm rms}$ being the noise level at each frequency (see Table \ref{tab:images_parameters}). The negative contour level is drawn with a dashed white line. We followed \cite{cassano+19} for the source labelling (their sources 1, 2, and 3 became D, E, and F1+F2, respectively). The dashed white circle in each map shows the $R=0.5R_{\rm SZ,500}$ region obtained from $M_{\rm SZ,500}$, {with the cross showing the cluster centre}. Bottom row: Spectral index map between 144 and 650 MHz at $18''$ resolution, and correspondent uncertainty map (left and right panels, respectively). uGMRT radio contours at 650 MHz are drawn in black, at the levels $3\sigma_{\rm rms}\times[-1, 1, 2, 4, 8, 16, 32]$, with $\sigma_{\rm rms}$ the noise level (see Table \ref{tab:images_parameters}).}\label{fig:images_099}
\end{figure*}

\begin{figure*}
\centering
{\includegraphics[width=0.45\textwidth]{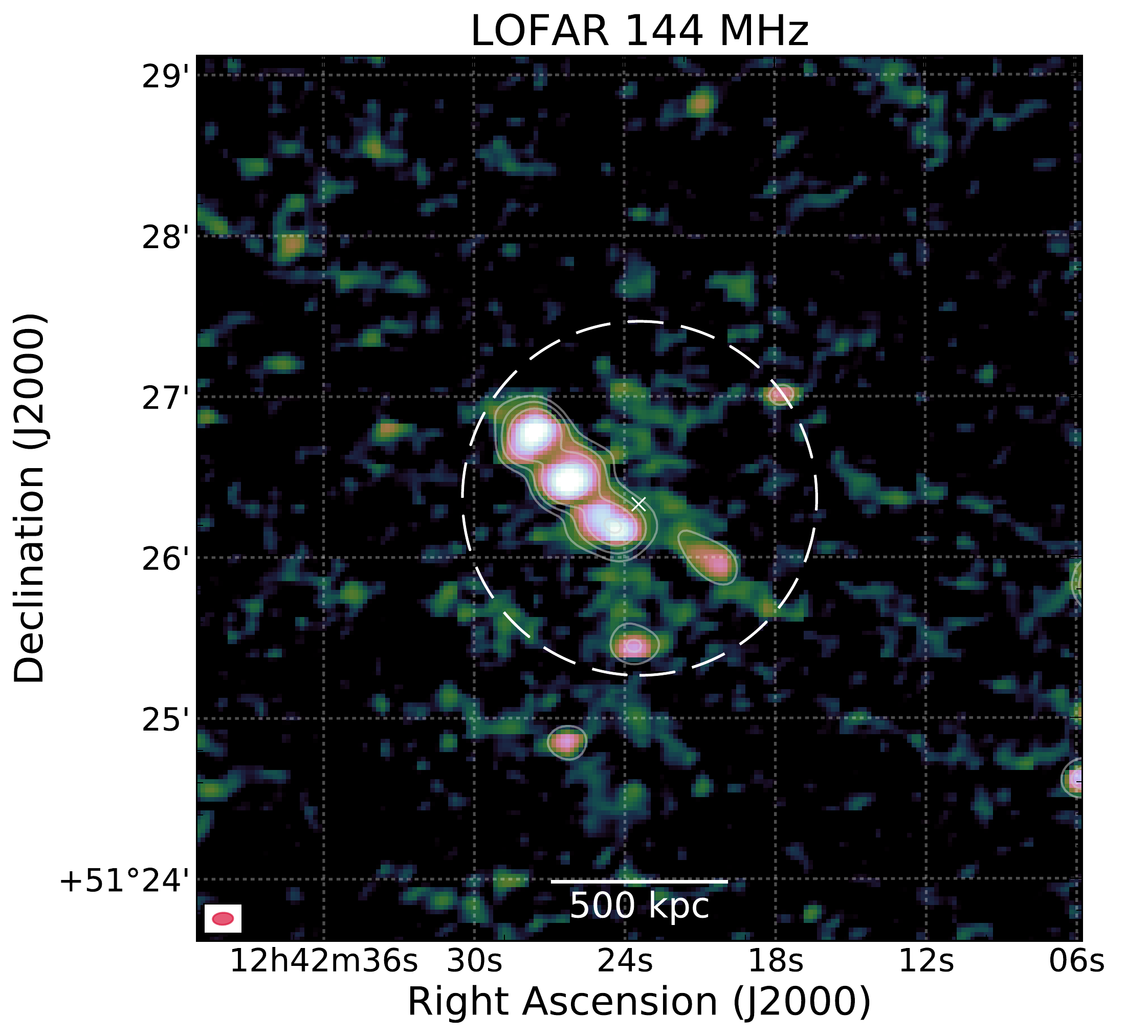}}
{\includegraphics[width=0.45\textwidth]{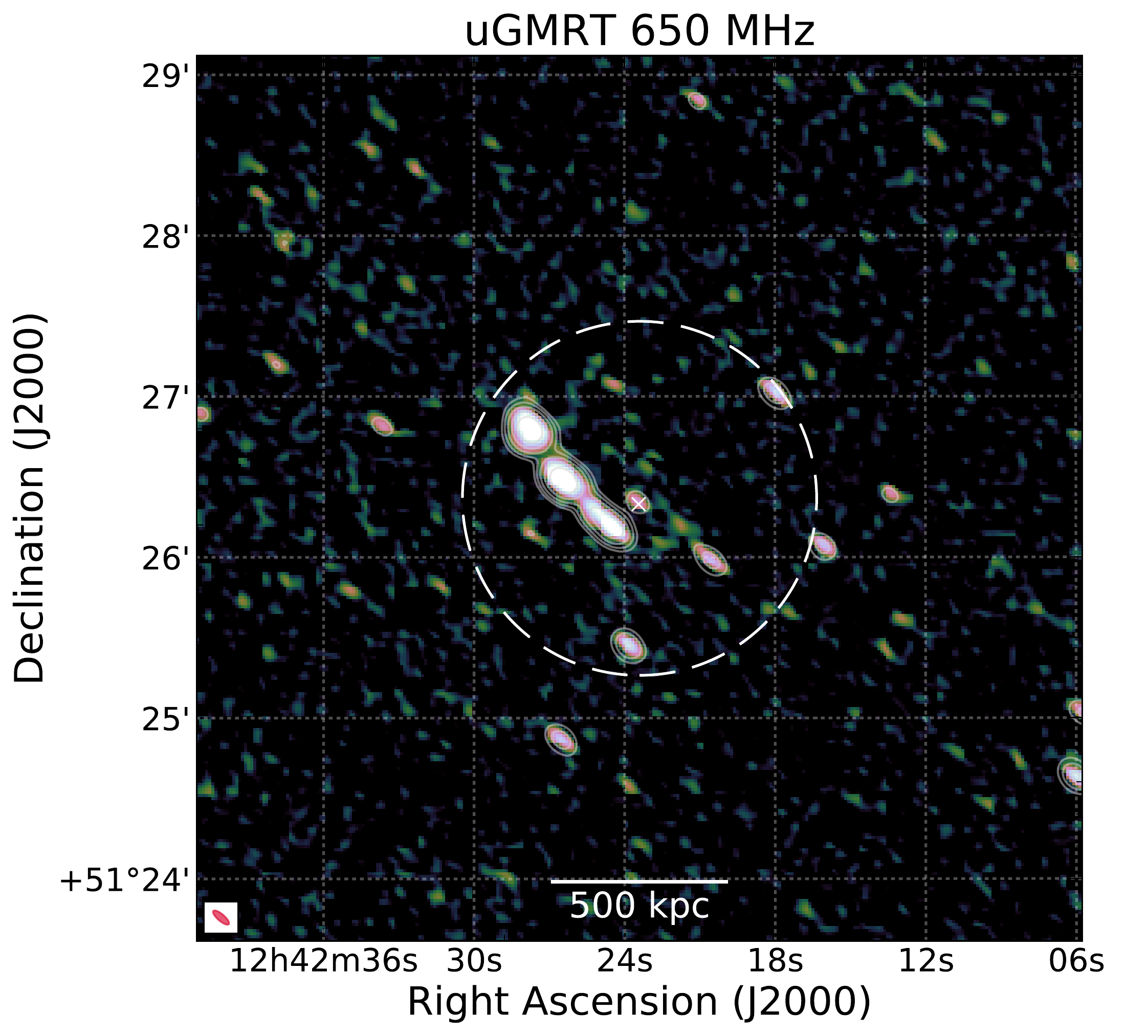}}\\
{\includegraphics[width=0.45\textwidth]{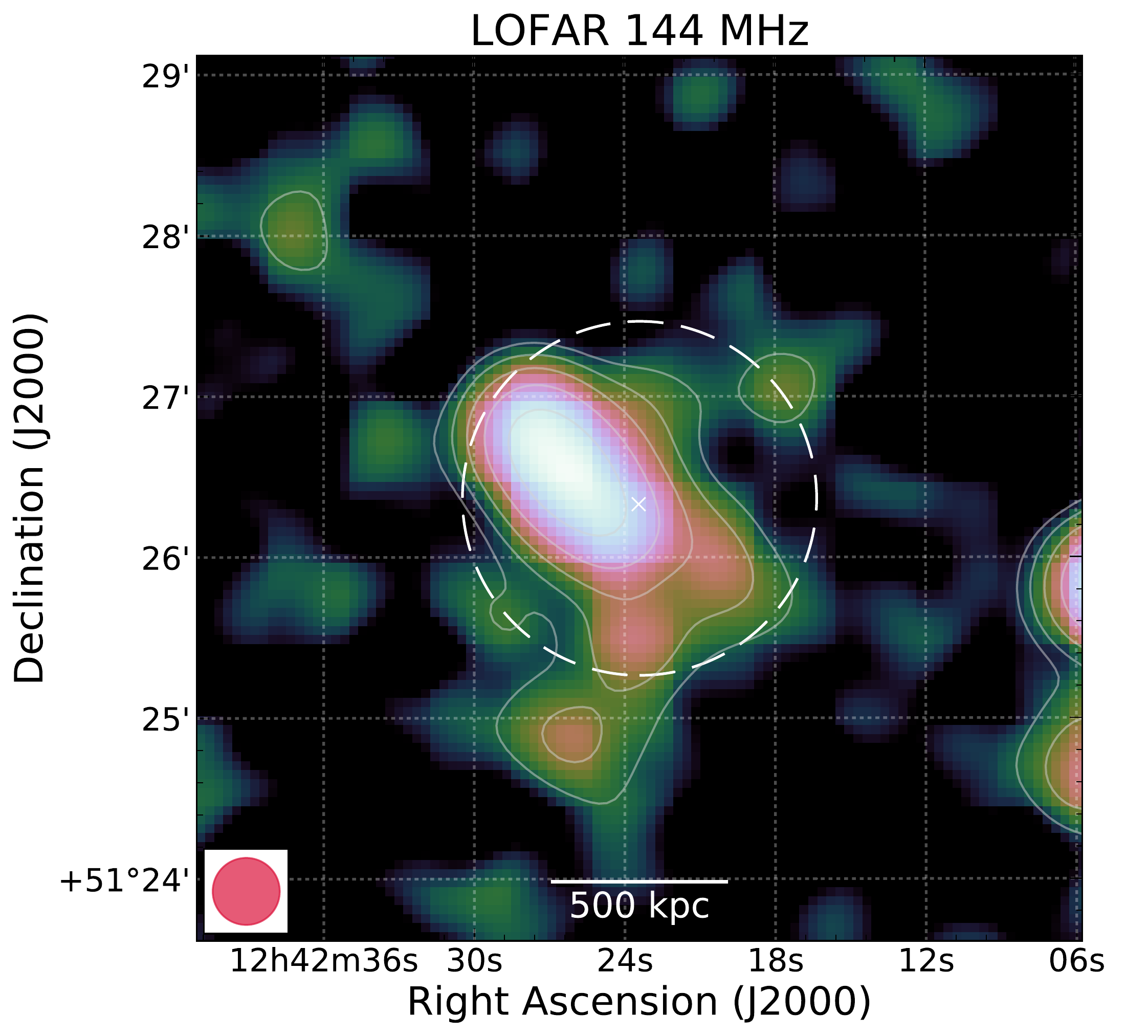}}
{\includegraphics[width=0.45\textwidth]{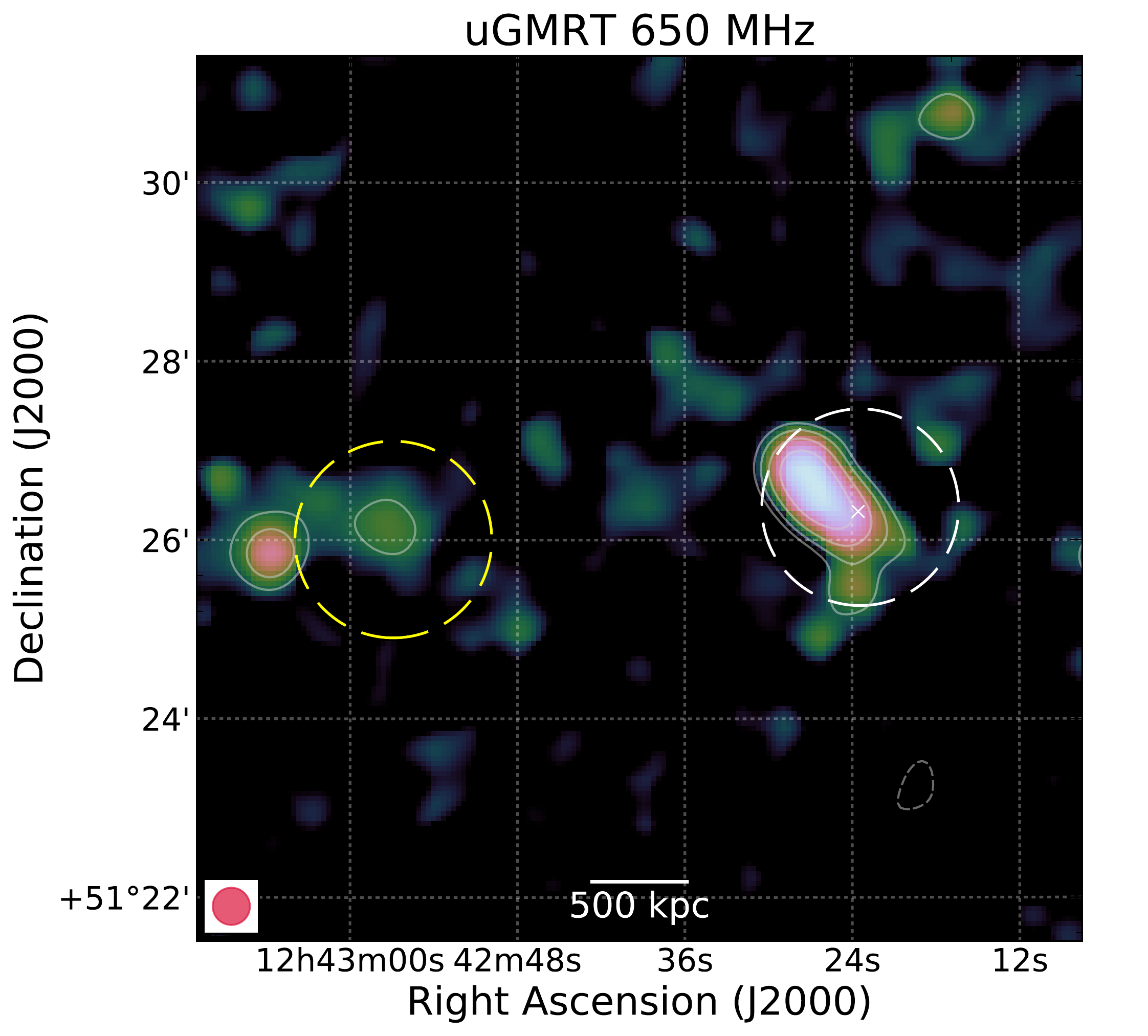}}
\caption{PSZ2\,G126.28+65.62.Top and bottom rows: Full-resolution and $25''$
images (\texttt{weighting=`Briggs'} and \texttt{robust=-0.5}) at 144 MHz (left) and 650 MHz (right). White-coloured radio contours are drawn at levels of $2.5\sigma_{\rm rms}\times[-1, 1, 2, 4, 8, 16, 32]$, with $\sigma_{\rm rms}$ being the noise level at each frequency (see Table \ref{tab:images_parameters}). The negative contour level is drawn with a dashed white line. The dashed white circle in each map shows the $R=0.5R_{\rm SZ,500}$ region obtained from $M_{\rm SZ,500}$, {with the cross showing the cluster centre}. The dashed yellow circle in the bottom right panels shows the position of the mock radio halo.}\label{fig:images_126}
\end{figure*}

\begin{figure*}
\centering
{\includegraphics[width=0.43\textwidth]{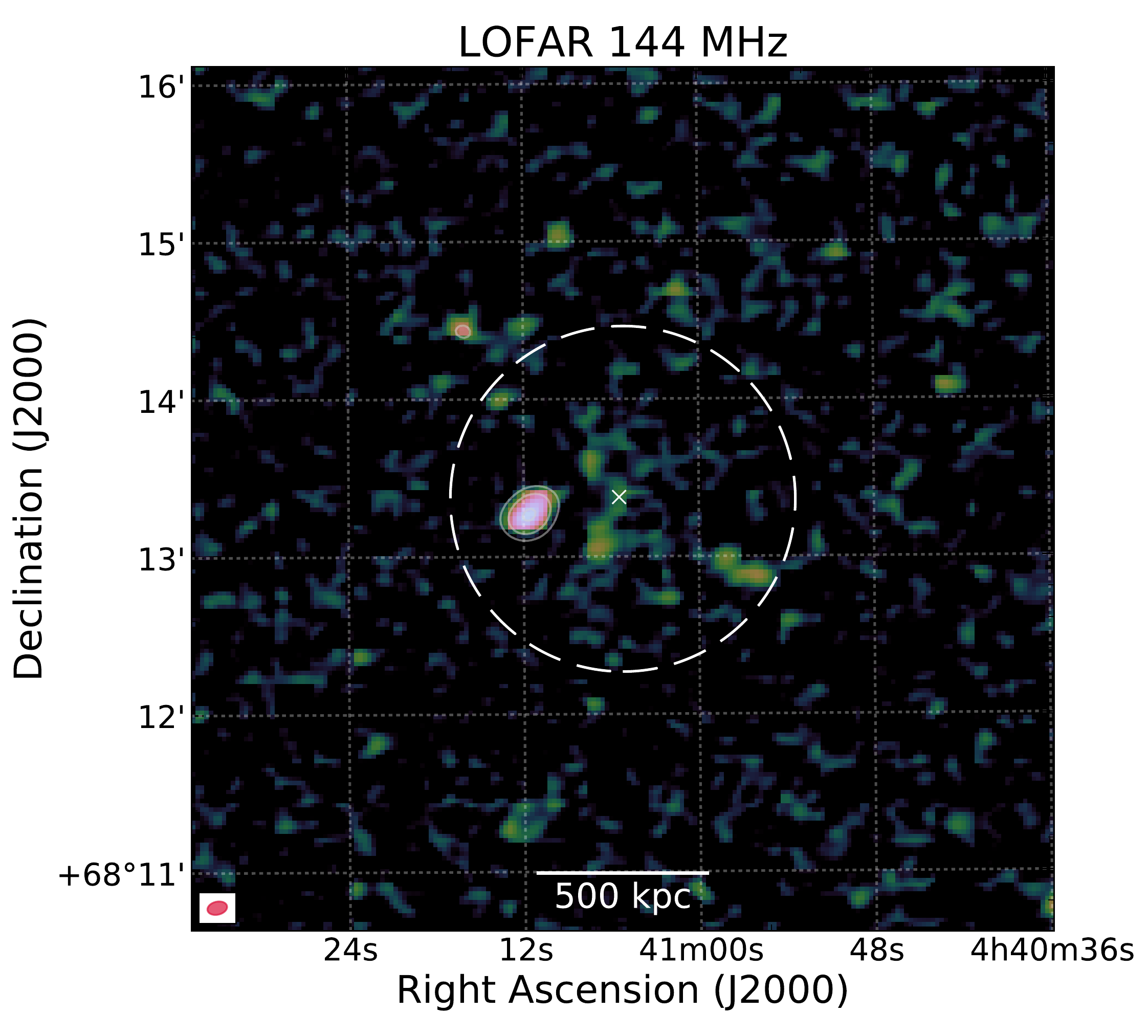}}
{\includegraphics[width=0.43\textwidth]{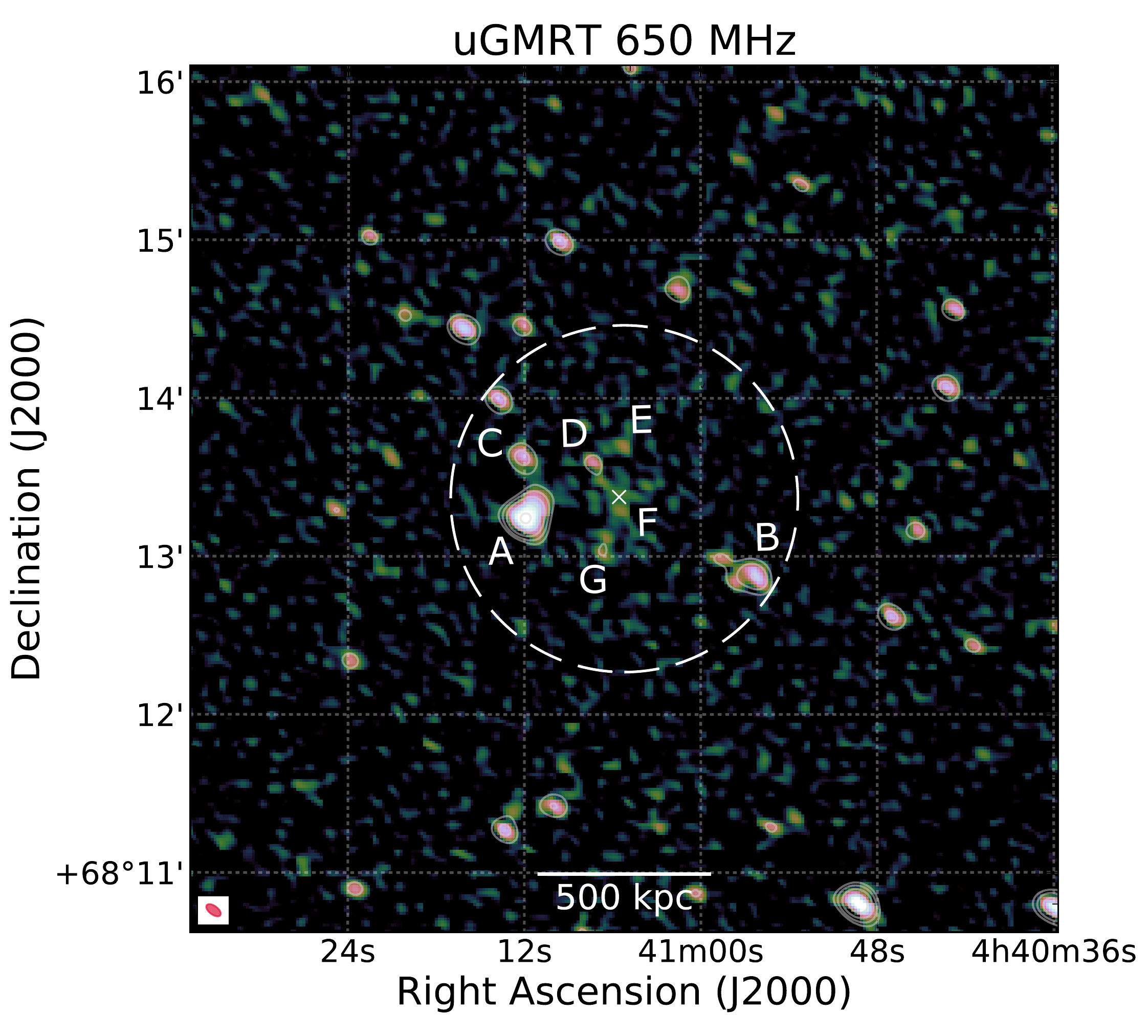}}\\
{\includegraphics[width=0.43\textwidth]{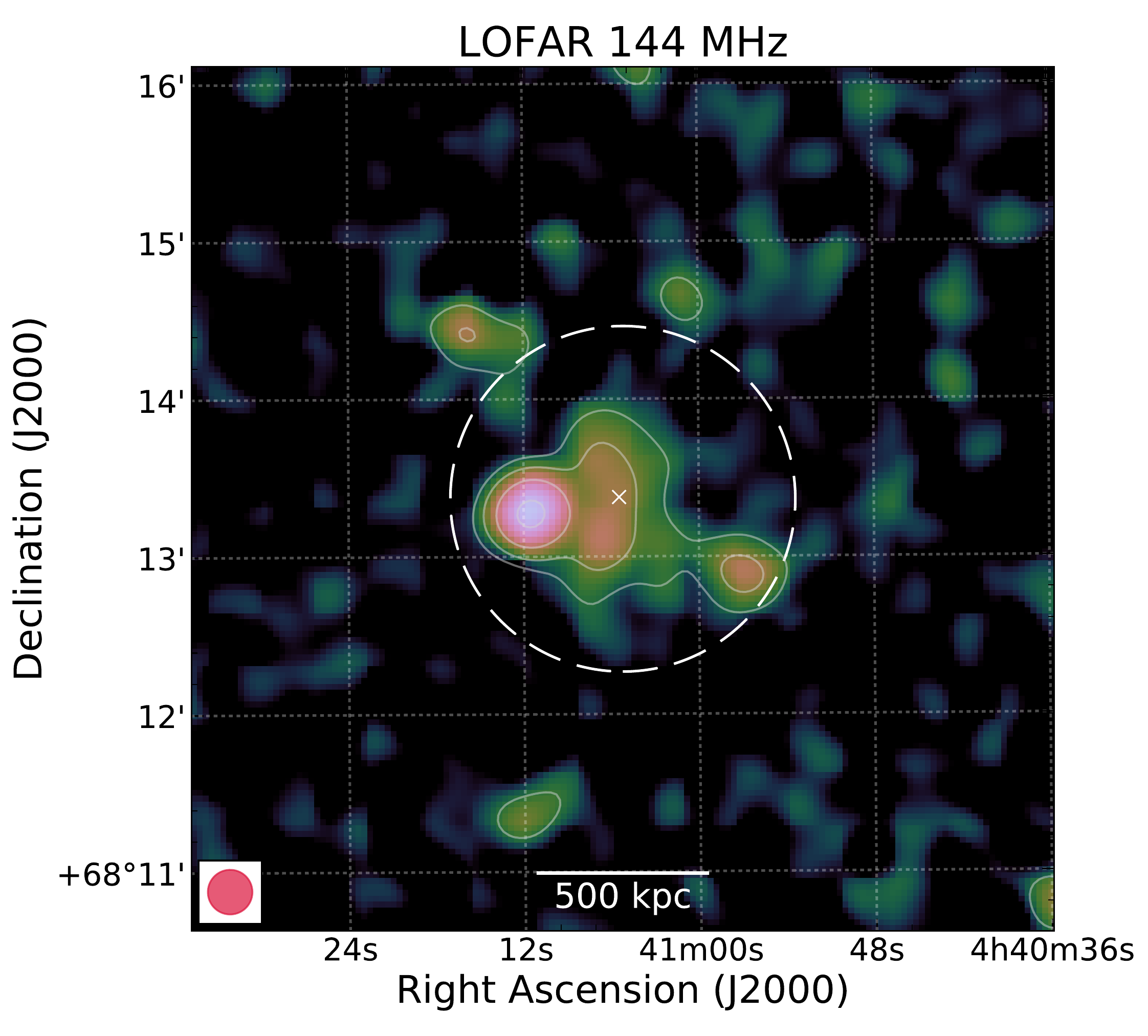}}
{\includegraphics[width=0.43\textwidth]{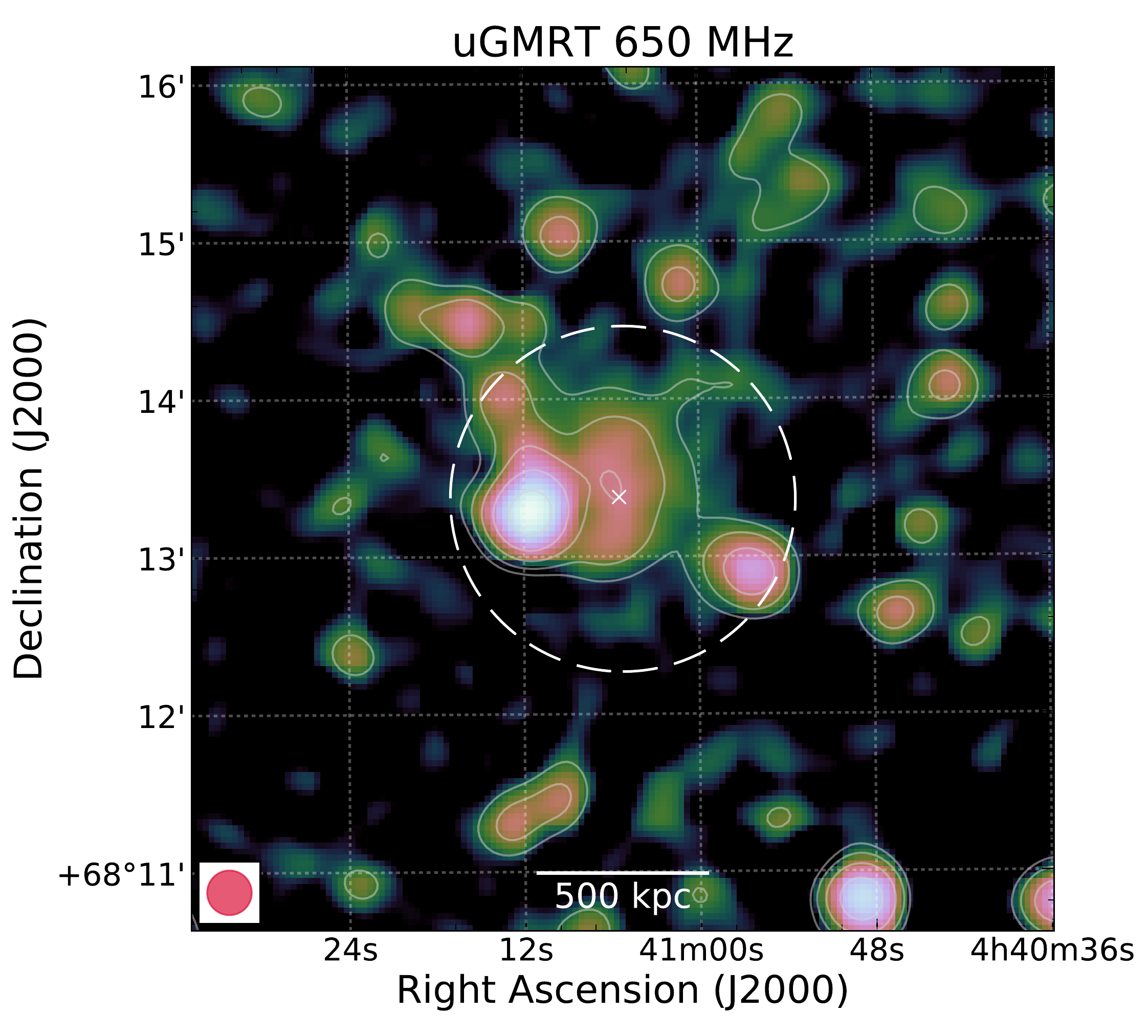}}
{\includegraphics[width=0.43\textwidth]{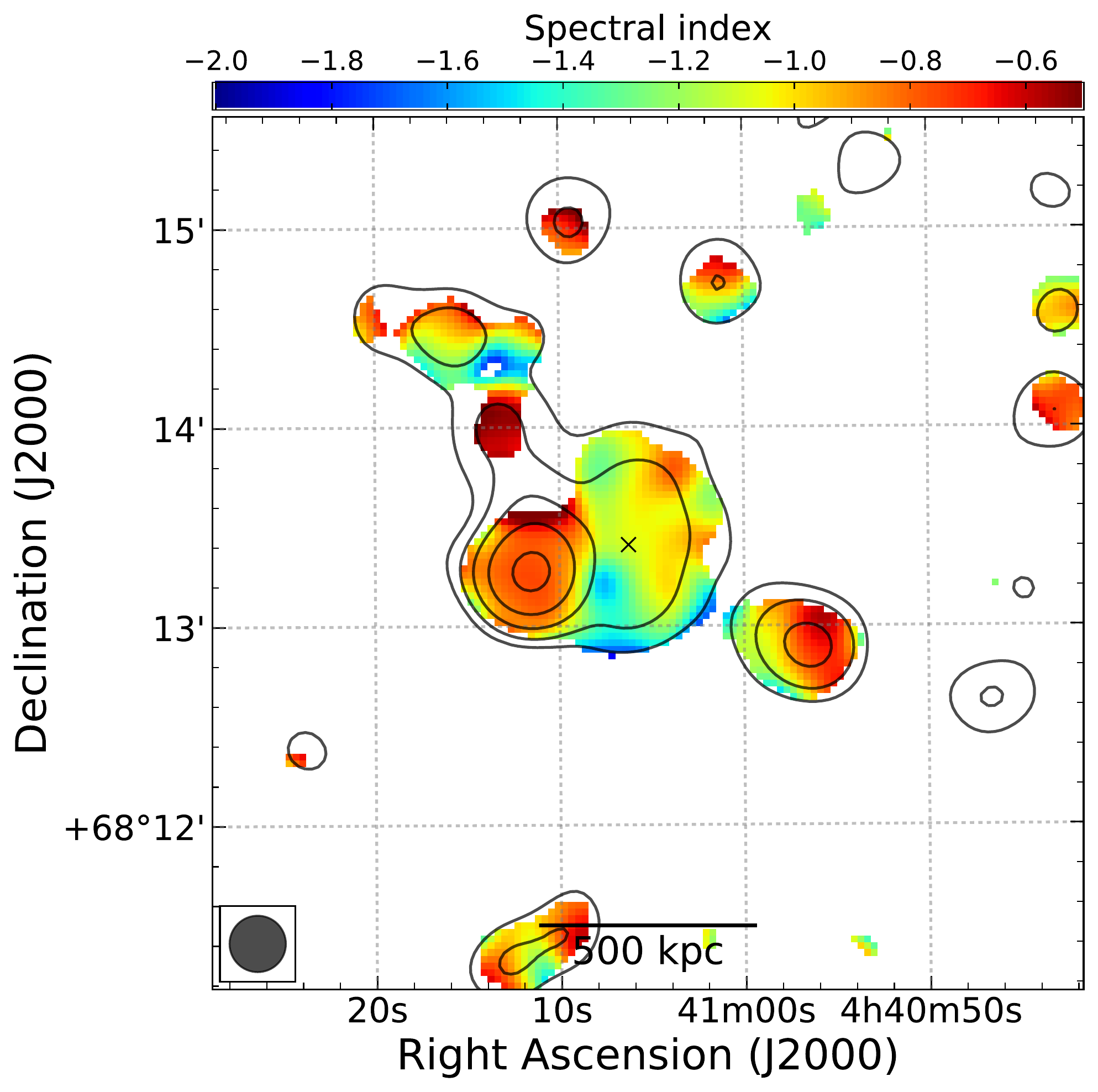}}
{\includegraphics[width=0.43\textwidth]{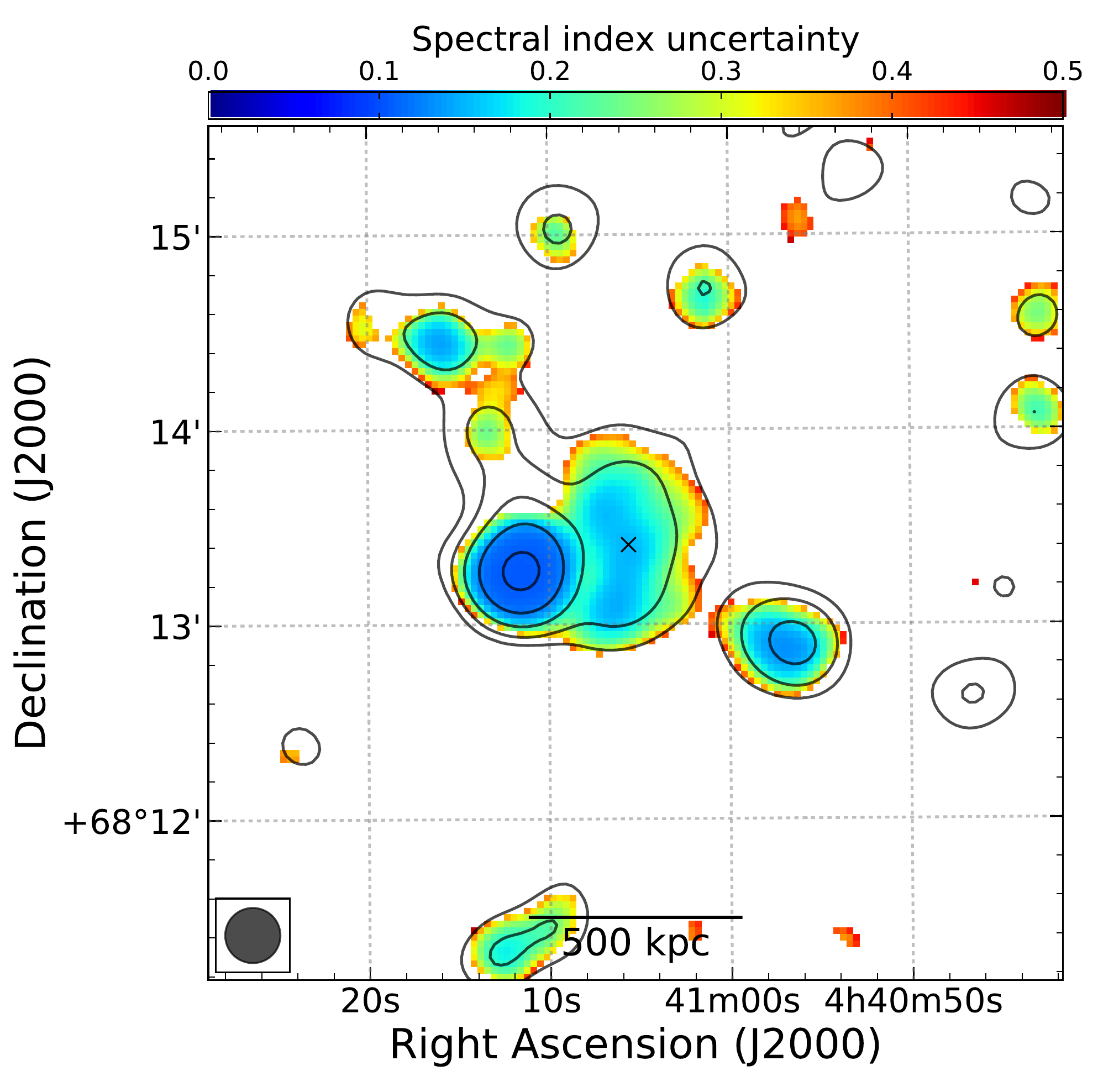}}
\caption{PSZ2\,G141.77+14.19 . Top and central rows: Full-resolution and $17''$ images (\texttt{weighting=`Briggs'} and \texttt{robust=-0.5}) at 144 MHz (left) and 650 MHz (right). 
White-coloured radio contours are drawn at levels of $2.5\sigma_{\rm rms}\times[-1, 1, 2, 4, 8, 16, 32]$, with $\sigma_{\rm rms}$ being the noise level at each frequency (see Table \ref{tab:images_parameters}). The negative contour level is drawn with a dashed white line. The dashed white circle in each map shows the $R=0.5R_{\rm SZ,500}$ region obtained from $M_{\rm SZ,500}$, {with the cross showing the cluster centre}. Bottom row: Spectral index map between 144 and 650 MHz at $17''$ resolution, and correspondent uncertainty map (left and right panels, respectively). uGMRT radio contours at 650 MHz are drawn in black, at the levels $3\sigma_{\rm rms}\times[-1, 1, 2, 4, 8, 16, 32]$, with $\sigma_{\rm rms}$ being the noise level (see Table \ref{tab:images_parameters}).}\label{fig:images_141}
\end{figure*}

\begin{figure*}
\centering
{\includegraphics[width=0.3\textwidth]{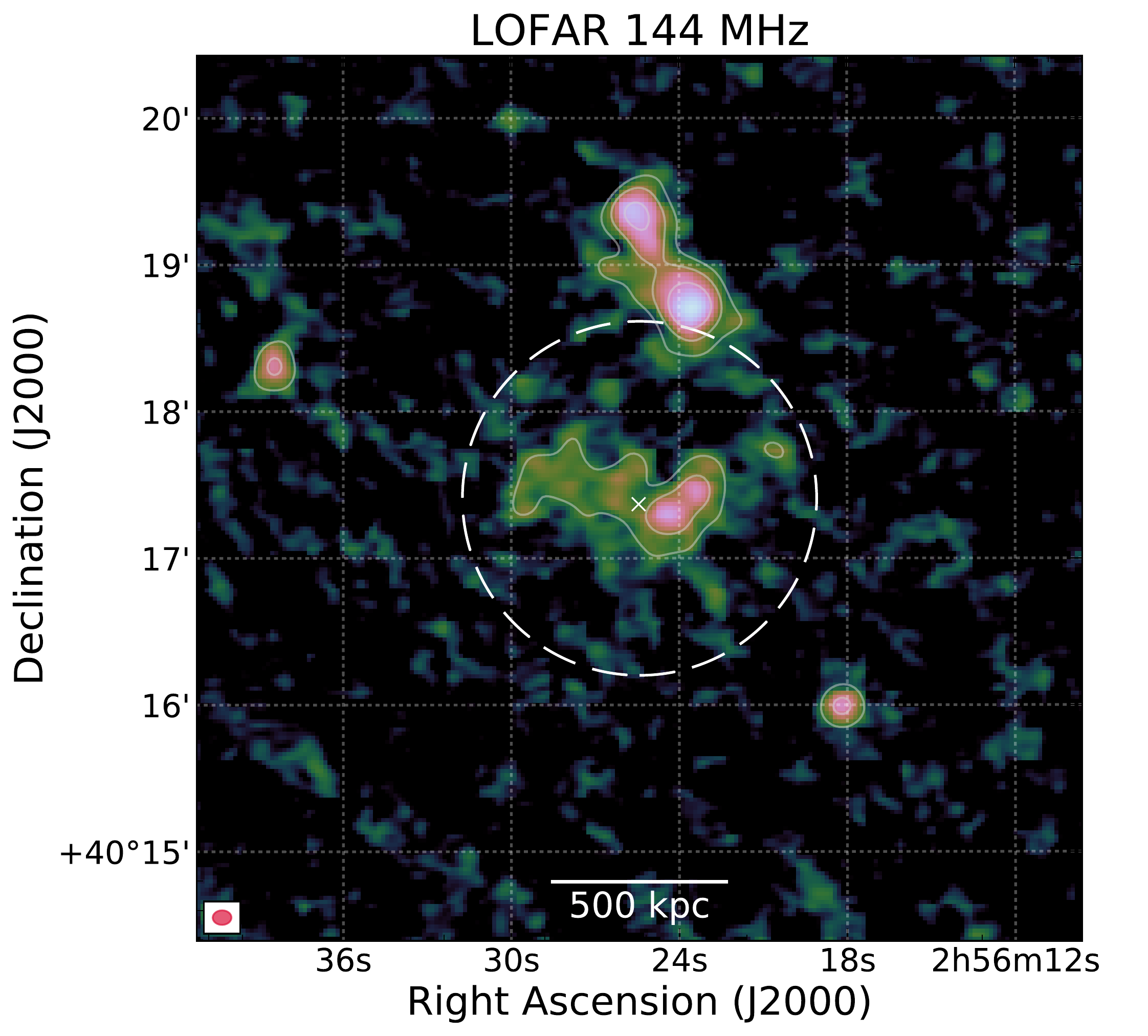}}
{\includegraphics[width=0.3\textwidth]{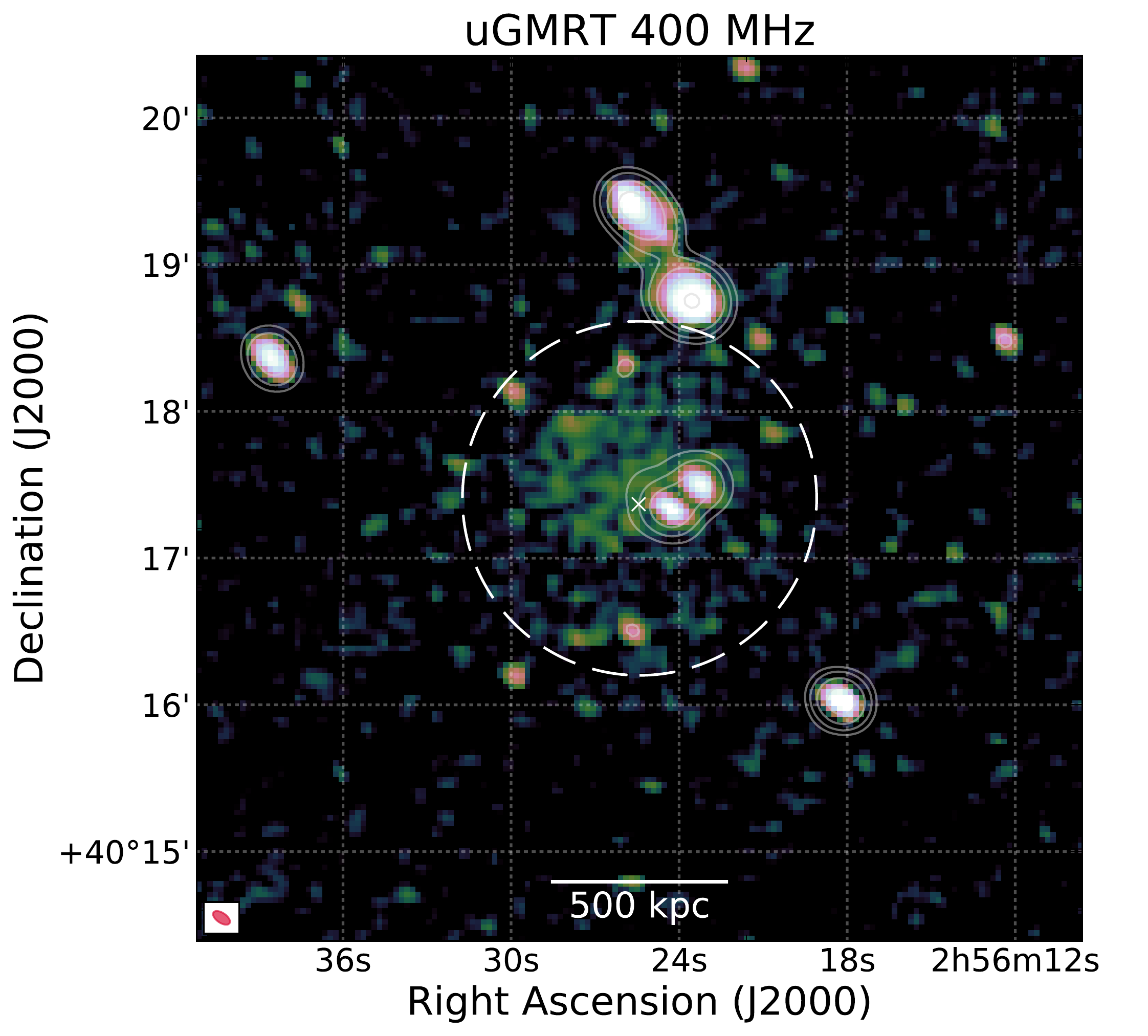}}
{\includegraphics[width=0.3\textwidth]{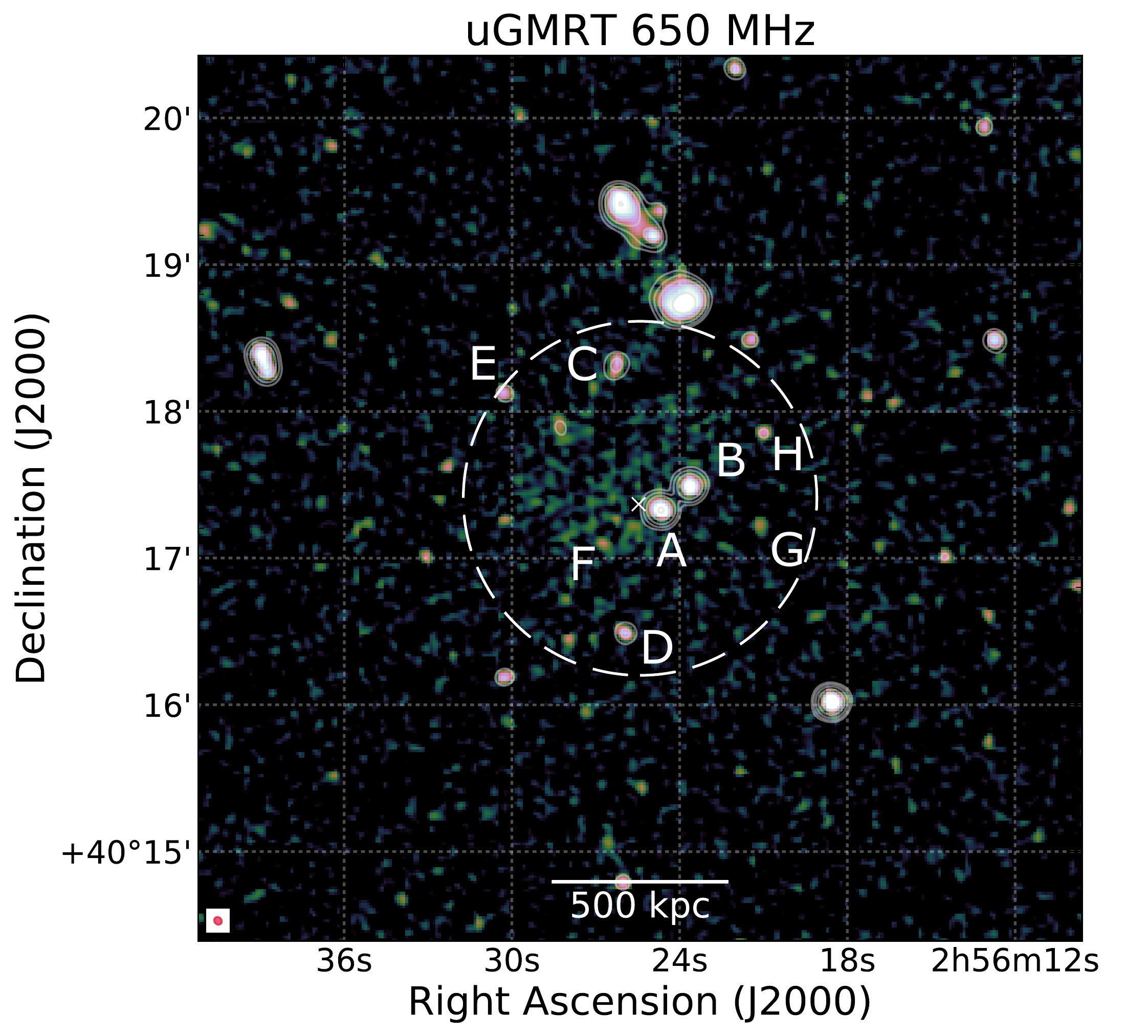}}
{\includegraphics[width=0.3\textwidth]{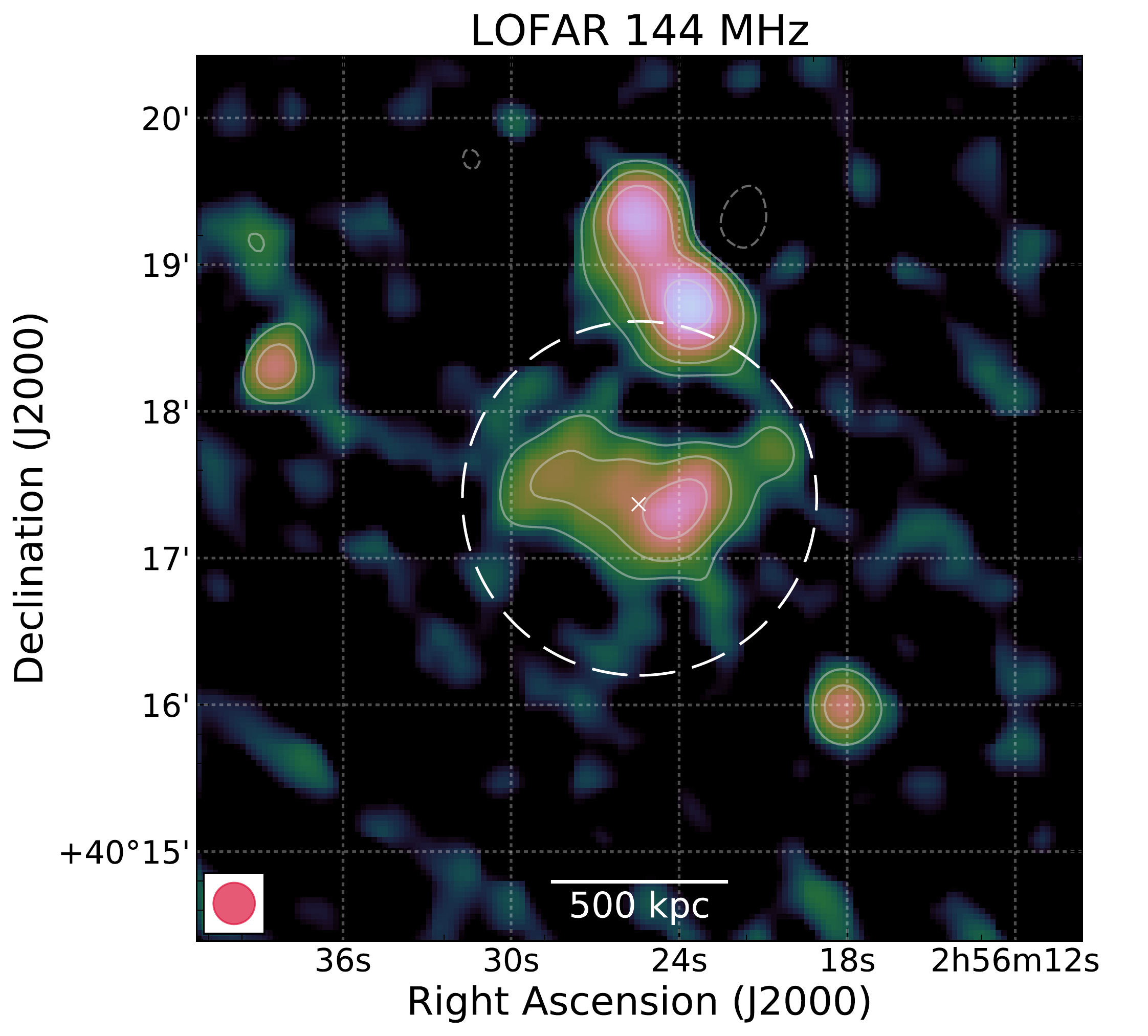}}
{\includegraphics[width=0.3\textwidth]{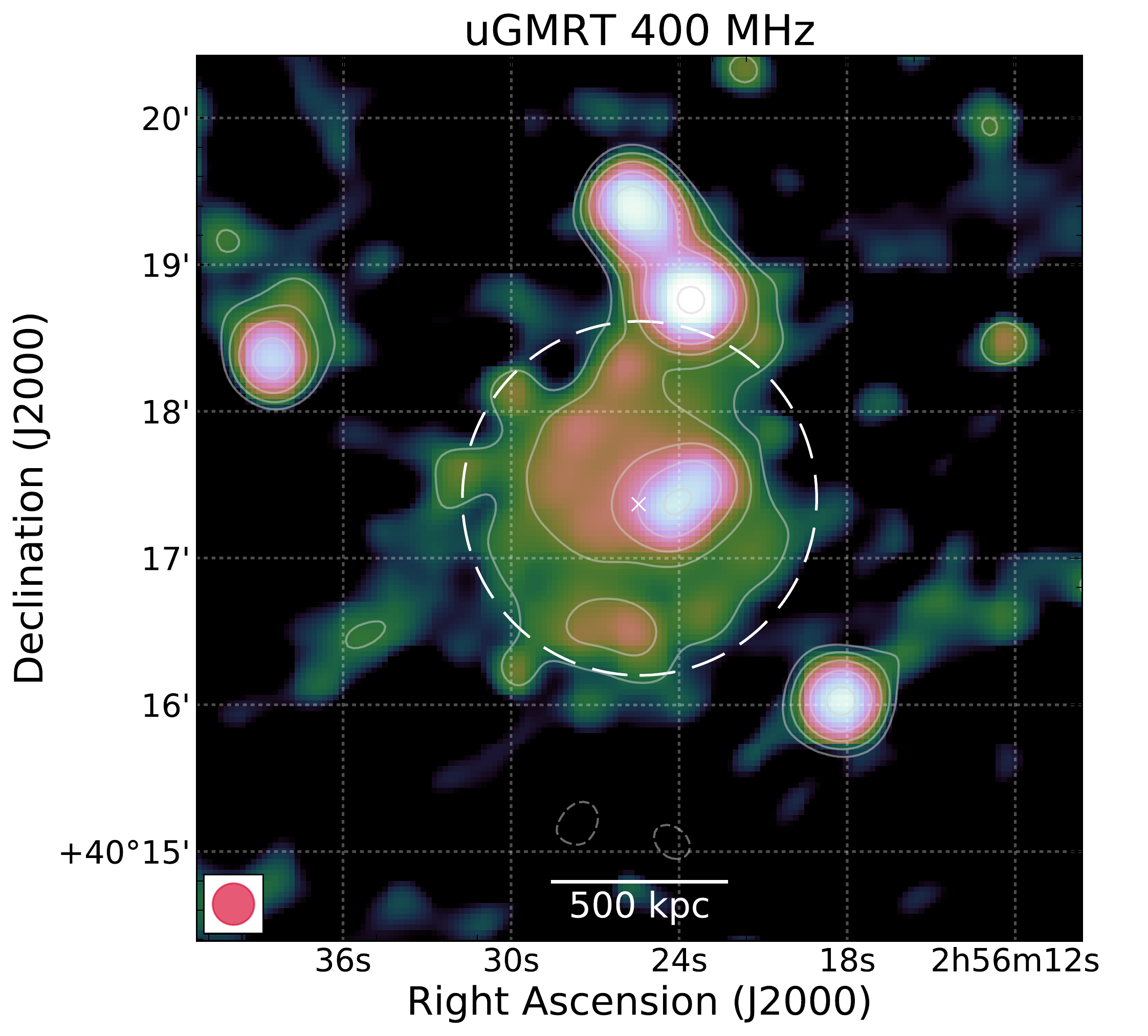}}
{\includegraphics[width=0.3\textwidth]{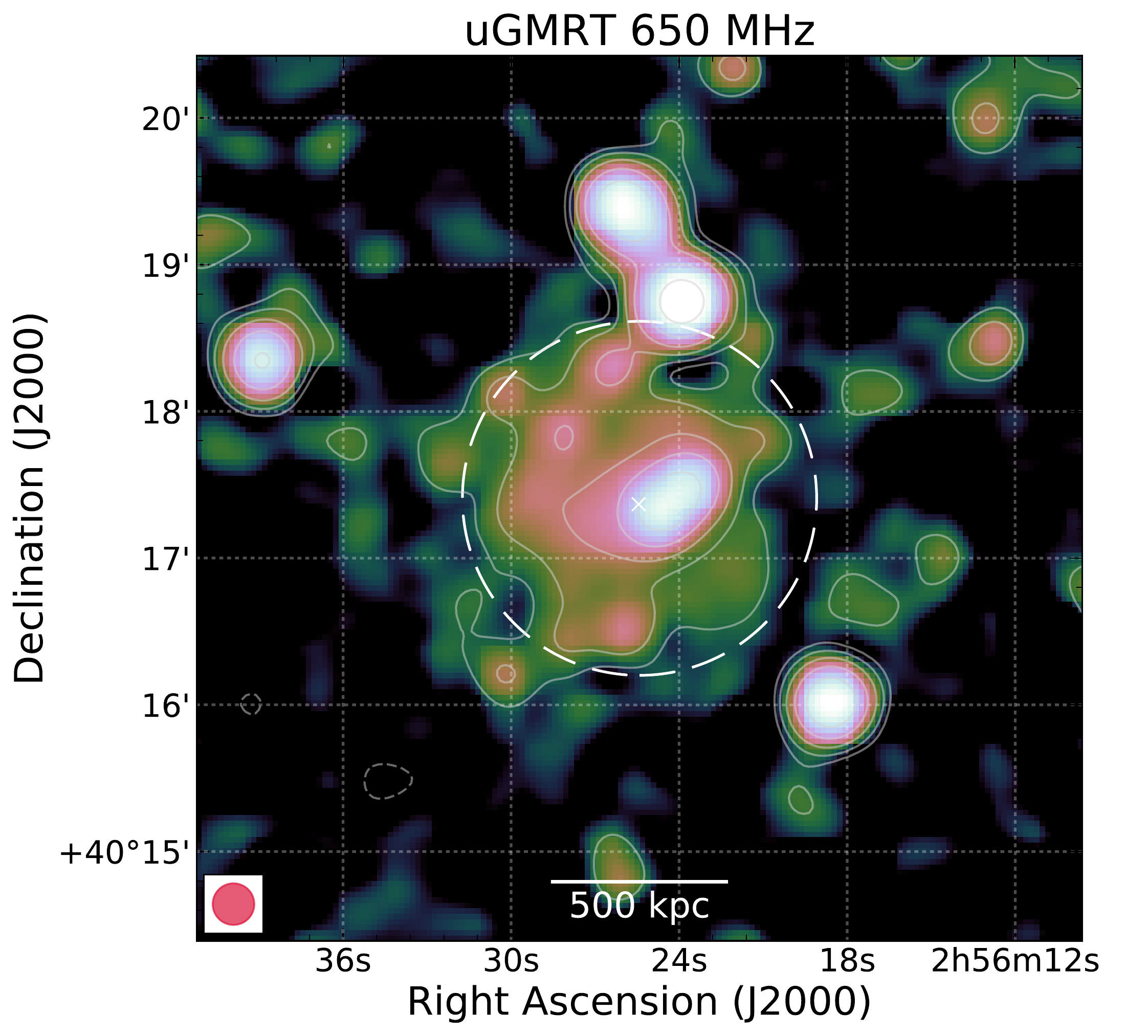}}
{\includegraphics[width=0.45\textwidth]{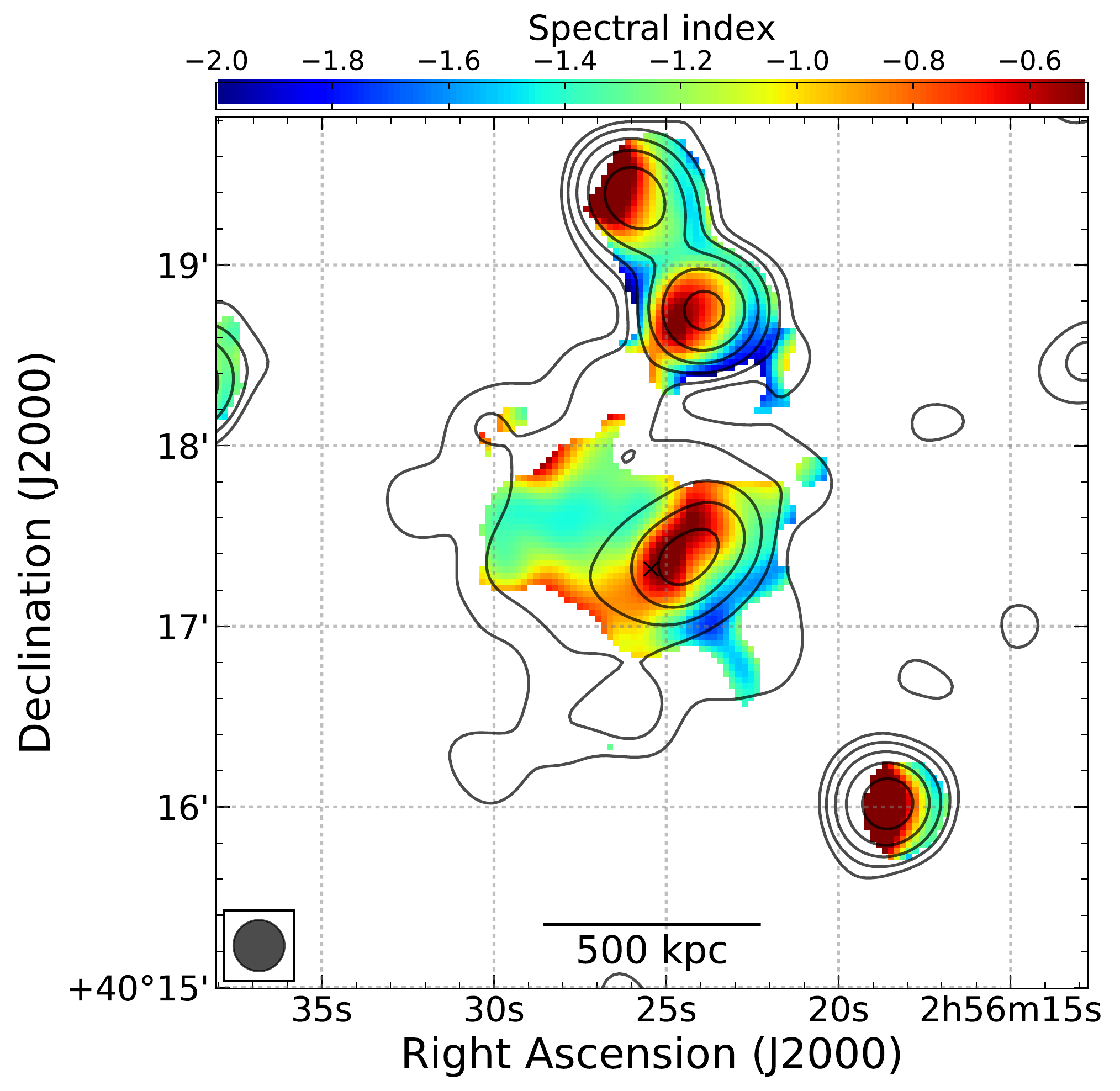}}
{\includegraphics[width=0.45\textwidth]{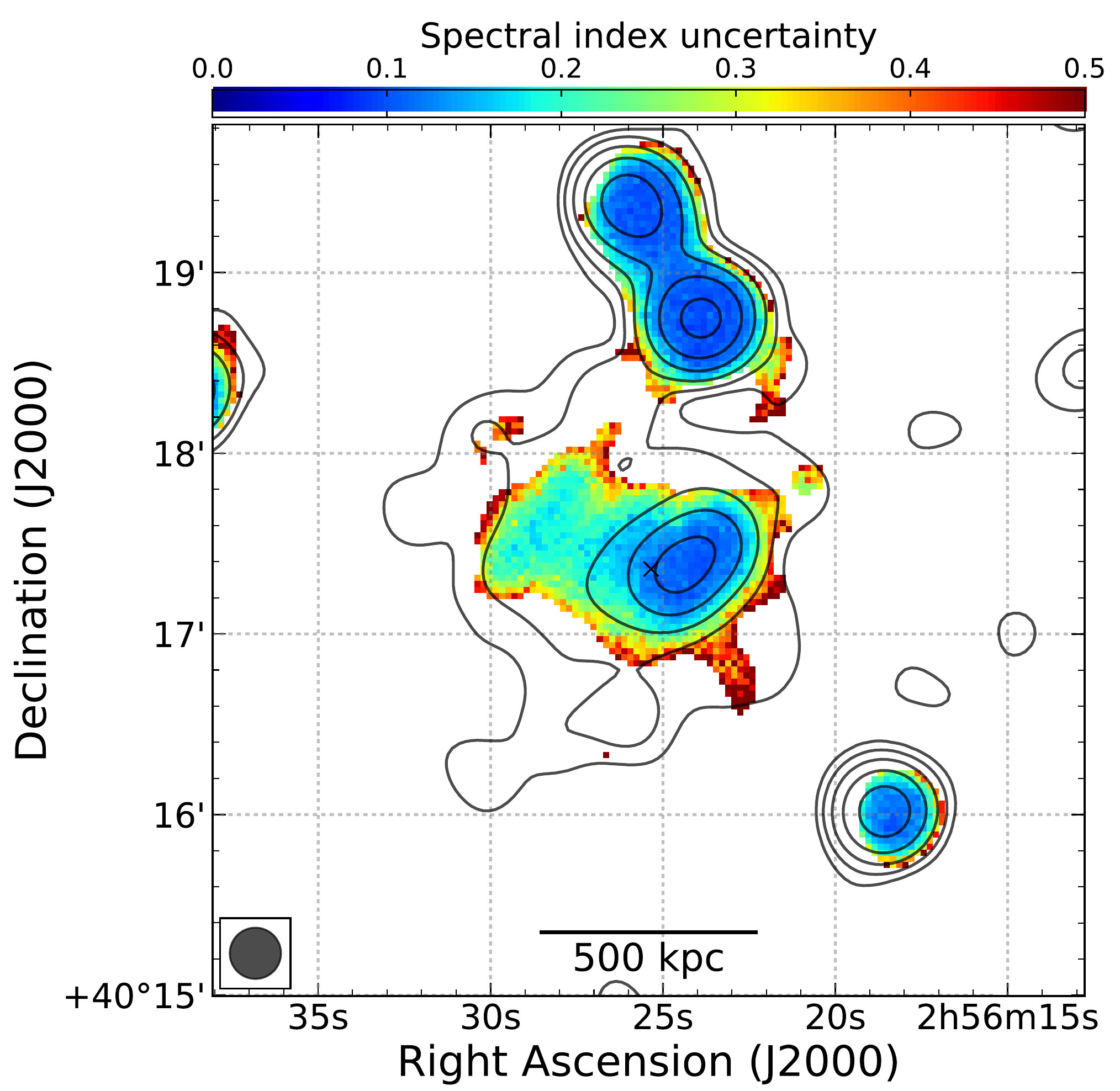}}
\caption{PLCK\,G147.3--16.6. Top and central rows: Full-resolution and $17''$ images (\texttt{weighting=`Briggs'} and \texttt{robust=-0.5}) at 144 MHz (left), 400 MHz (middle) and 650 MHz (right). Radio contours are drawn in white at levels of $2.5\sigma_{\rm rms}\times[-1, 1, 2, 4, 8, 16, 32]$, with $\sigma_{\rm rms}$ being the noise level at each frequency (see Table \ref{tab:images_parameters}). The negative contour level is drawn with a dashed white line. We followed \cite{vanweeren+14} for the source labelling. The dashed white circle in each map shows the $R=0.5R_{\rm SZ,500}$ region, obtained from $M_{\rm SZ,500}$, {with the cross showing the cluster centre}. Bottom row: Spectral index map at 400 MHz (see Sect. \ref{sec:spix}), at $17''$ resolution, and correspondent uncertainty map (left and right panels, respectively). uGMRT radio contours at 650 MHz are drawn in black, at the levels $3\sigma_{\rm rms}\times[-1, 1, 2, 4, 8, 16, 32]$, with $\sigma_{\rm rms}$ the noise level (see Table \ref{tab:images_parameters}).}\label{fig:images_plck147}
\end{figure*}

\begin{figure*}
\centering
{\includegraphics[width=0.43\textwidth]{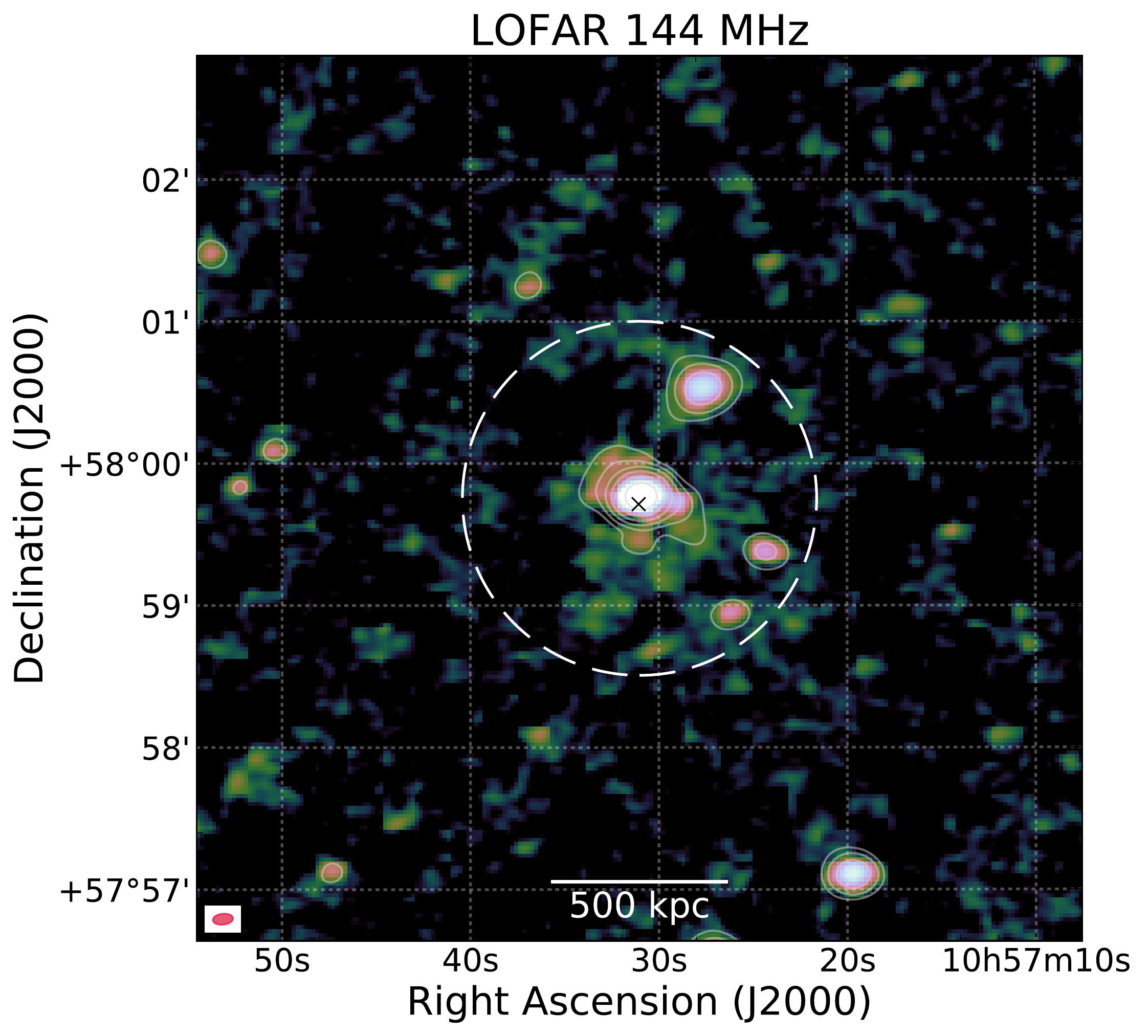}}
{\includegraphics[width=0.43\textwidth]{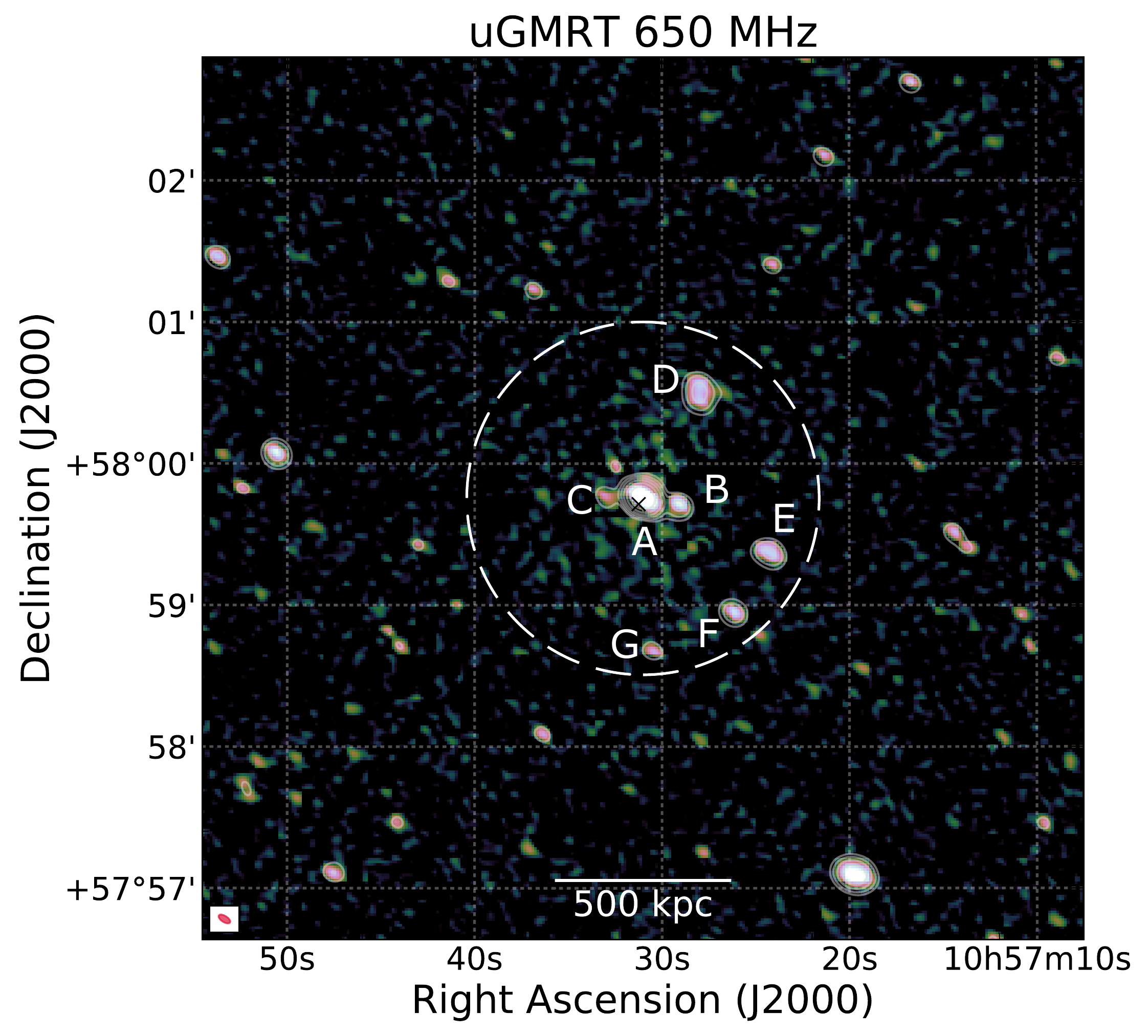}}\\
{\includegraphics[width=0.43\textwidth]{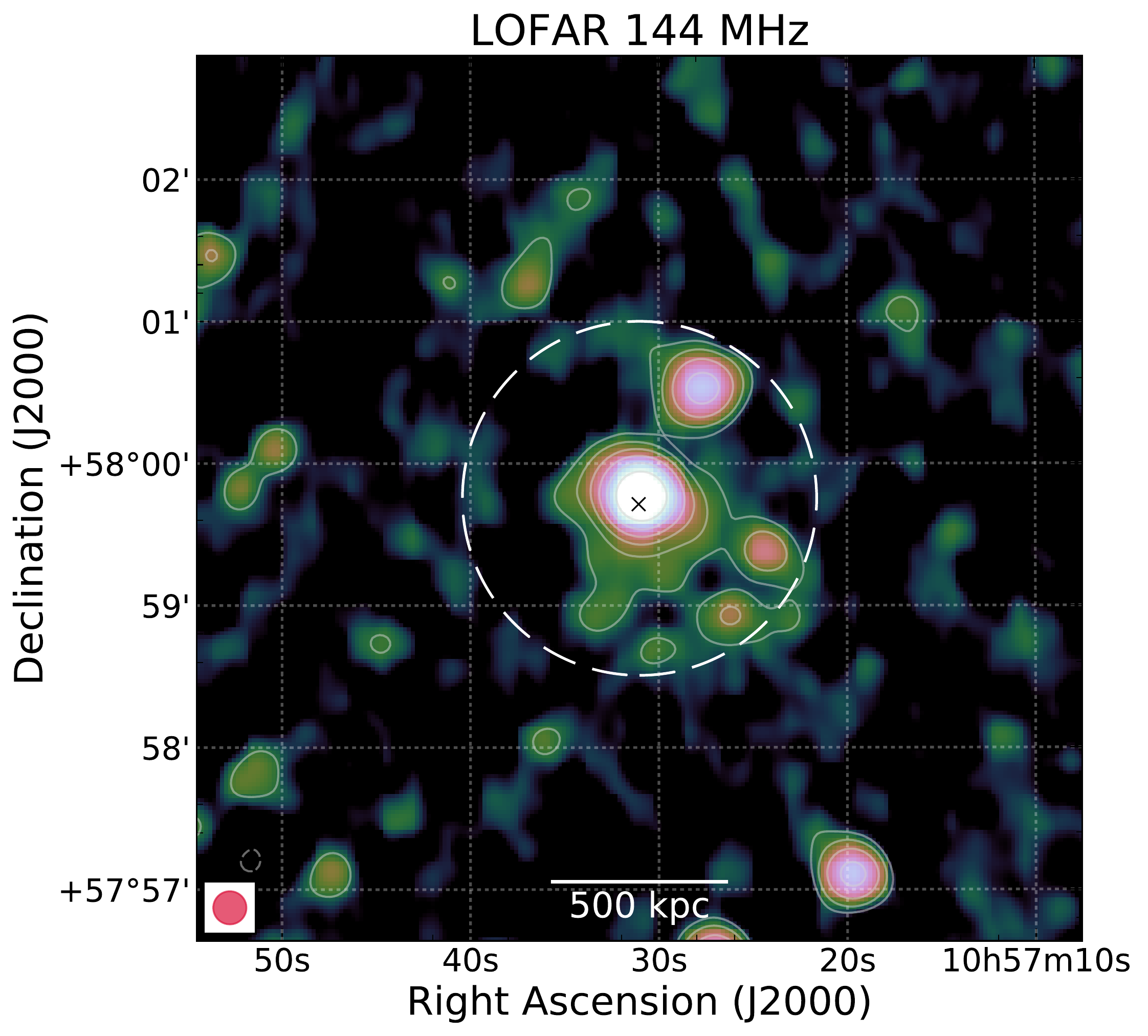}}
{\includegraphics[width=0.43\textwidth]{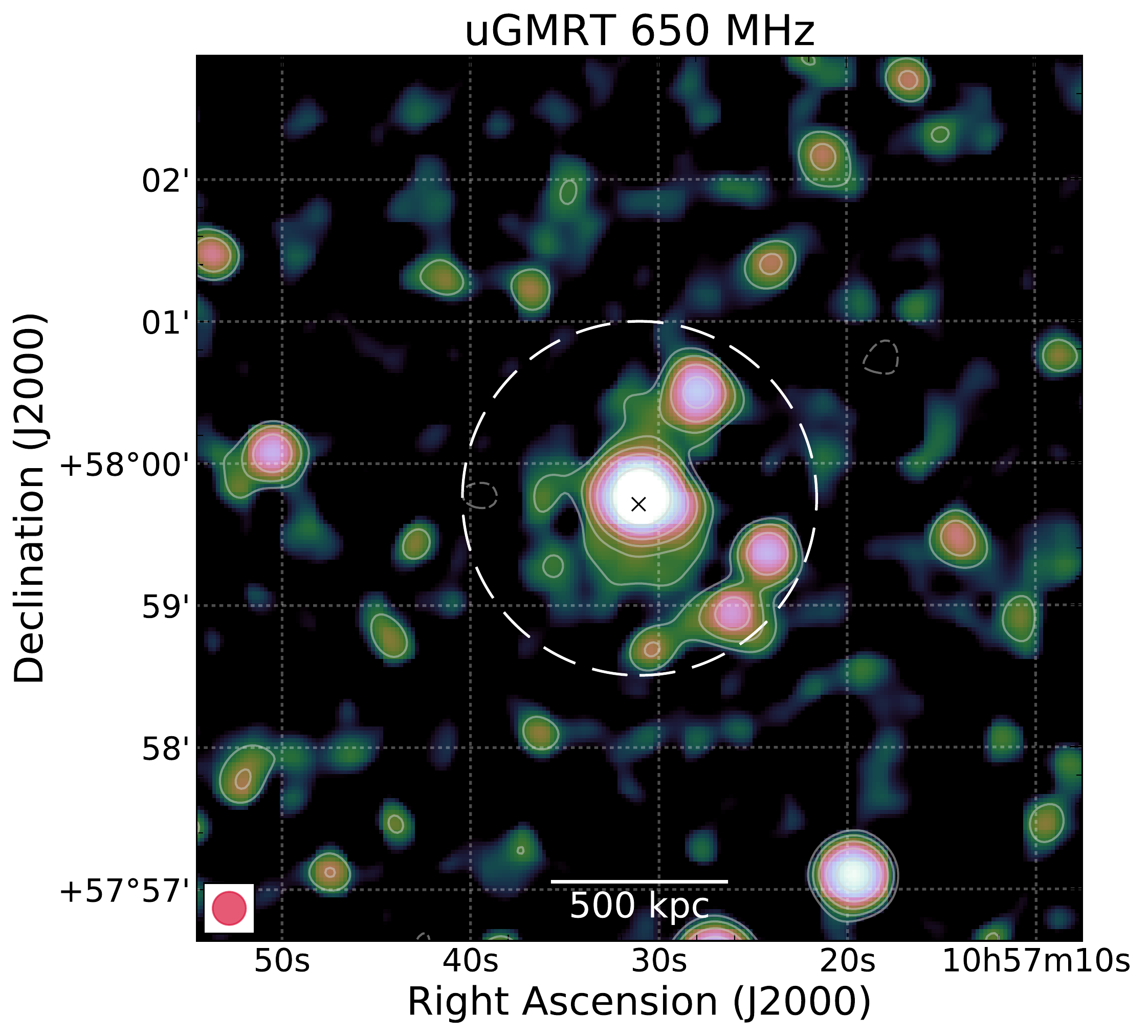}}
{\includegraphics[width=0.43\textwidth]{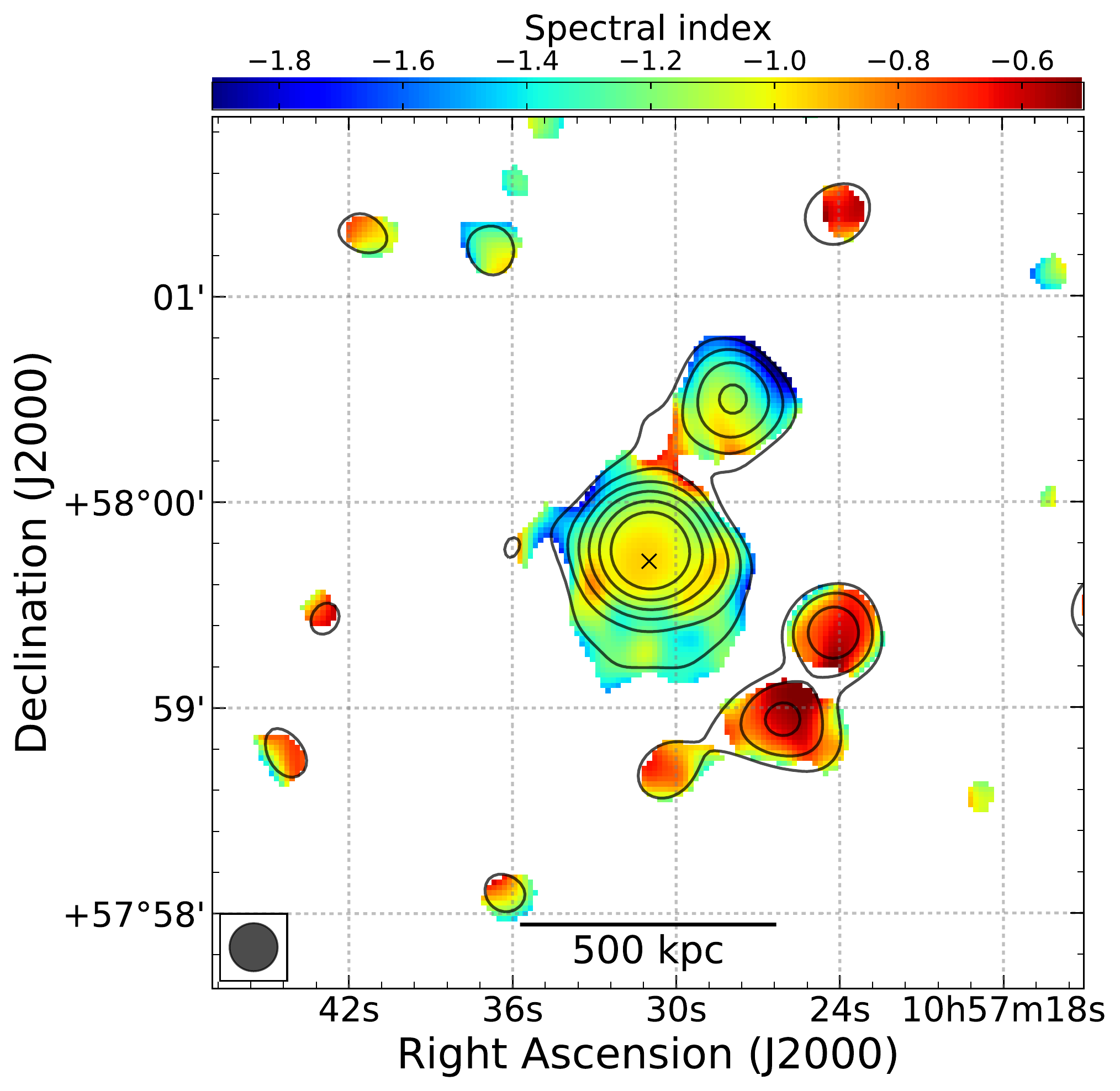}}
{\includegraphics[width=0.43\textwidth]{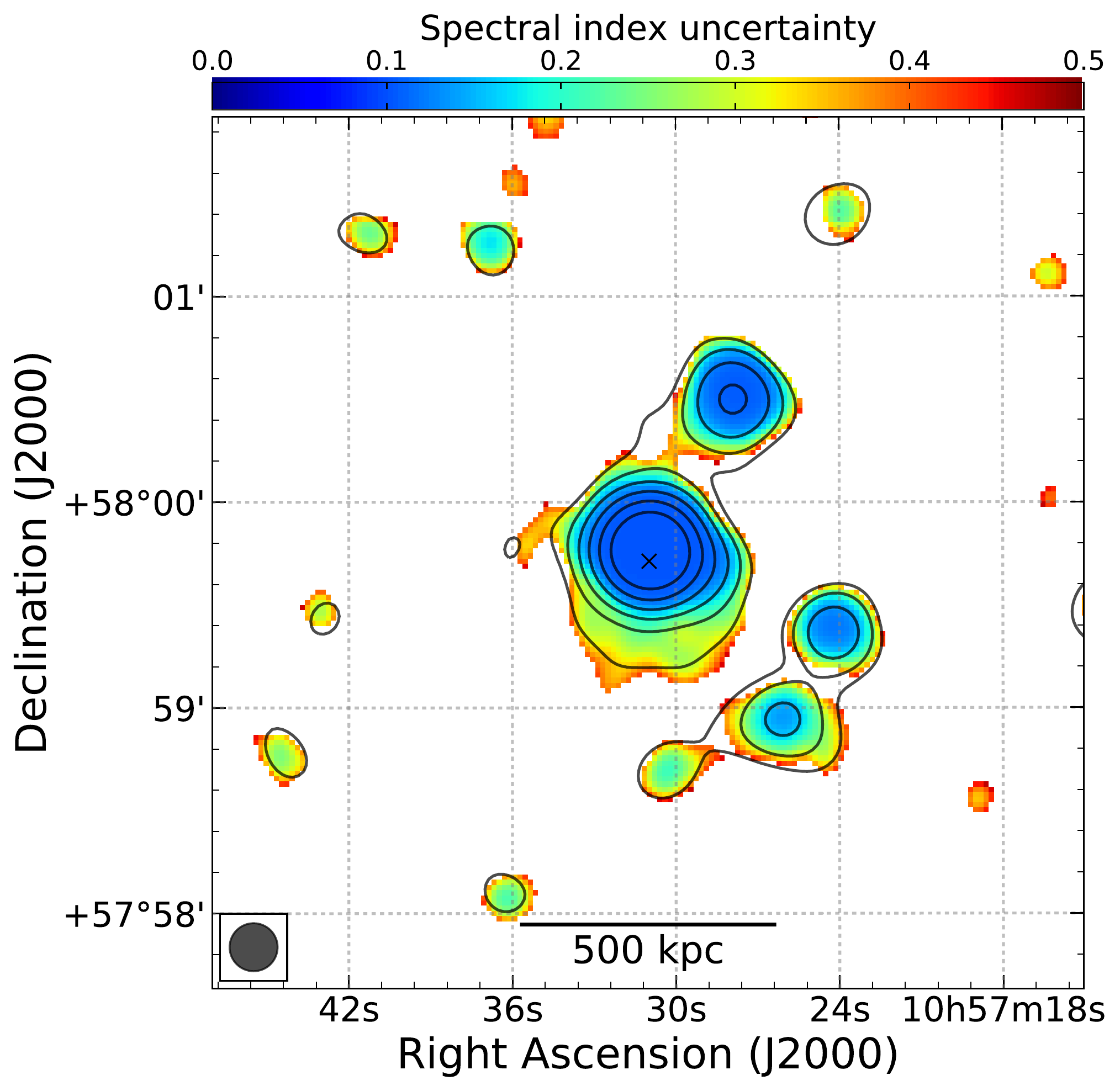}}
\caption{PSZ2\,G147.88+53.24. Top and central rows: Full-resolution and $14''$ images (\texttt{weighting=`Briggs'} and \texttt{robust=-0.5}) at 144 MHz (right) and 650 MHz (left). White-coloured radio contours are drawn at levels of $2.5\sigma_{\rm rms}\times[-1, 1, 2, 4, 8, 16, 32]$, with $\sigma_{\rm rms}$ being the noise level at each frequency (see Table \ref{tab:images_parameters}). The negative contour level is drawn with a dashed white line. The dashed white circle in each map shows the $R=0.5R_{\rm SZ,500}$ region, obtained from $M_{\rm SZ,500}$, {with the cross showing the cluster centre}. Bottom row: Spectral index map between 144 and 650 MHz at $14''$ resolution, and corresponding uncertainty map (left and right panels, respectively). uGMRT radio contours at 650 MHz are drawn in black, at the levels $3\sigma_{\rm rms}\times[-1, 1, 2, 4, 8, 16, 32]$, with $\sigma_{\rm rms}$ being the noise level (see Table \ref{tab:images_parameters}).}\label{fig:images_147}
\end{figure*}

\begin{figure*}
\centering
{\includegraphics[width=0.45\textwidth]{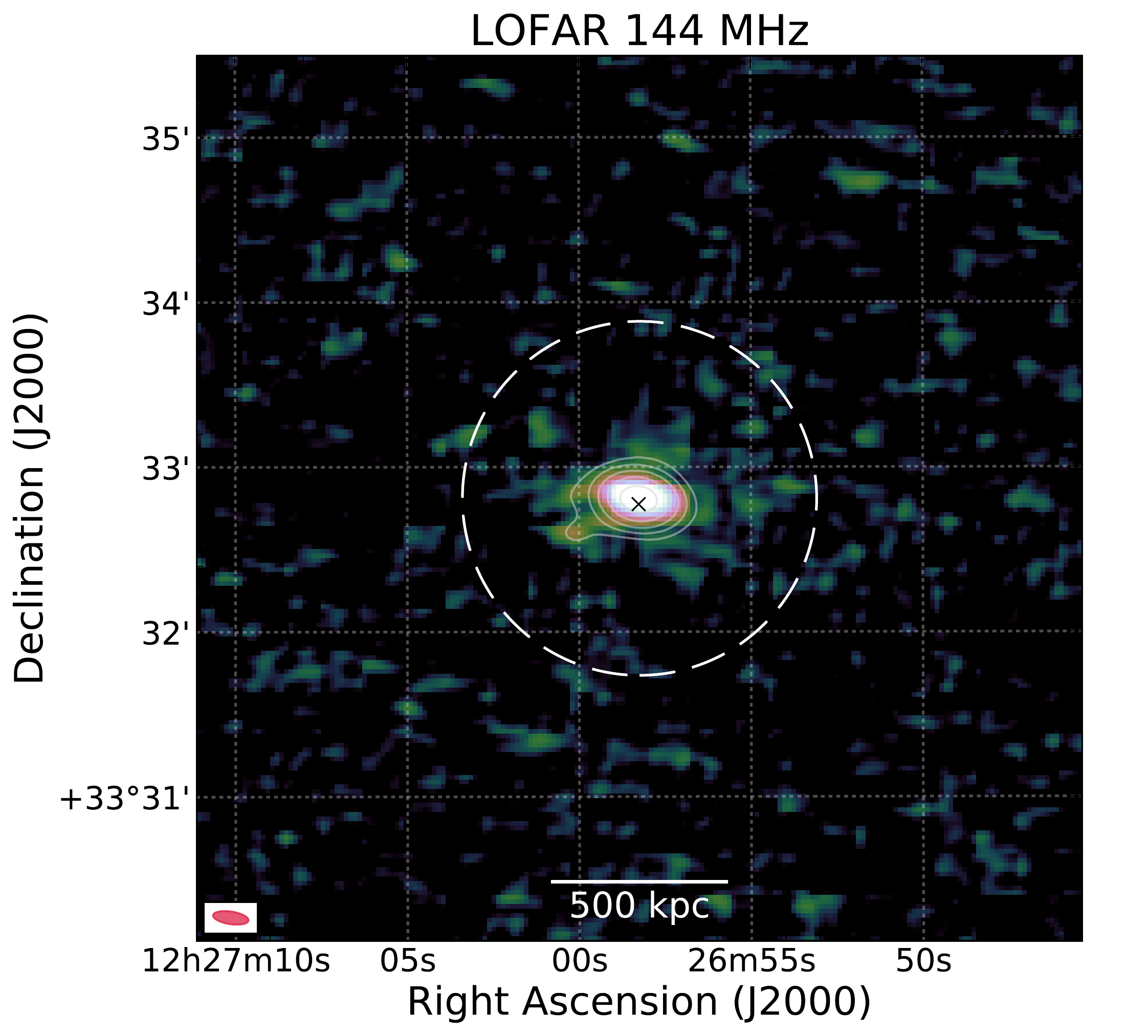}}
{\includegraphics[width=0.45\textwidth]{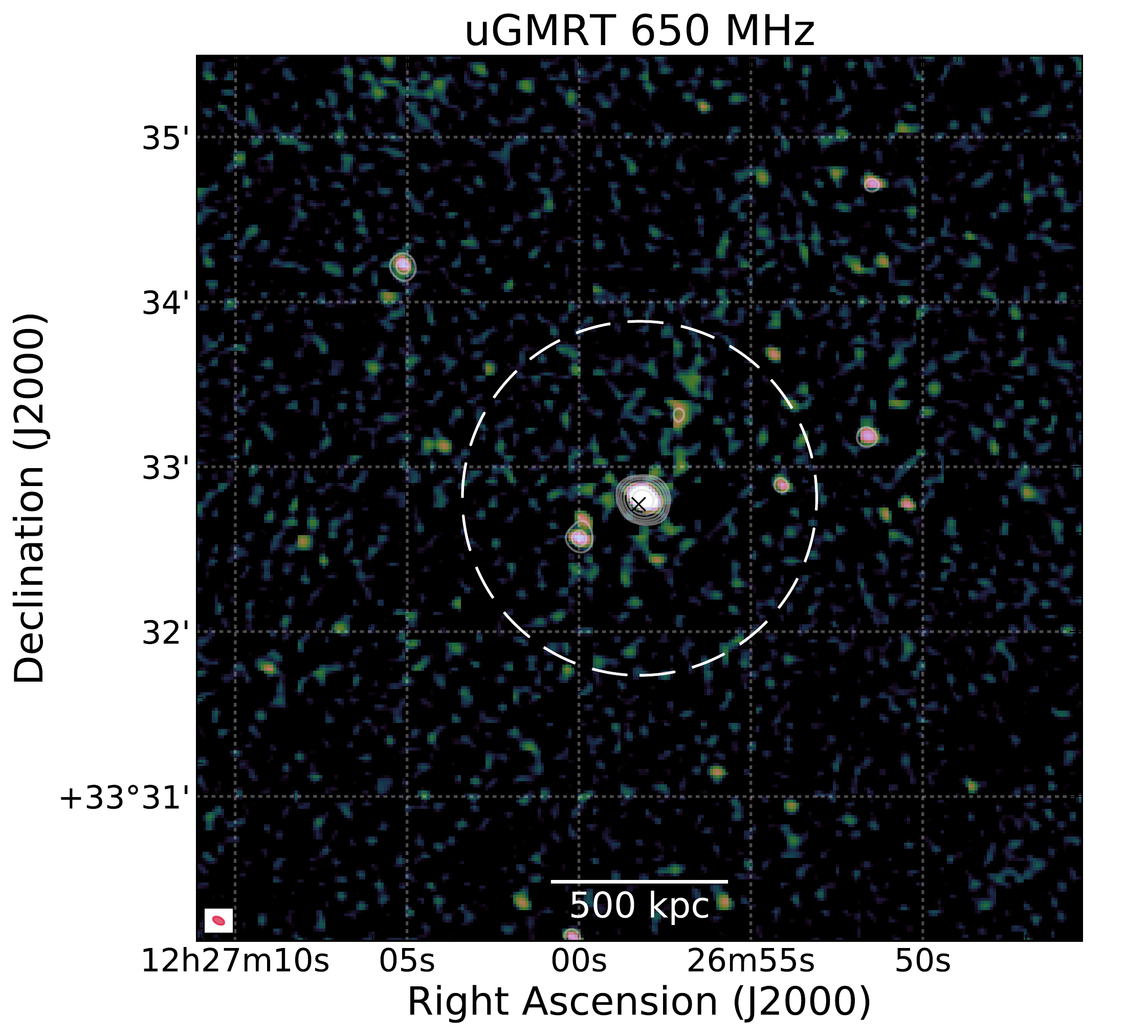}}\\
{\includegraphics[width=0.45\textwidth]{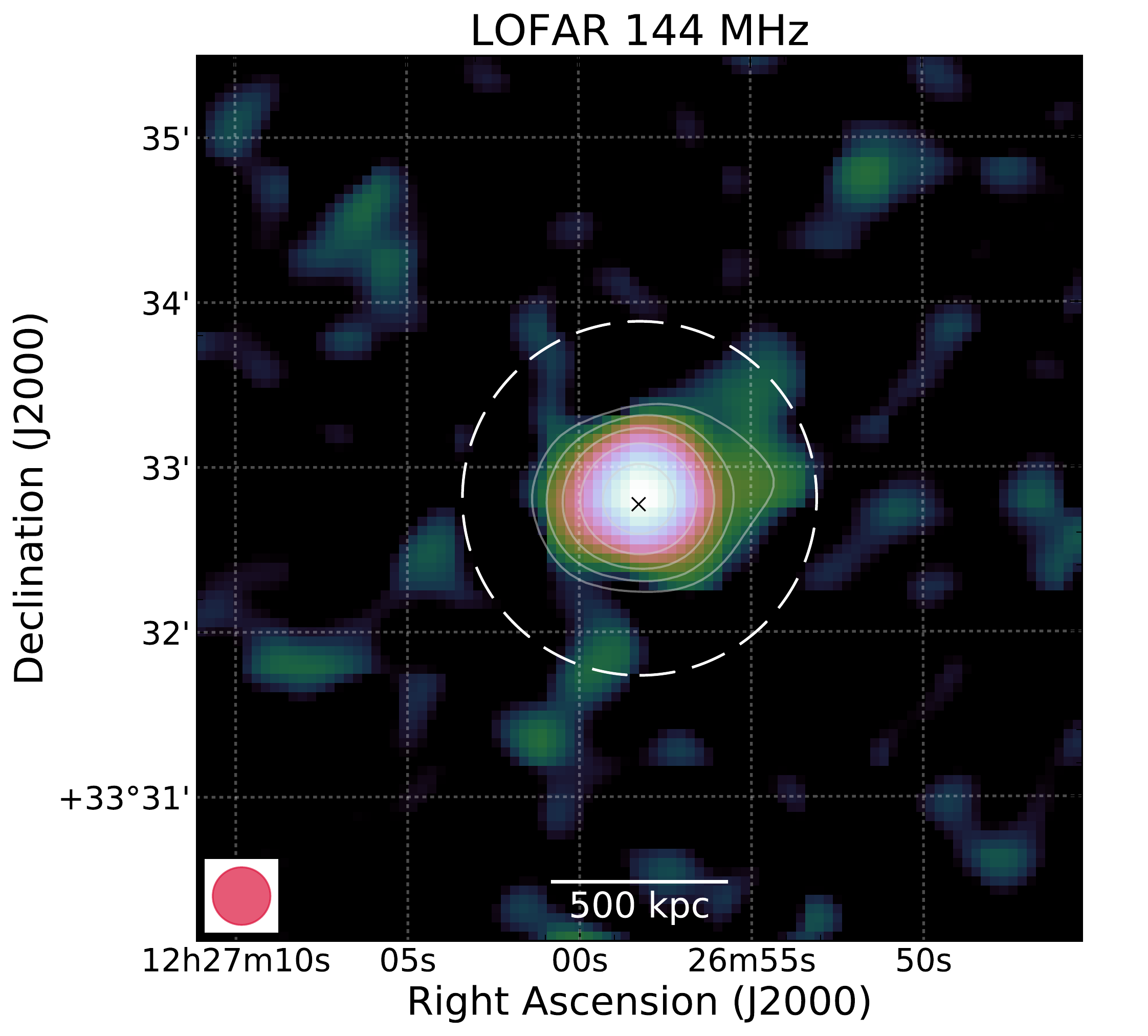}}
{\includegraphics[width=0.45\textwidth]{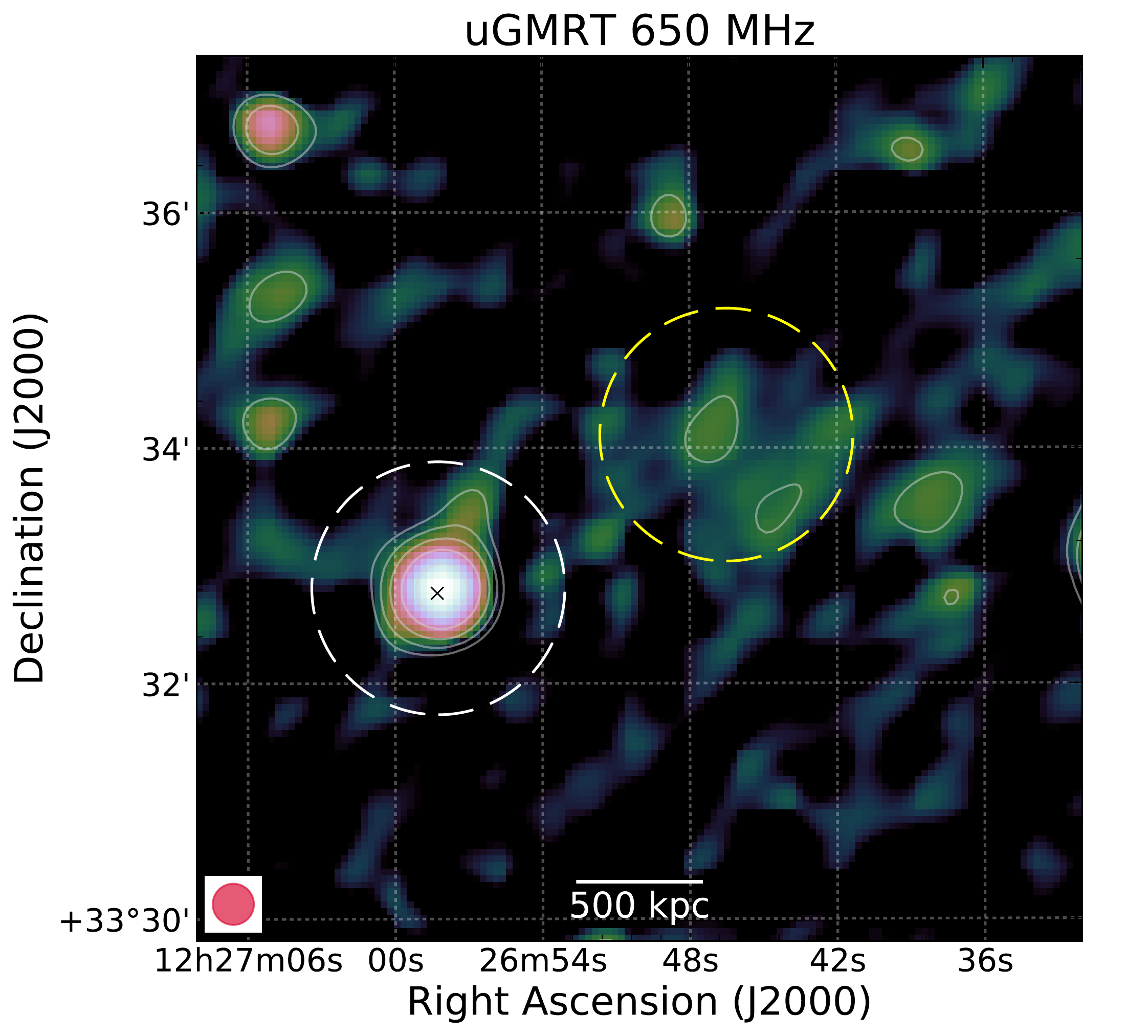}}\\
\caption{PSZ2\,G160.83+81.66. Top and bottom rows: Full-resolution and $21''$
images (\texttt{weighting=`Briggs'} and \texttt{robust=-0.5}) at 144 MHz (left) and 650 MHz (right). White-coloured radio contours are drawn at levels of $2.5\sigma_{\rm rms}\times[-1, 1, 2, 4, 8, 16, 32]$, with $\sigma_{\rm rms}$ being the noise level at each frequency (see Table \ref{tab:images_parameters}. The negative contour level is drawn with a dashed white line. The dashed white circle in each map shows the $R=0.5R_{\rm SZ,500}$ region, obtained from $M_{\rm SZ,500}$, {with the cross showing the cluster centre}. The dashed yellow circle in the bottom right panels shows the position of the mock radio halo.}\label{fig:images_160}
\end{figure*}

\subsection{Individual clusters}
In this section we provide  a brief description of each cluster, at all the observing frequencies. The results are summarised in Table \ref{tab:fluxes}.

\subsubsection{PSZ2\,G086.93+53.18}\label{sec:086}

This is the faintest radio halo found in the LOFAR observations ($S_{144}=6.9\pm1.3$ mJy), with a largest linear size $\rm LLS_{144}=0.4-0.5$ Mpc. No diffuse radio emission is visible at 650~MHz, despite the better depth of the observation (see Fig. \ref{fig:images_086}). Using the mock halo injection, we are able to detect diffuse radio emission with a flux density of $S_{650}<0.7$~mJy ($I_0=1.4~\mu{\rm Jy~arcsec}^{-2}$ and $r_e=60$ kpc), corresponding to an integrated spectral index of about $-1.5$. 

\subsubsection{PSZ2\,G089.39+69.36}\label{sec:089}
A megaparsec-scale radio halo is found in the LOFAR observations ($S_{144}=10.0\pm1.6$ mJy, $\rm LLS_{144}=1$ Mpc). No diffuse radio emission is observed in the 400~MHz image (Fig. \ref{fig:images_089}).  Using the mock halo injection, we are able to detect diffuse radio emission with a flux density of $S_{400}<1.9$~mJy ($I_0=3.4~\mu{\rm Jy~arcsec}^{-2}$ and $r_e=160$ kpc), corresponding to an integrated spectral index of about $-1.6$.

\subsubsection{PSZ2\,G091.83+26.11}
The radio halo in this cluster is the brightest in our sample at all three frequencies (see Fig. \ref{fig:images_091}). The largest linear size of the radio halo is the same in both the LOFAR and uGMRT observations, namely about 1.2~Mpc. The 650 MHz observation is the deepest among the others, assuming $\alpha=-0.8$. At this frequency, we also see substructures in the halo. 
We measure a flux density of $S_{144}=65.4\pm9.9$ mJy, $S_{400}=23.9\pm2.0$ mJy, and $S_{650}=14.5\pm0.8$ mJy for the LOFAR, uGMRT Band 3, and uGMRT Band 4 observations. This corresponds to integrated spectral indices of $\alpha_{650}^{144}=-1.00\pm0.11$, $\alpha_{400}^{144}=-0.99\pm0.17,$ and $\alpha_{650}^{400}=-1.03\pm0.21$. These values are consistent with a single power-law spectral shape. 
The spectral index map shows a steeper spectral index in the central part of the cluster, with $\alpha_{400}$ between $\sim-1.2$ and $\sim-1.4$. Northward, the spectral index gets flatter, with $\alpha_{650}^{144}\sim-0.75$, in correspondence with sources C and D.

In the east and south-east directions, the elongated  source that was classified as a candidate radio relic in \cite{digennaro+20} maintains its morphology. This source can be divided into two pieces: One (R1) located east of the cluster centre, is a faint patchy filament that is {$80''$ wide and $8''$ long}, corresponding to $640\times60$ kpc$^2$ at the cluster redshift; the other (R2), located southeast of cluster centre, is brighter and extends for about $40''$ (i.e. 300 kpc at the cluster redshift). Interestingly, its morphology resembles a double-lobe radio galaxy at 650 MHz. However, no optical counterpart is visible from the available PanSTARRS  \citep[Panoramic Survey Telescope and Rapid Response System;][]{panstarss16} optical image (see Appendix \ref{apx:optical_images}). However, we note that the optical image is rather shallow, and might miss faint galaxies. It is possible that R2 combines the emission of a radio relic and a double-lobe radio galaxy. Observations with the Karl-Jansky Very Large Telescope (VLA) in the 1--4 GHz band will help in the  classification of this elongated piece of emission. In particular, the polarisation characteristics will be crucial and will be presented in a forthcoming paper (Di Gennaro et al., in prep). 
We measure the flux density for the candidate radio relic from the low-resolution images (i.e. $14''\times14''$), considering the full length of the source (i.e. R1+R2). 
At this resolution, the compact sources B, C, and D are embedded in the candidate relic, and so we measured their flux densities from the full-resolution image and we subtracted them arithmetically from the total flux density. We obtain $S_{144}=274.8\pm45.6$ mJy, $S_{400}=76.6\pm6.5$ mJy, and $S_{650}=36.3\pm2.1$ mJy, corresponding to $\alpha_{650}^{144}=-1.34\pm0.12$ ($\alpha_{400}^{144}=-1.25\pm0.18$ and $\alpha_{650}^{400}=-1.54\pm0.21$). The spectral index map shows hints of steepening for R2 (up to $\alpha_{400}\sim-2$), which is typical of of radio relics \citep[e.g.][]{digennaro+18,rajpurohit+17}. This is not observed for R1. 
Next to the candidate relic, source B is characterised by a very steep spectrum ($\alpha_{400}\sim-2$).

\subsubsection{PSZ2\,G099.86+58.45}
We detect diffuse radio emission in the 650 MHz observations, similar to what is visible at 144~MHz (Fig. \ref{fig:images_099}, $\rm LLS_{144}=1.2$ and $\rm LLS_{650}=0.95$ Mpc). 
We measure a flux density of $S_{144}=18.1\pm2.9$ mJy and $S_{650}=4.0\pm0.4$ mJy for the LOFAR and uGMRT Band 4 observations. This corresponds to an integrated spectral index of $\alpha_{650}^{144}=-1.00\pm0.13$. The spectral index map in Fig. \ref{fig:images_099} shows steeper values at the cluster centre (i.e. $\alpha_{650}^{144}\sim-1.6$) and flatter at the cluster outskirts (i.e. $\alpha_{650}^{144}\sim-0.9$).
For this cluster,  L-band VLA observations are also available \citep{cassano+19}. In these observations, hints of a halo are present only at the $2\sigma_{\rm rms}$ level (with $\rm \sigma_{rms,VLA} = 20~\mu Jy~beam^{-1}$). We repeated the flux measurement, covering the same region as the LOFAR halo and finding a flux of $S_{1500}\sim1.0\pm0.4$ mJy, in agreement with the flux density reported in \cite{cassano+19}. This VLA flux density suggests a steepening towards GHz frequencies, with $\alpha_{1500}^{650}\sim-1.7\pm0.5$. 

As for PSZ2\,G091.83+26.11,  we also detect ultra-steep spectra from source A ($\alpha_{600}^{144}\sim-1.7$) and source C ($\alpha^{144}_{650}\sim-2.5$) in this cluster, as was also mentioned by \cite{cassano+19}.

\subsubsection{PSZ2\,G126.28+65.62}\label{sec:126}
No diffuse radio emission is found in the uGMRT 650 MHz observations (see right panel Fig. \ref{fig:images_126}, $\rm LLS_{144}=0.8$ Mpc). We measure $S_{144}=8.9\pm1.2$ mJy for the radio halo. Using the mock halo injection, we are able to detect diffuse radio emission with a flux density of $S_{650}<1.0$~mJy ($I_0=1.0~\mu{\rm Jy~arcsec}^{-2}$ and $r_e=130$ kpc). This corresponds  to a spectral index of about $-1.5$. 

\subsubsection{PSZ2\,G141.77+14.19}
Hints of the presence of diffuse emission are present in the uGMRT 650 MHz data around sources D, E, F, and G (see right panel Fig. \ref{fig:images_141}, $\rm LLS_{144}=0.6$ and $\rm LLS_{650}=0.55$ Mpc).
We measure a flux density of $S_{144}=6.5\pm1.2$ mJy and $S_{650}= 1.2\pm0.1$ mJy from the same halo region. This corresponds to a spectral index of $\alpha_{650}^{144}=-1.12\pm0.13$, and it agrees with the values found in the spectral index map.

\subsubsection{PLCK\,G147.3--16.6}
Observations at 610 MHz with the GMRT were published by \cite{vanweeren+14}, where a radio halo was discovered. With the new wide-band GMRT observations we confirm the presence of a megaparsec-size radio halo at both 400 and 650 MHz (Fig. \ref{fig:images_plck147}). Interestingly, in the two uGMRT observations, the diffuse emission appears to be larger than on the LOFAR image ($\rm LLS_{144}=0.8$, $\rm LLS_{400}=1$ and $\rm LLS_{650}=1$ Mpc). However, we note that this observation is less deep, which is probably due to a bad ionosphere. 
The flux densities encompassed in the area covered by the halo in the LOFAR images (i.e. solid yellow region, see Appendix \ref{apx:sub_images}) are $S_{144}=21.0\pm3.7$ mJy, $S_{400}=5.4\pm0.6$ mJy, and $S_{650}=2.8\pm0.3$ mJy for the 144 MHz, 400 MHz, and 650 MHz observations. 
Increasing the region to cover the full extension of the halo in the uGMRT images (i.e. dashed yellow region, see Appendix \ref{apx:sub_images}), we obtain $S_{144}=26.6\pm4.4$ mJy, $S_{400}=10.0\pm0.9$ mJy, and $S_{650}=5.6\pm0.4$ mJy.
We note that the 650 MHz flux we report is slightly below the one found by \cite{vanweeren+14}. This is probably due to a better subtraction of the contribution of the compact sources with the deeper wide-band observations. Given the integrated flux densities, we obtain spectral indices of $\alpha_{650}^{144}=-1.34\pm0.15$, $\alpha_{400}^{144}=-1.33\pm0.20,$ and $\alpha_{650}^{400}=-1.35\pm0.37$, in the small area, and $\alpha_{650}^{144}=-1.03\pm0.12$, $\alpha_{400}^{144}=-0.96\pm0.19,$ and $\alpha_{650}^{400}=-1.19\pm0.23$, in the big area.  
The spectral index map in Fig. \ref{fig:images_plck147} shows steeper spectral index in the halo centre, with $\alpha_{400}\sim-1.4$, in agreement with the integrated spectral indices in the small halo region.

\subsubsection{PSZ2\,G147.88+53.24}
Diffuse radio emission at 144 MHz was reported in \cite{digennaro+20}, with $\rm LLS_{144}=0.6$ Mpc. This is also confirmed by deep observations at the same frequency \citep{osinga+20}. Hints of diffuse radio emission are visible in the 650~MHz image (see Fig. \ref{fig:images_147}, $\rm LLS_{650}=0.5$ Mpc). Excluding sources D, E, F, and G from the halo region, we measure $S_{144}=8.2\pm1.3$ mJy and $S_{650}=1.1\pm0.2$ mJy for the LOFAR and uGMRT Band 4 observation respectively. This corresponds to a spectral index of $\alpha_{650}^{144}=-1.33\pm0.17$. 
The spectral index map for PSZ2\,G147.88+53.24 is mostly dominated by the central compact source (i.e. source A, see Fig. \ref{fig:images_147}), which is characterised by a spectral index $\alpha_{650}^{144}\sim-1$. Just south of this, we detect steeper spectral index values ($\alpha_{650}^{144}\sim-1.3$) that can be associated with the radio halo.

\subsubsection{PSZ2\,G160.83+81.66}\label{sec:160}
This cluster represents the most distant radio halo found so far, at a redshift of 0.888 ($S_{144}=9.5\pm1.5$ mJy, $\rm LLS_{144}=0.7$ {see Fig. \ref{fig:images_160}}). No diffuse radio emission is visible at 650~MHz. Using the mock halo injection, we are able to detect diffuse radio emission with a flux density of $1.0$~mJy ($I_0=0.8~\mu{\rm Jy~arcsec}^{-2}$ and $r_e=110$ kpc). This corresponds to a spectral index of about $-1.5$.

\section{Discussion}
Investigating the spectral index properties of distant radio halos is crucial to understanding the mechanism of particle acceleration in these radio sources. 
So far, spectral studies in radio halos have been carried out starting from high-frequency (i.e. $\sim$ GHz) observations, which were then followed up at lower frequencies. However, this approach tends to miss a large fraction of steep-spectra sources simply because they are not detected in the GHz observations. This issue is particularly important for high-redshift clusters, as a large fraction of these halos, especially the low-mass systems, should have a steep spectral index \citep[$\alpha<-1.5$, see][]{cassano+brunetti05,cassano+06}. This is due to the larger contribution of inverse Compton energy losses (i.e. $dE/dt_{\rm IC}\propto(1+z)^4$), which is expected to hamper the acceleration of high-energy electrons. As a consequence, the ensuing synchrotron luminosity should be reduced. 
The LoTSS survey  will help to avoid this bias towards flat-spectra halos. This survey at low frequencies is expected to observe a large number of previously undiscovered radio halos \citep[e.g.][]{cassano+10,vanweeren+20}, which can be followed up at higher frequencies.

\begin{figure*}
\centering
{\includegraphics[width=0.32\textwidth]{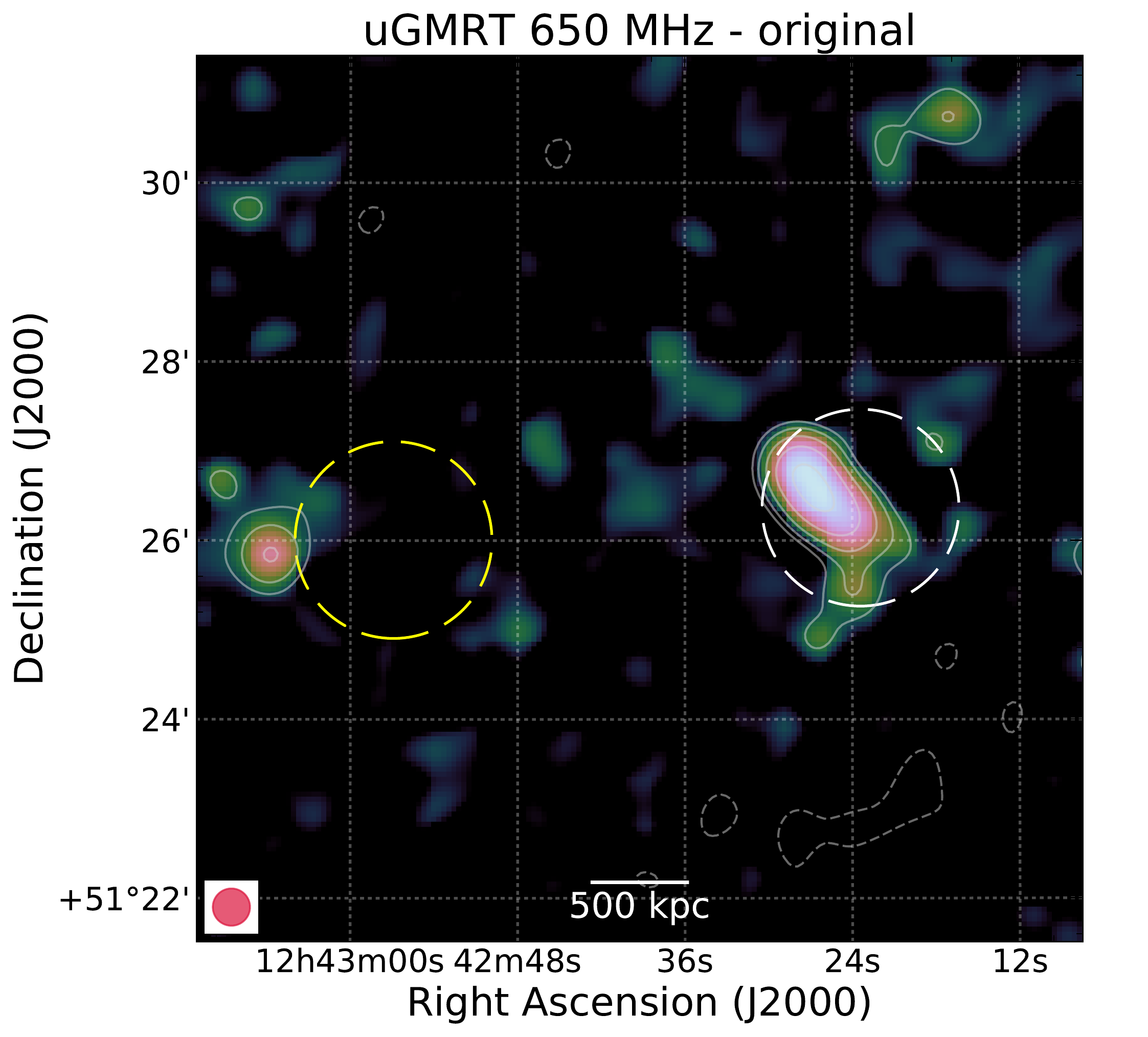}}
{\includegraphics[width=0.32\textwidth]{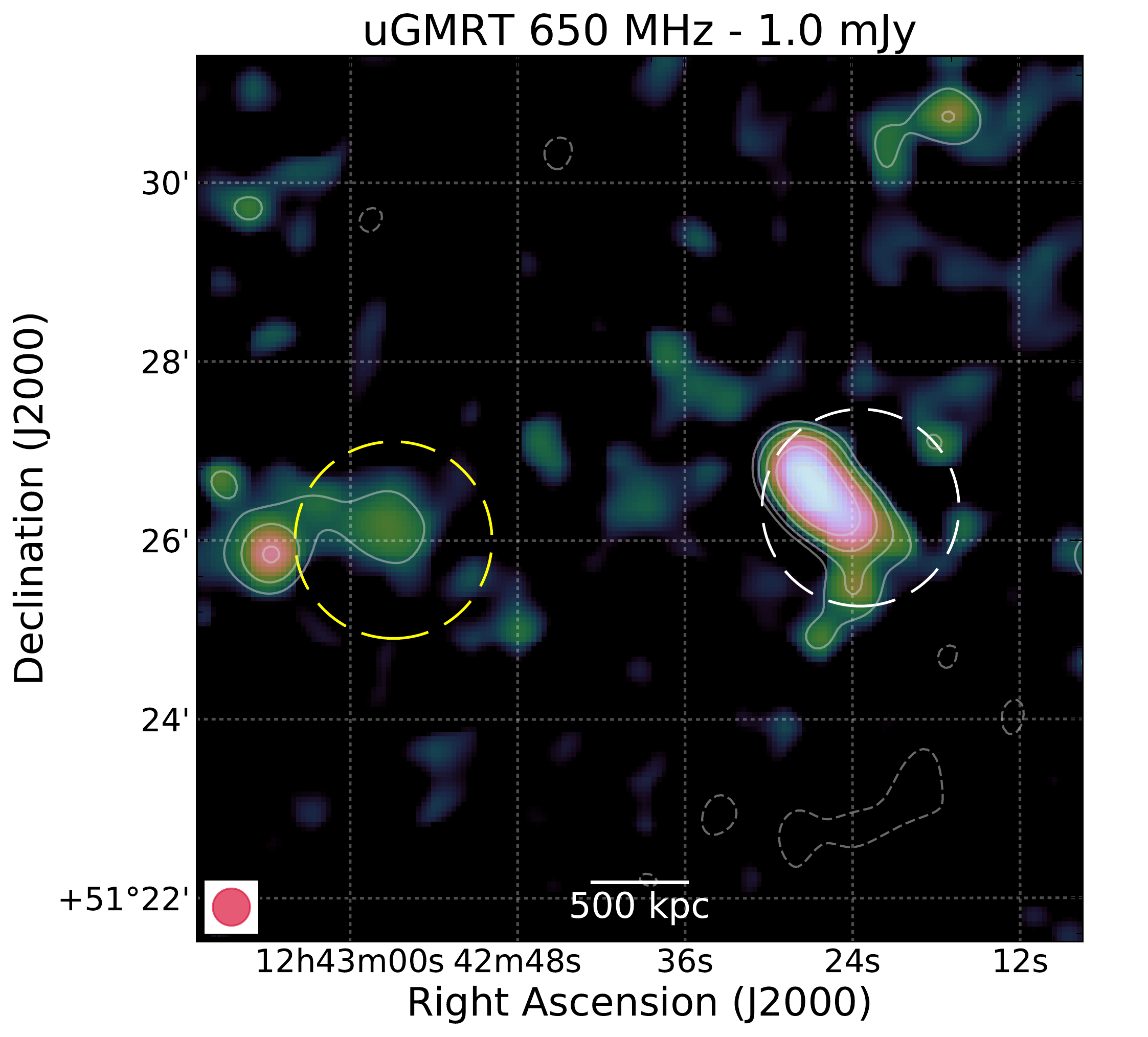}}
{\includegraphics[width=0.32\textwidth]{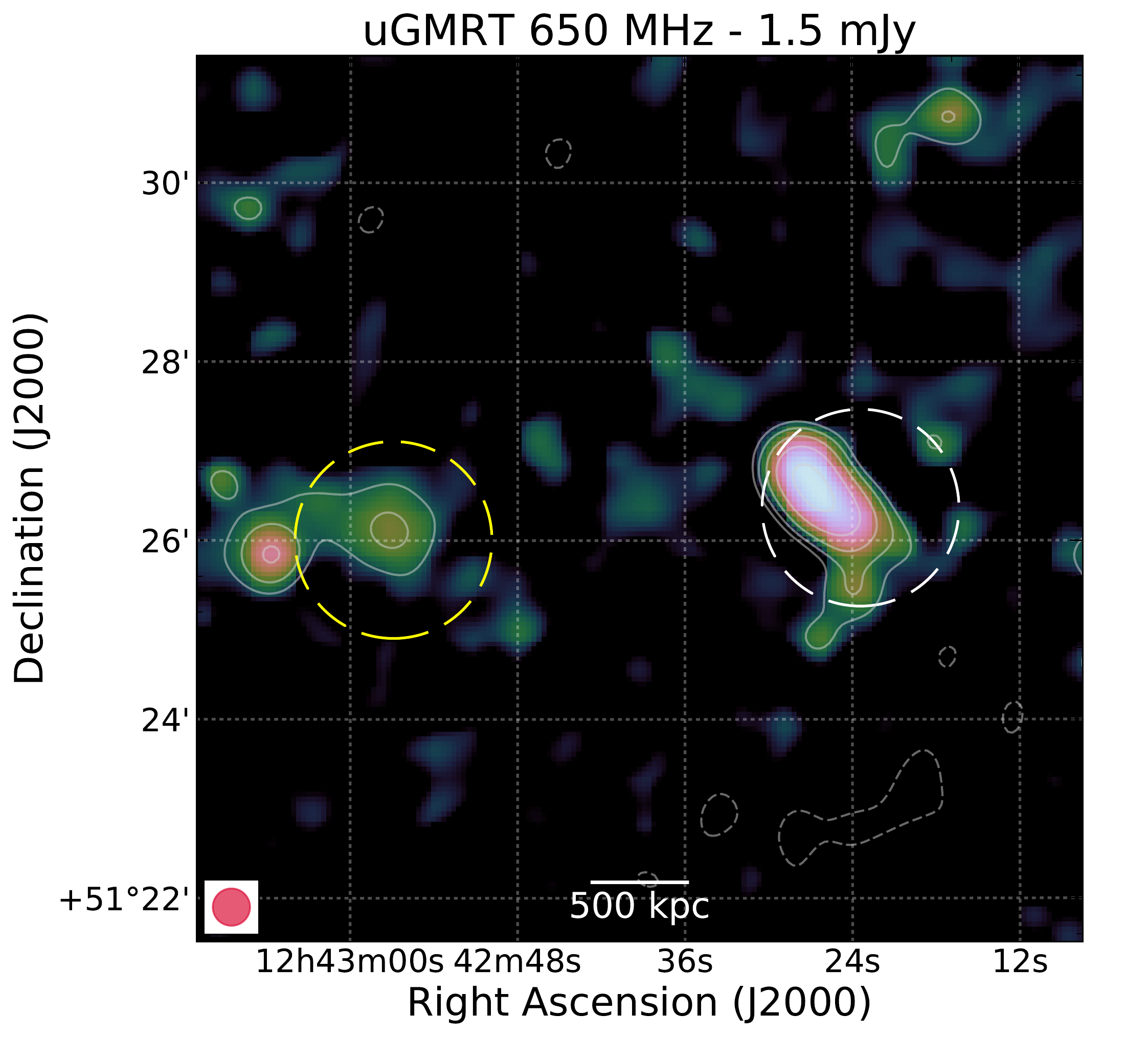}}
\caption{Example of injection of mock radio halos in PSZ2\,G126.28+65.62. Left: Original image. Middle: Injected radio halo with a flux density of 1.0 mJy. Right: Injected radio halo with a flux density of 1.5 mJy. Radio contours are displayed starting from the $2\sigma_{\rm rms}$ level. The two circles display the $R_{500}$ area, with the white one centred on the cluster {(see cross)} and the yellow one centred on the injected halo.}
\label{fig:mock}
\end{figure*}

\begin{figure}
\centering
\includegraphics[width=0.48\textwidth]{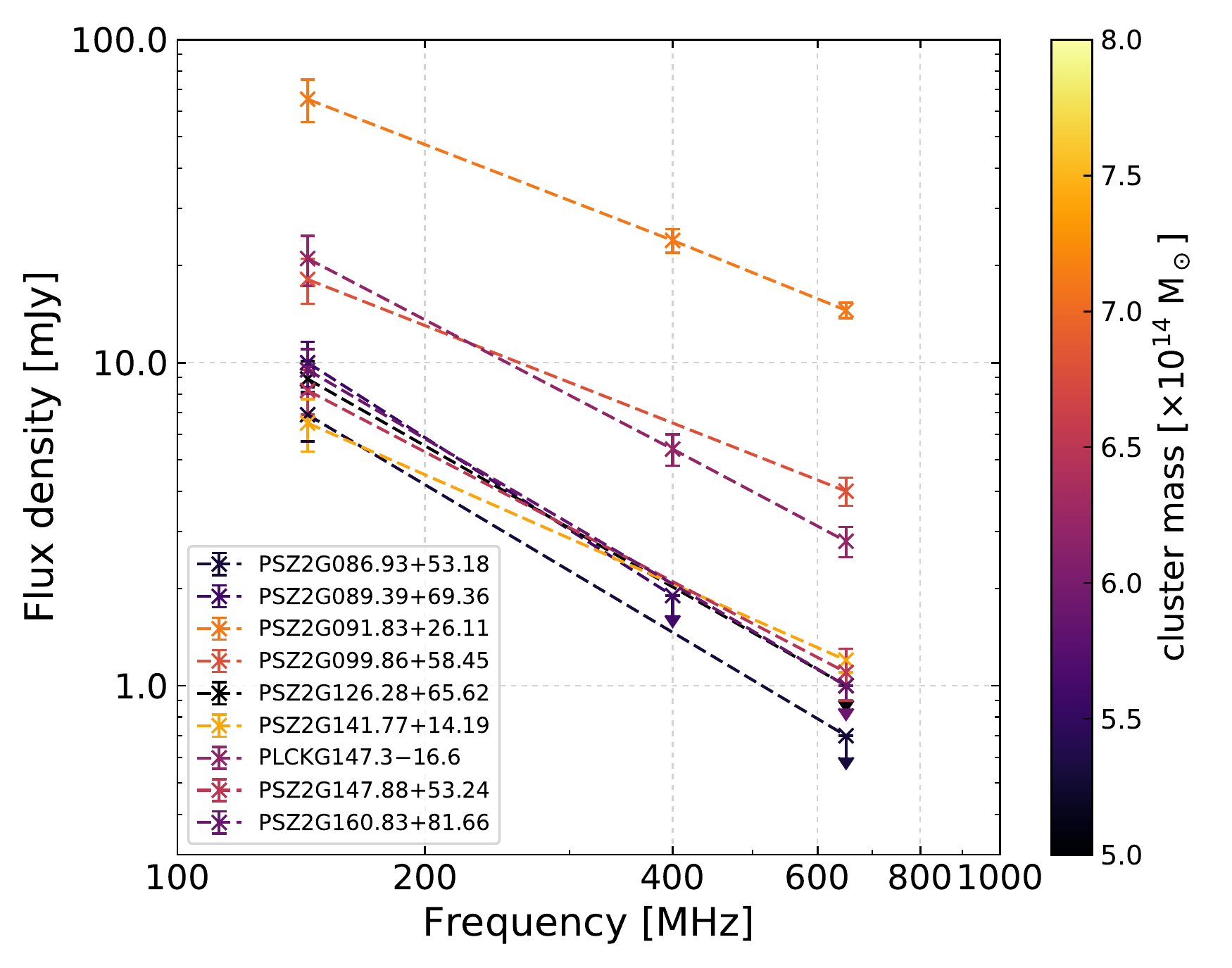}
\caption{Radio spectra of the clusters in our sample (Table \ref{tab:fluxes}). The arrows show the upper limits on the flux densities.}
\label{fig:fluxes}
\end{figure}

In \cite{digennaro+20} we presented the first statistical study of diffuse radio emission with LOFAR (120--168 MHz) of a sample of 19 distant ($z\geq0.6$) galaxy clusters selected from the Planck SZ catalogue \citep{planckcoll16}.  
In the present work, we present a follow-up study at higher frequencies of the nine radio halos detected in the LOFAR observations.
Our observations were carried out with the uGMRT, mainly at 550--900 MHz (Band 4), but for two clusters (i.e. PSZ2\,G091.83+26.11 and PLCK\,G147.3--16.6) we also obtained observations at 250--550 MHz (Band 3). At these higher frequencies, we find the presence of diffuse radio emission in five of the nine clusters discovered in LOFAR.
These are PSZ2\,G091.83+26.11, {PSZ2\,G099.86+58.45}, PSZ2\,G141.77+14.19, PLCK\,G147.3--16.6, and PSZ2\,G147.88+53.24. While we note that these clusters are also the most massive objects in our sample, with $M_{\rm 500,SZ}\sim6-8\times10^{14}$ M$_\odot$, 
the low-mass clusters in our sample (i.e. $M_{\rm 500,SZ}=5-6\times10^{14}~{\rm M_\odot}$; PSZ2\,G086.93+53.18, PSZ2\,G089.39+69.36, PSZ2\,G126.28+65.62 and PSZ2\,G160.83+81.66) do not show diffuse radio emission at the uGMRT frequencies and we could only derive upper limits on the flux density (Fig. \ref{fig:fluxes}).

For the uGMRT-detected radio halos, we measure integrated spectral indices between $-1$ and $-1.4$ (see Table \ref{tab:fluxes}). These values are similar to those found for classical halos in local clusters \citep[e.g.][]{feretti+12,vanweeren+19}. 
We also note that for those targets with three observing frequencies (i.e. PSZ2\,G091.83+26.11 and PLCK\,G147.3--16.6), no spectral curvature is present. This means that the steepening frequency $\nu_s$ has to be at higher frequencies (i.e. $\nu_s>650$ MHz). Hints of spectral steepening are indeed suggested with archival 1.4 GHz VLA observations of PSZ2\,G099.86+58.45 \citep{cassano+19}. 
For the non-detections, we tested the sensitivity of our uGMRT data by injecting mock halos with spectral indices of $-1.5$ (Sects. \ref{sec:086}, \ref{sec:126} and \ref{sec:160}) or $-1.6$ (Sect. \ref{sec:089}). In this case, we found that diffuse radio emission is observed above the $2\sigma_{\rm rms}$ level. As these halos are not detected in our uGMRT images, this implies that they should have a steeper spectral index.

\begin{table*}
\caption{Flux densities and largest linear sizes (LLS) of the radio halos in our sample. In the last column, we report the measured integrated spectral index of the radio halos between 144 MHz and 650 MHz. 
}
\vspace{-5mm}
\begin{center}
\resizebox{0.8\textwidth}{!}{
\begin{tabular}{lccccccccc}
\hline\hline
Cluster name &  $S_{144}$ & ${\rm LLS_{144}}$ & $S_{400}$ & ${\rm LLS_{400}}$ & $S_{650}$ & ${\rm LLS_{650}}$ & $\alpha_{650}^{144}$ \\
& & [mJy] & [Mpc] & [mJy]& [Mpc] & [mJy]& [Mpc] \\
\hline
PSZ2\,G086.93+53.18 & $6.9\pm1.3$ & 0.4--0.5 & -- & -- & $<0.7$ & N/A & -- \\  
PSZ2\,G089.39+69.36 &   $10.0\pm1.6$ & 1.0 & $<1.9$ & N/A & -- & -- & -- \\ 
PSZ2\,G091.83+26.11 &  $65.4\pm9.9$ & 1.2 & $23.9\pm2.0$ & 1.2 & $14.5\pm0.8$ & 1.2 & $-1.00\pm0.11$ \\
PSZ2\,G099.86+58.45 & $18.1\pm2.9$ & 1.2 & -- & -- & $4.0\pm0.4$ & 1.0 & $-1.00\pm0.13$ \\
PSZ2\,G126.28+65.62 & $8.9\pm1.2$ & 0.8 & -- & -- & $<1.0$ & N/A & -- \\ 
PSZ2\,G141.77+14.19 & $6.5\pm1.2$ & 0.6 & -- & -- & $1.2\pm0.1$ & 0.55 & $-1.12\pm0.13$ \\
PLCK\,G147.3–16.6 & $21.0\pm3.7$ & 0.8 & $5.4\pm0.6$ & 1.0 & $2.8\pm0.3$ & 1.0 & $-1.34\pm0.20$ \\
PSZ2\,G147.88+53.24 & $8.2\pm1.3$ & 0.6 & -- & -- & $1.1\pm0.2$ & 0.5 & {$-1.33\pm0.17$} \\
PSZ2\,G160.83+81.66 & $9.5\pm1.5$ & 0.7 & -- & -- & $<1.0$ & N/A & -- \\ 
\hline
\end{tabular}}
\end{center}
\vspace{-5mm}
\tablefoot{The LOFAR flux densities agree within $1\sigma$ with the values reported in \cite{digennaro+20}. The flux densities reported for PLCK\,G147.3–16.6 refer to the small halo region (see Appendix \ref{apx:sub_images}). 
}
\label{tab:fluxes}
\end{table*}

\subsection{Comparison with theoretical models}
\begin{figure}
\centering
\includegraphics[width=0.48\textwidth]{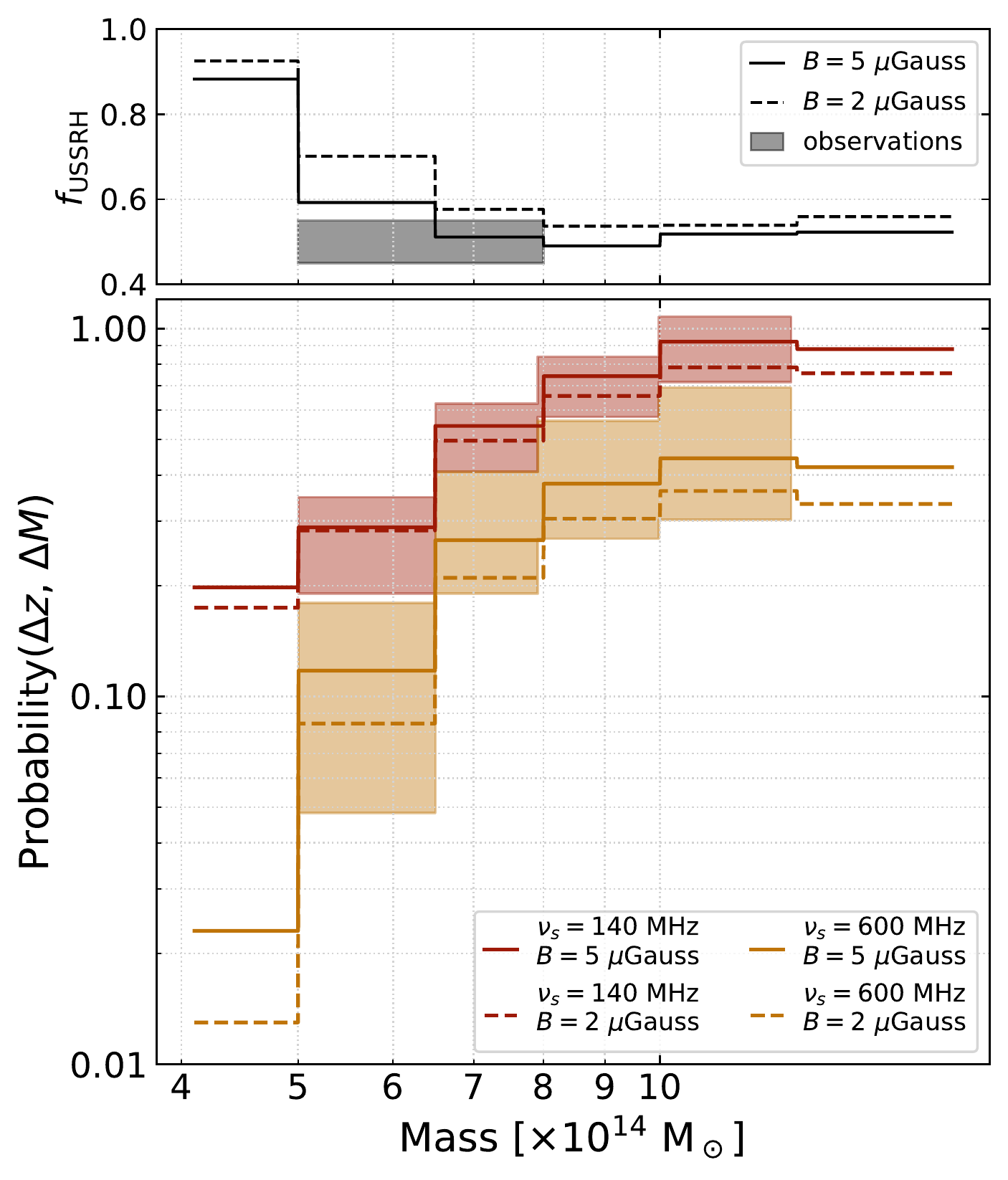}
\caption{Probability of forming radio halos with $\nu_s>140$ and $\nu_s>600$ MHz (red and yellow lines, respectively) as a function of the cluster virial mass in the redshift range 0.6-0.7. Magnetic fields of $B=5 ~\mu$Gauss and $B=2 ~\mu$Gauss are assumed (solid and dashed lines, respectively). The shadowed regions represent the $1\sigma$ uncertainty derived through Monte Carlo calculations.
Top panel: Expected fraction of USSRH visible at 144 MHz with steepening frequency $\nu_s<600$ MHz as a function of the cluster virial mass, assuming $B=5 ~\mu$Gauss and $B=2 ~\mu$Gauss (solid and dashed lines, respectively). The shadowed area indicates the observed fraction of USSRH in our sample.}
\label{fig:halo_prob}
\end{figure}

Although our sample is not designed to test the occurrence of radio halos in high-redshift systems (because of its small size and low completeness), we can compare our results with model expectations.
According to re-acceleration models, the possibility to detect diffuse radio emission depends on the steepening frequency $\nu_s$ of the synchrotron spectra. Therefore, halos can be observed only at  $\nu\leq\nu_s,$ with $\nu$ being the observing frequency. 
Following the procedure in \cite{cassano+19}, we estimated the probability of forming a radio halo as a function of cluster mass at a median redshift of $z=0.7$ and with a steepening frequency of $\nu_s\geq140$ MHz and $\nu_s\geq600$ MHz (see Fig. \ref{fig:halo_prob}). The formation probability takes into account the cluster merger history (merger trees), the generation of turbulence, particle acceleration (including energy losses) and the resulting cluster synchrotron spectrum. In these models, the turbulent energy, acceleration rate, and magnetic field per volume unit are considered constant \citep[i.e. homogeneous models,][]{cassano+10}. 
Here we consider magnetic field strengths of 2 and 5 $\mu$Gauss (solid and dashed lines in Fig. \ref{fig:halo_prob}), which matches the range of results in \cite{digennaro+20}. The highest value for the magnetic field strength is based on the value that maximises the lifetime of relativistic electrons at the system redshift, that is, $B=B_{\rm CMB}/\sqrt{3}$ (where $B_{\rm CMB}=3.25(1+z)^2~\mu$Gauss is the magnetic field strength equivalent to the energy density of the cosmic microwave background). We assumed that the radio emission encompasses a region $R_H=400$ kpc, which is the median size of the halos in our sample (see Table \ref{tab:fluxes}). {The value of $R_H$ is smaller than the typical radio halos found in the local Universe (i.e. $z\sim0.2$). This is probably due to the fact that, for the same mass, distant clusters have smaller virial radii \citep{kitayama+suto96}.} The uncertainty of the estimated fraction of halos is obtained via 1000 Monte Carlo extractions of galaxy cluster samples from the pool of simulated merger trees \citep[see red and yellow shaded areas in Fig. \ref{fig:halo_prob}]{cassano+10}.
We find that, in the mass interval of our sample $M_{500}=5.0-8.0\times10^{14}~\rm M_\odot$,  
and assuming $B=5~\mu$Gauss, the probability of observing a radio halo with steepening frequency $\nu_s\geq140$ MHz is between 60\% and 30\% (Fig.~\ref{fig:halo_prob} red line), while it decreases down to 13--30\% for $\nu_s\geq600$ MHz (Fig.~\ref{fig:halo_prob} yellow line), with a clear dependence on cluster mass.
This agrees with our observations, where we detect a radio halo in about 47\% of the total sample (9/19) at 144 MHz \citep{digennaro+20} and in about 26\% (5/19) at 650 MHz.
Assuming $B=2~\mu$Gauss, the expected fractions of halos are consistent with the uncertainties given the Monte Carlo simulations (see dashed lines in Fig. \ref{fig:halo_prob}).

These expectations imply the presence of a population of ultra-steep spectra radio halos (USSRHs) with $140\leq\nu_s<600$ MHz, which will be missed by observations at frequencies higher than 600 MHz because of their ultra-steep radio spectra.
In the top panel of Fig. \ref{fig:halo_prob}, we show the fraction of these USSRH with respect to the total number of radio halos expected at 140 MHz, $f_{\rm USSRH}$, as a function of the cluster mass, assuming $B=5~\mu$Gauss and $B=2~\mu$Gauss (solid and dashed lines, respectively). Given the mass range of our clusters, namely $M_{500}=5.0-8.0\times10^{14}~\rm M_\odot$, we expect 50--60\% ($B=5~\mu$Gauss) or 65--60\% ($B=2~\mu$Gauss) of these to be ultra steep (i.e. $\alpha<-1.4$). Despite the low statistics, the estimate using higher magnetic fields is in good agreement with our observations, where we estimate about 45\%  USSRHs (see Table \ref{tab:fluxes}). Increasing the number of distant radio halos will help to better determine the magnetic field levels in these clusters.

\subsection{Occurrence of radio relics at high redshift}
According to the predictions by \cite{nuza+12}, about 800 radio relics should have been observed at $0.5<z<1$ in the full LoTSS survey. However, \citet{nuza+17} showed in a follow-up work that the majority of these relics, if physical and not artefacts due to a simplified modelling of the ICM, would have small angular extensions ($\lesssim 2'$ when the image is smoothed with a $45''$ beam). Therefore, they would be challenging to detect and classify correctly. 
The results by \citet{nuza+17} confirm that relics as extended as the one in PSZ2\,G091.83+26.11 and sufficiently bright to be detected are rare in distant clusters.

In our total sample of 19 clusters at $z\geq0.6$ we found one candidate radio relic, in PSZ2\,G091.83+26.11. This source has a linear size of larger than 1 Mpc. Although our sample is not complete and is rather small, we are likely observing the most violent mergers in the distant Universe (i.e. $M>5\times10^{14}~{\rm M_\odot}$) and we would have expected to detect more relics. However, a reliable prediction which includes a proper analysis of the probability of identifying and classifying relics, and their number to be detected in LoTSS, is still missing.  
At $z>0.5$, five additional radio relics, namely PLCK\,G004.5-19.5 \citep[$z=0.52, \rm DEC=-33^\circ$, ][]{albert+17}, MACS\,J1149.5+2223 \citep[$z=0.544$,][]{bonafede+12},
MACS\,J0717.5+3745 \citep[$z=0.546$,][]{bonafede+09,vanweeren+09},
MACS\,J0025.4-1222 \citep[$z=0.584, \rm DEC=-12^\circ$,][]{riseley+17}, and ACT-CL\,J0102-4915 \citep[``el Gordo''; $z=0.87, \rm DEC=-49^\circ$;][]{lindner+14}, and two candidate radio relics, namely ACT-CL\,J0014.9-0057 \citep[$z = 0.533, \rm DEC=-1^\circ$,][]{knowles+19} and ACT-CL\,J0046.4-3912 \citep[$z=0.592, \rm DEC=-39^\circ$,][]{knowles+20}, are known in the literature\footnote{Only MACS\,J1149.5+2223 and MACS\,J0717.5+3745 are observed with LOFAR \citep[respectively]{bruno+21,bonafede+18}.}. Among these, MACS\,J0717.5+3745, MACS\,J1149.5+2223, and ACT-CL\,J0046.4-3912 host relics with linear sizes of about 1 Mpc. The small number of detections available to date does not allow a clear comparison with the current models. 
Despite that, the findings we present in this paper {and future observations of high-$z$ relics} will be crucial to developing more stringent predictions of the magnetic properties of the ICM at the location of the shock, acceleration efficiency, and seed populations when the first structure formed \citep{bruggen+vazza20}.

\section{Conclusions}
In this paper, we present follow-up observations of the high-redshift Planck-SZ clusters hosting diffuse radio sources presented in \cite{digennaro+20}. Our observations were taken with the upgraded GMRT (uGMRT) in Band 4 (550--900 MHz) and, for two clusters (i.e. PSZ2\,G091.83+26.11 and PLCK\,G147.3--16.6) also in Band 3 (250--500 MHz)\footnote{PSZ2\,G089.39+69.36 was observed only in Band 3.}. These observations were combined with LOFAR data at 144 MHz. Below we summarise our findings.

\begin{itemize}
\item About 50\% (5/9) of the clusters presented show the presence of a radio halo in the uGMRT observations, up to $\nu=650$ MHz. 

\medskip

\item For these systems, we measure integrated spectral indices of between $-1$ and $-1.4$. 
We note that these clusters are also the most massive in our sample, and at these redshifts ($M>6\times10^{14}$ M$_\odot$). This also implies more energetic merger events.

\medskip

\item  The injection of mock radio halos reveals that diffuse radio emission with $\alpha\sim-1.5$ is detectable in our uGMRT data. Therefore, for those clusters with a radio halo in the LOFAR images but not in the uGMRT ones, we estimate that they should have integrated spectral indices steeper than $-1.5$, in line with the predictions of re-acceleration models \citep{cassano+10}.

\medskip

\item Although our sample is not complete, the fraction of clusters hosting halos and the spectral indices agree with expectations from theoretical models of re-acceleration \citep{cassano+brunetti05,cassano+06,cassano+10}.

\medskip

\item We confirm the presence of a single candidate radio relic in this sample of distant clusters, in PSZ2\,G091.83+26.11, as reported in \cite{digennaro+20}. This is possibly contaminated by a double radio galaxy, although no optical counterparts have been observed in the PanSTARRS data. A future polarisation analysis with the VLA will provide further information on the nature of this radio source.
\end{itemize}

Observing distant diffuse radio emission is particularly important for the investigation of the evolution of magnetic fields  over cosmic time and Universe magnetogenesis. Given the small size of the sample presented in this work, comparison with cosmological simulations is difficult. Upcoming observations with the X-ray satellite eROSITA \citep[Extended Roentgen Survey Imaging Telescope Array;][]{merloni+12,merloni+20} will help in finding new distant galaxy clusters that can be easily followed up by low-frequency  radio observations.

\medskip

\begin{acknowledgements}
We thank the anonymous referee for useful comments which have improved the quality of the manuscript. 
This paper is based on data obtained with the Giant Metrewave Radio Telescope (GMRT). GMRT is run by the National Centre for Radio Astrophysics of the Tata Institute of Fundamental Research. The National Radio Astronomy Observatory is a facility of the National Science Foundation operated under cooperative agreement by Associated Universities, Inc.
We thank the staff of the GMRT that made these observations possible.
LOFAR \citep{vanhaarlem+13} is the Low Frequency Array designed and constructed by ASTRON. It has observing, data processing, and data storage facilities in several countries, which are owned by various parties (each with their own funding sources), and which are collectively operated by the ILT foundation under a joint scientific policy. The ILT resources have benefited from the following recent major funding sources: CNRS-INSU, Observatoire de Paris and Universit\'e d'Orl\'eans, France; BMBF, MIWF-NRW, MPG, Germany; Science Foundation Ireland (SFI), Department of Business, Enterprise and Innovation (DBEI), Ireland; NWO, The Netherlands; The Science and Technology Facilities Council, UK; Ministry of Science and Higher Education, Poland; The Istituto Nazionale di Astrofisica (INAF), Italy.  This research made use of the Dutch national e-infrastructure with support of the SURF Cooperative (e-infra 180169) and the LOFAR e-infra group. The J\"ulich LOFAR Long Term Archive and the German LOFAR network are both coordinated and operated by the J\"ulich Supercomputing Centre (JSC), and computing resources on the supercomputer JUWELS at JSC were provided by the Gauss Centre for Supercomputing e.V. (grant CHTB00) through the John von Neumann Institute for Computing (NIC). This research made use of the University of Hertfordshire high-performance computing facility and the LOFAR-UK computing facility located at the University of Hertfordshire and supported by STFC [ST/P000096/1], and of the Italian LOFAR IT computing infrastructure supported and operated by INAF, and by the Physics Department of Turin university (under an agreement with Consorzio Interuniversitario per la Fisica Spaziale) at the C3S Supercomputing Centre, Italy.
The National Radio Astronomy Observatory is a facility of the National Science Foundation operated under cooperative agreement by Associated Universities, Inc. 
GDG and RJvW acknowledge support from the ERC Starting Grant ClusterWeb 804208. RC and GB acknowledge support from INAF through the mainstream project ``Cluster science with LOFAR''. HJAR acknowledge support from the ERC Advanced Investigator programme NewClusters 321271. AB and RJvW acknowledge support from the VIDI research programme with project number 639.042.729, which is financed by the Netherlands Organisation for Scientific Research (NWO). VC acknowledges support from the Alexander von Humboldt Foundation.
\end{acknowledgements}

\appendix

\section{Source-subtracted images}\label{apx:sub_images}
In this section we present the source-subtracted images and the region used to determine the radio halo flux densities reported in Table \ref{tab:fluxes} (solid yellow polygons). The positions of the subtracted compact sources are shown with a red cross. 

\begin{figure*}
\centering
{\includegraphics[width=0.45\textwidth]{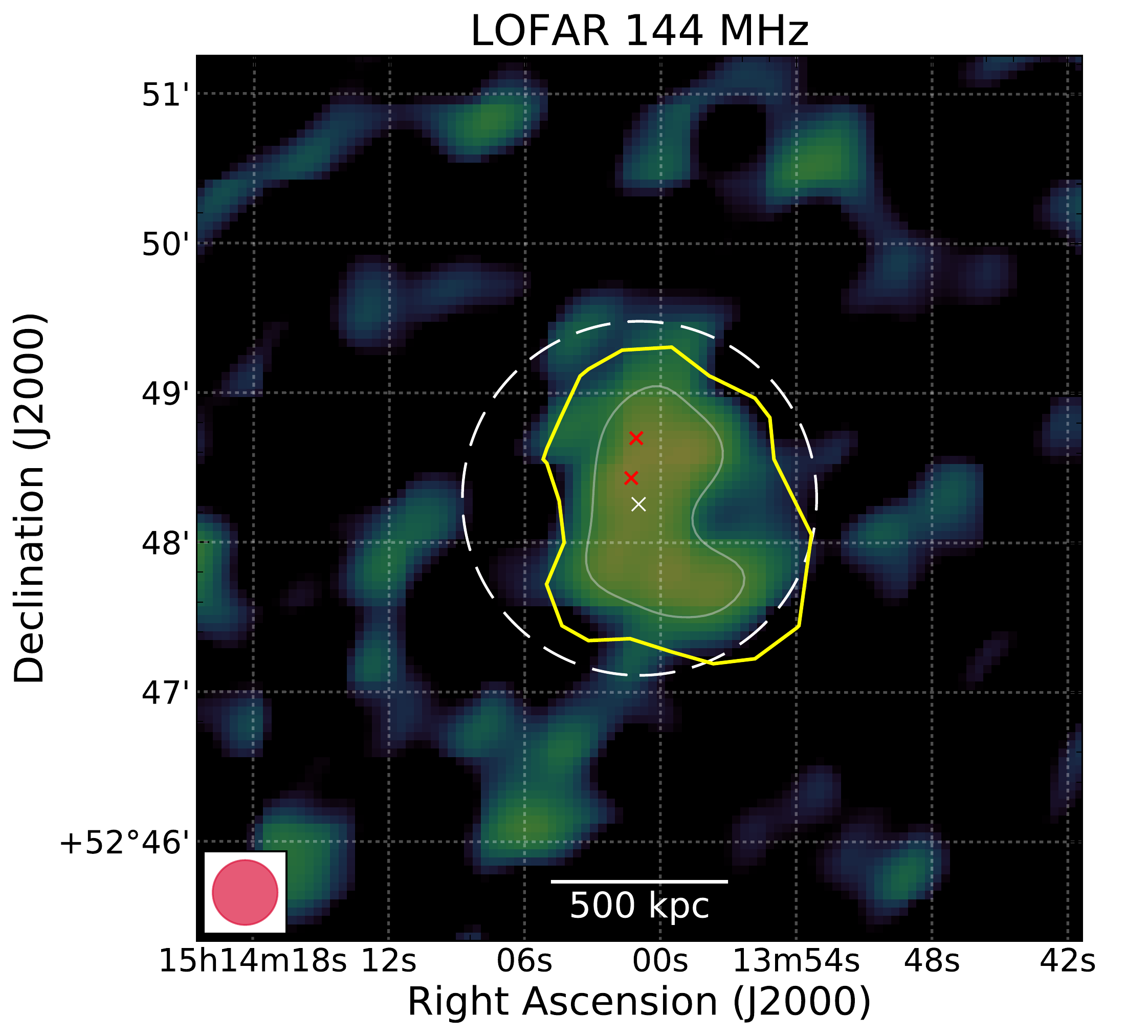}}
{\includegraphics[width=0.45\textwidth]{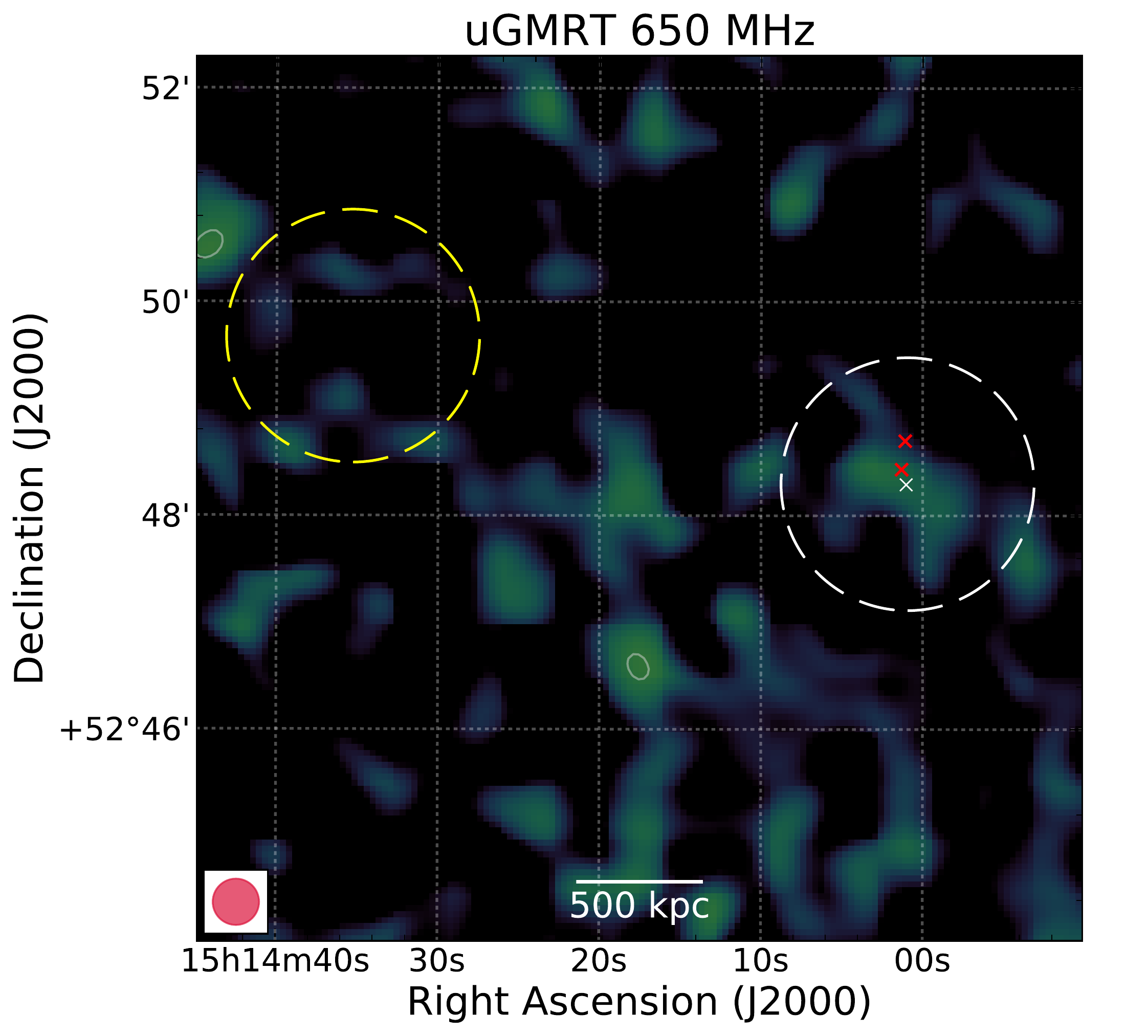}}
\caption{PSZ2\,G086.93+53.18. $26''$-resolution compact source-subtracted LOFAR and uGMRT images at 144 MHz (left) and 650 MHz (right). White-coloured radio contours at the same resolution are drawn at levels of $2.5\sigma_{\rm rms}\times[-1,1,2,4,8,16,32]$, with $\rm \sigma_{rms,144}=260~\mu Jy~beam^{-1}$ and $\rm \sigma_{rms,650}=50~\mu Jy~beam^{-1}$ the maps noise. The negative contour level is drawn with a dashed white line. The dashed white circle in each map shows the $R= 0.5 R_{\rm SZ,500}$ region obtained from $M_{\rm SZ,500}$, {with the white cross showing the cluster centre}, while the dashed yellow circle shows the position of the injected mock halo. The positions of the subtracted sources in the cluster region are highlighted with red crosses. The yellow polygon represents the area where the flux densities were measured.}
\end{figure*}

\begin{figure*}
\centering
{\includegraphics[width=0.45\textwidth]{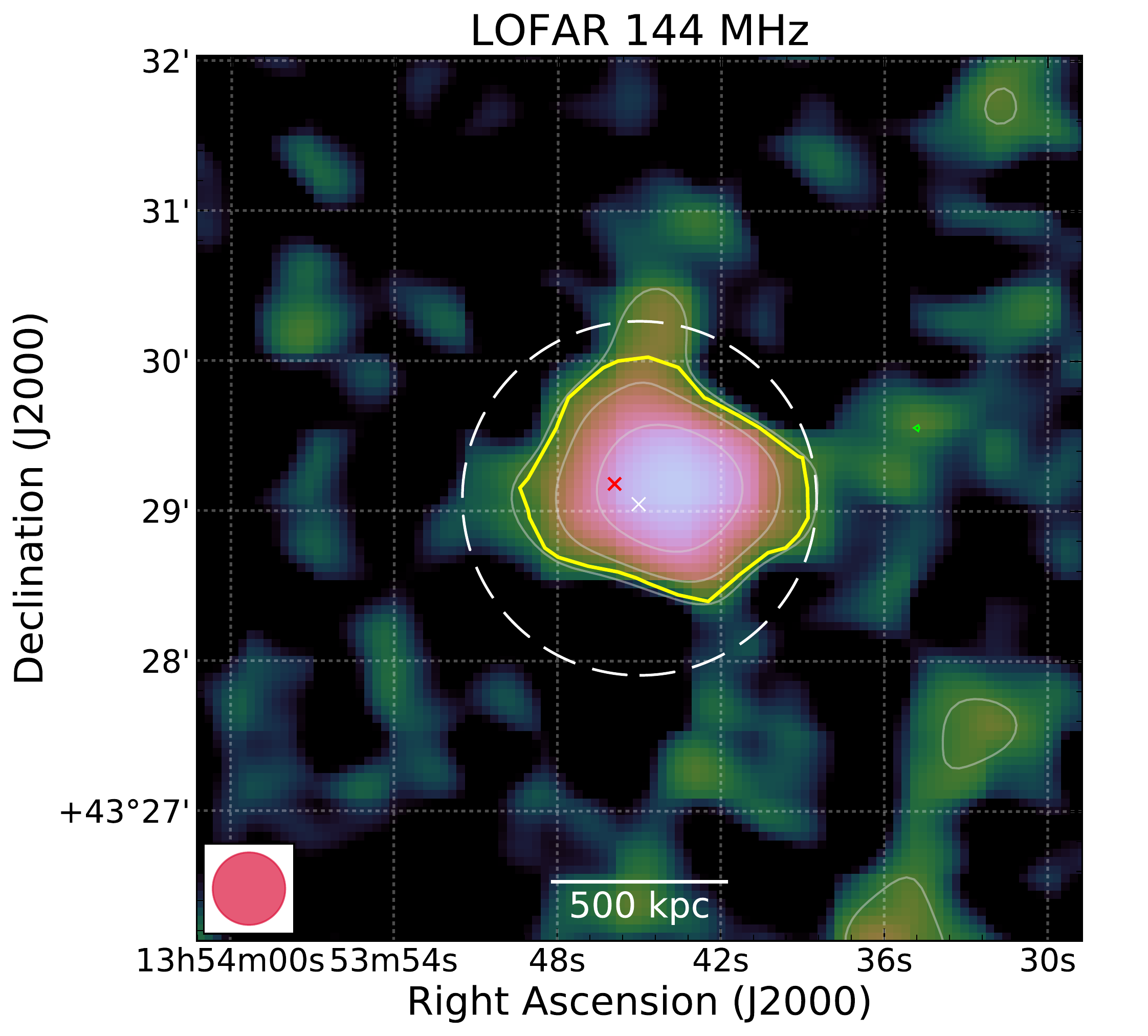}}
{\includegraphics[width=0.45\textwidth]{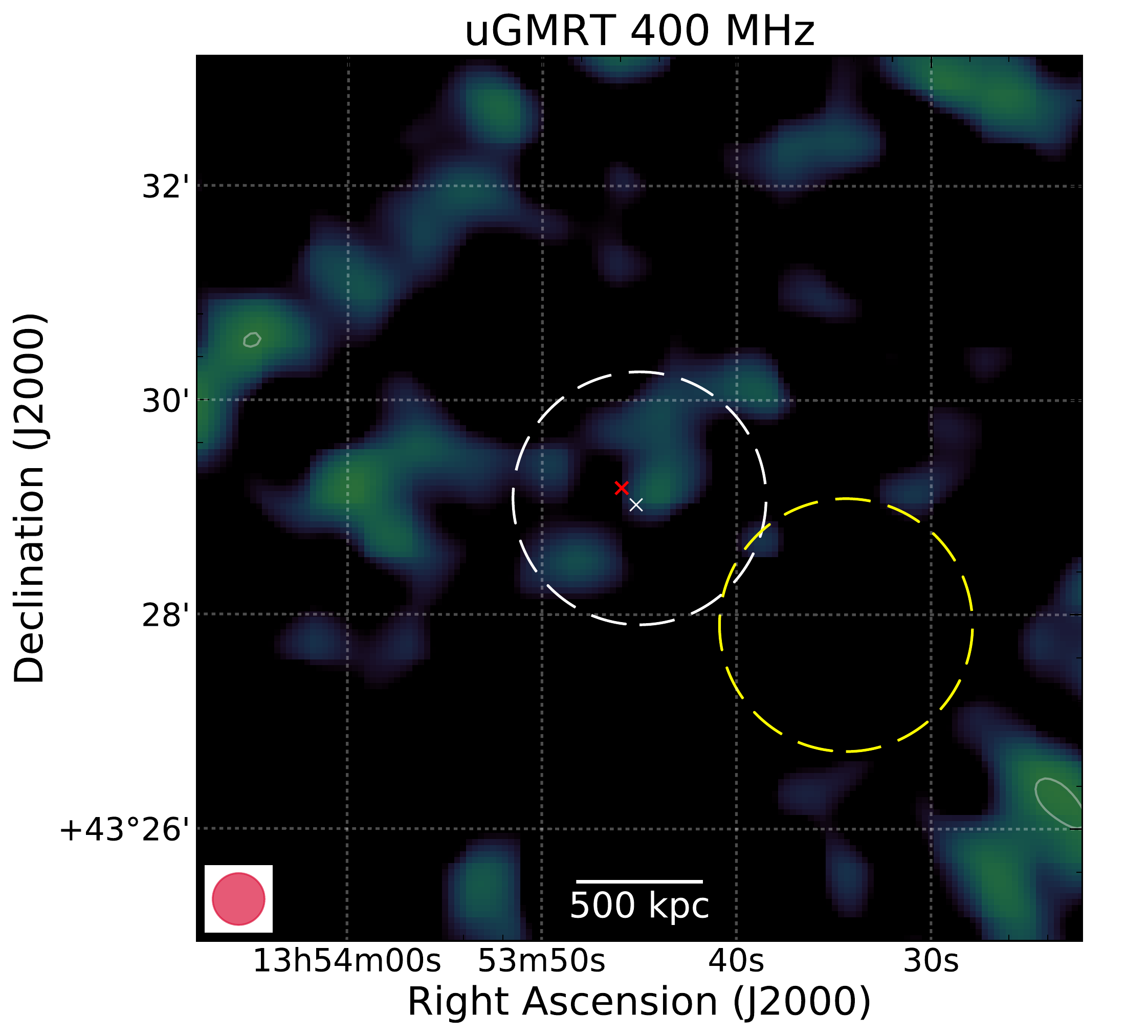}}
\caption{PSZ2\,G089.39+69.36. $29''$-resolution compact source-subtracted LOFAR and uGMRT images at 144 MHz (left) and 650 MHz(right). White-coloured radio contours at the same resolution are drawn at levels of $2.5\sigma_{\rm rms}\times[-1,1,2,4,8,16,32]$, with $\rm \sigma_{rms,144}=150~\mu Jy~beam^{-1}$ and $\rm \sigma_{rms,400}=416~\mu Jy~beam^{-1}$ the maps noise. The negative contour level is drawn with a dashed white line. The dashed white circle in each map shows the $R= 0.5 R_{\rm SZ,500}$ region obtained from $M_{\rm SZ,500}$, {with the white cross showing the cluster centre}, while the dashed yellow circle shows the position of the injected mock halo. The positions of the subtracted sources in the cluster region are highlighted with red crosses. The yellow polygon represents the area where the flux densities were measured.}
\end{figure*}

\begin{figure*}
\centering
{\includegraphics[width=0.3\textwidth]{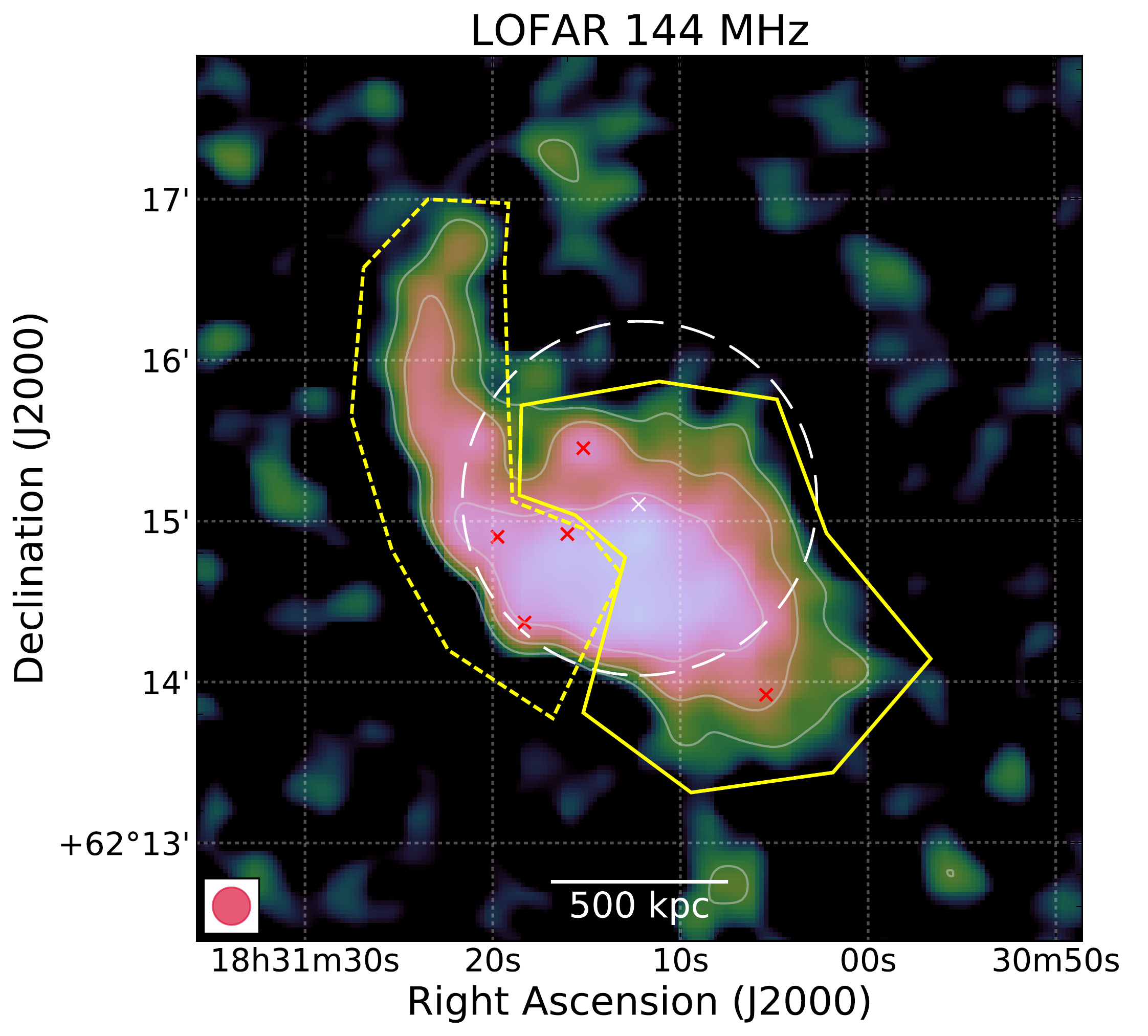}}
{\includegraphics[width=0.3\textwidth]{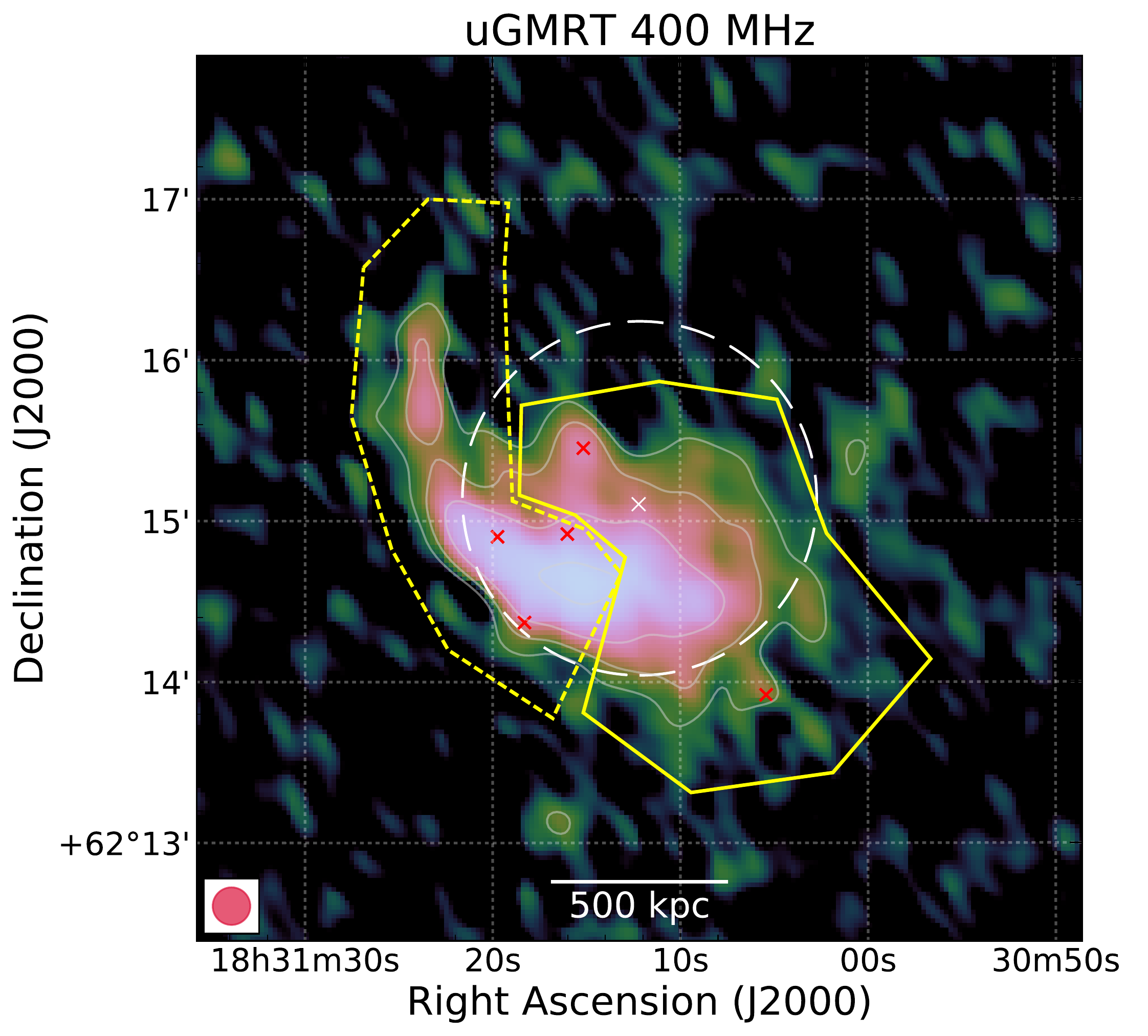}}
{\includegraphics[width=0.3\textwidth]{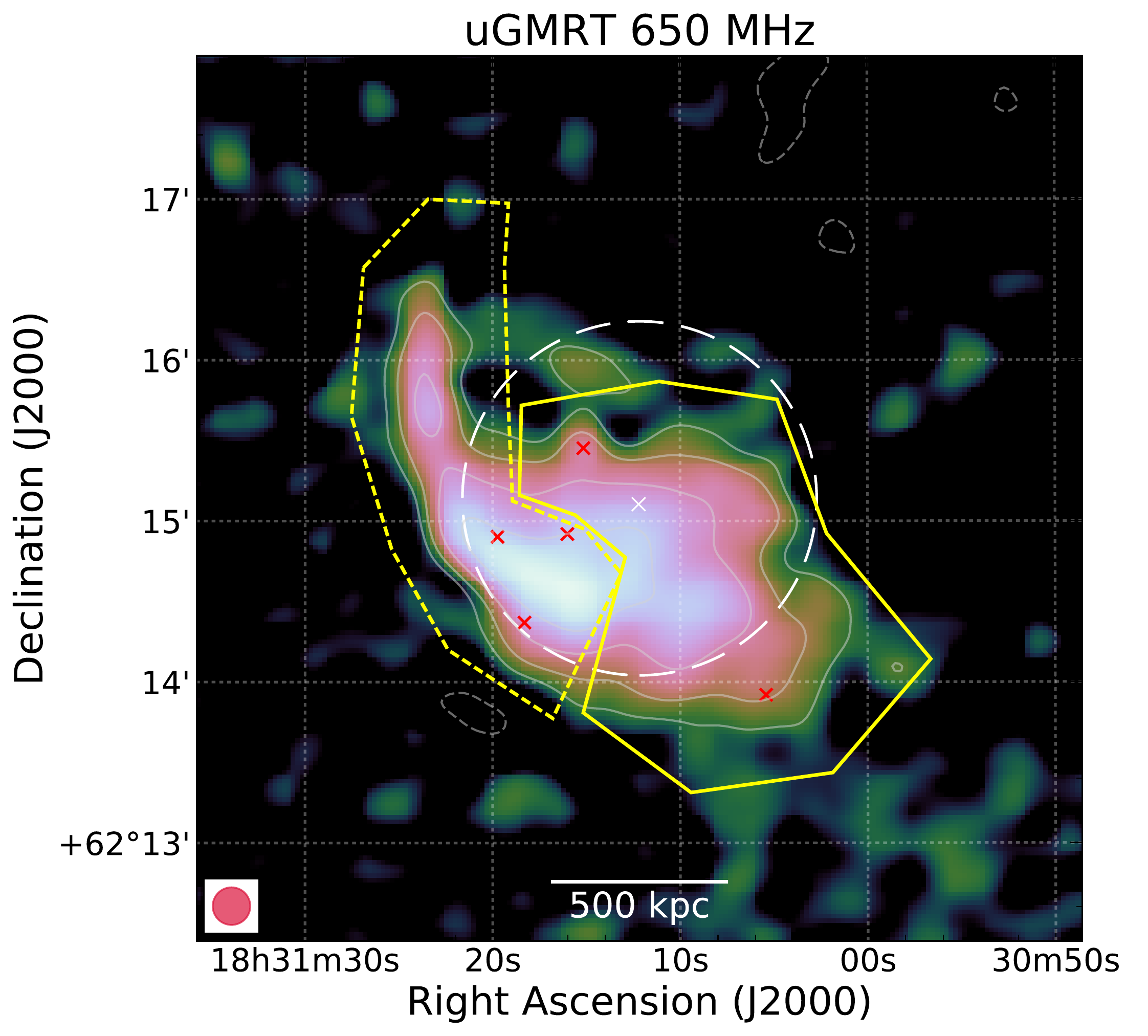}}
\caption{PSZ2\,G091.83+26.11. $14''$-resolution compact source-subtracted LOFAR and uGMRT images at 144 MHz (left), 400 MHz (middle) and 650 MHz(right). White-coloured radio contours at the same resolution are drawn at levels of $2.5\sigma_{\rm rms}\times[-1,1,2,4,8,16,32]$, with $\rm \sigma_{rms,144}=180~\mu Jy~beam^{-1}$, $\rm \sigma_{rms,400}=90~\mu Jy~beam^{-1}$ and $\rm \sigma_{rms,650}=38~\mu Jy~beam^{-1}$ the maps noise. The negative contour level is drawn with a dashed white line. The dashed white circle in each map shows the $R= 0.5 R_{\rm SZ,500}$ region obtained from $M_{\rm SZ,500}$, {with the white cross showing the cluster centre}. The positions of the subtracted sources in the cluster region are highlighted with red crosses. The yellow polygons represent the area where the flux densities were measured for the halo (solid) and the relic (dashed).}
\end{figure*}

\begin{figure*}
\centering
{\includegraphics[width=0.45\textwidth]{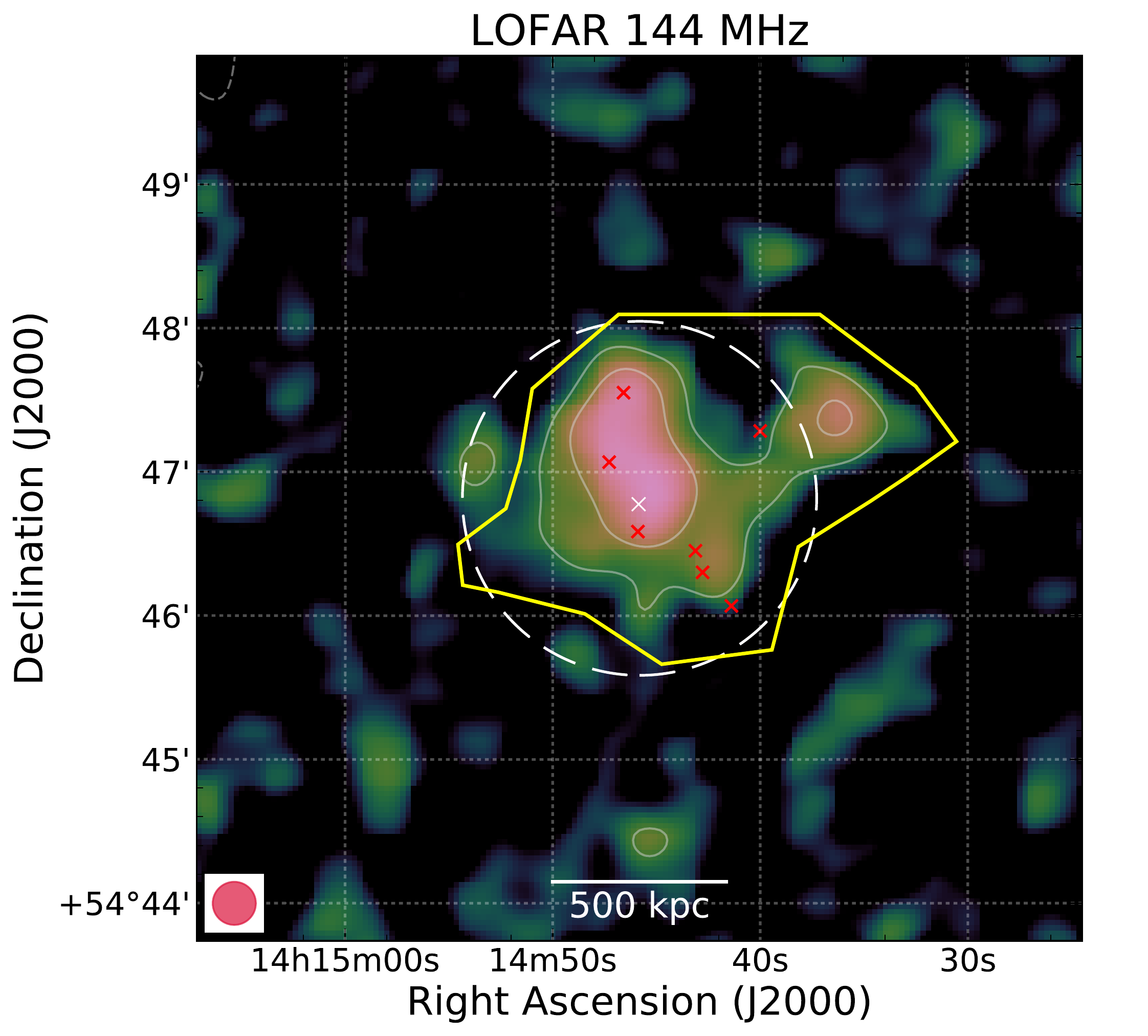}}
{\includegraphics[width=0.45\textwidth]{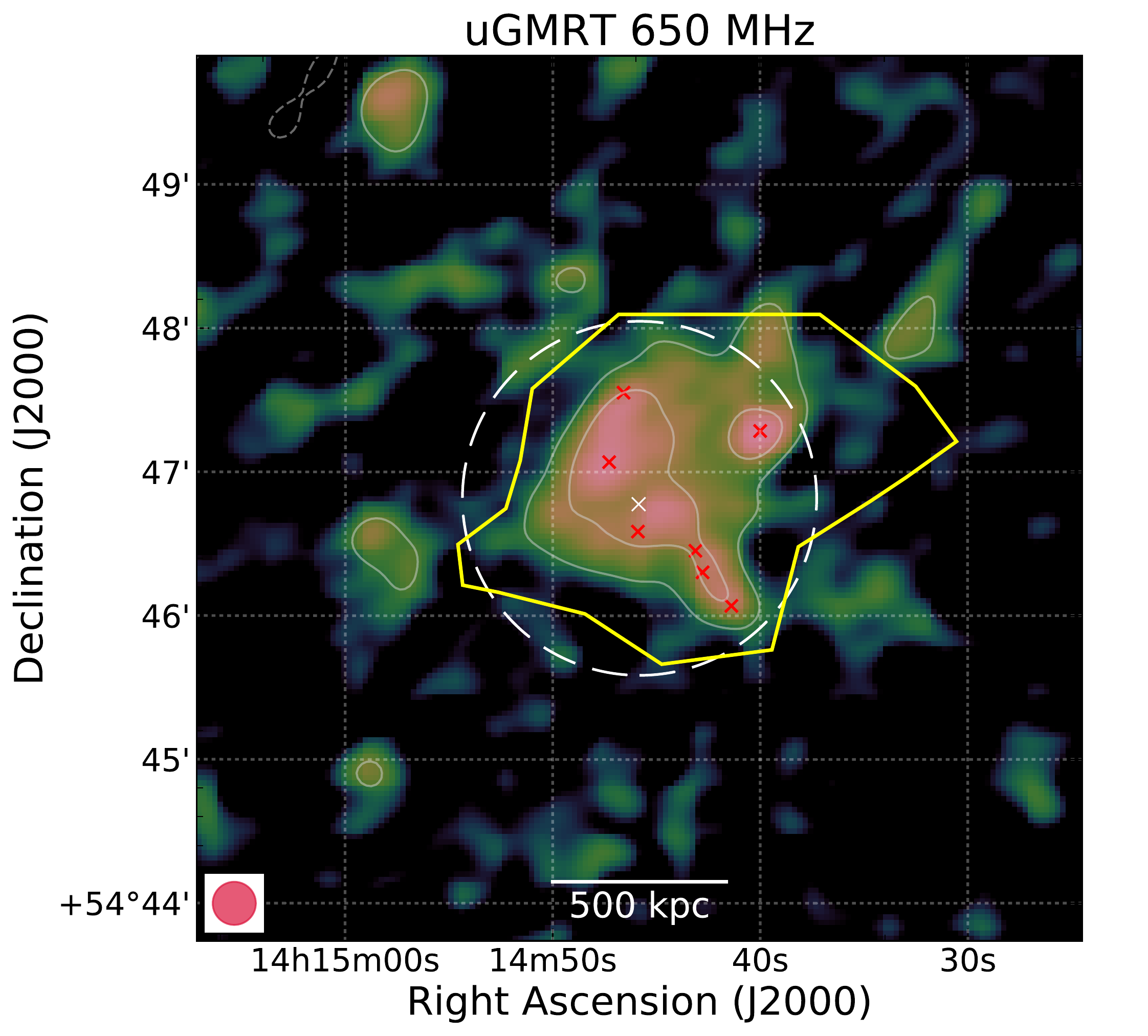}}
\caption{PSZ2\,G099.86+58.45 $18''$-resolution compact source-subtracted LOFAR and uGMRT images at 144 MHz (left) and 650 MHz(right). White-coloured radio contours at the same resolution are drawn at levels of $2.5\sigma_{\rm rms}\times[-1,1,2,4,8,16,32]$, with $\rm \sigma_{rms,144}=140~\mu Jy~beam^{-1}$ and $\rm \sigma_{rms,650}=32~\mu Jy~beam^{-1}$ the maps noise. The negative contour level is drawn with a dashed white line. The dashed white circle in each map shows the $R= 0.5 R_{\rm SZ,500}$ region obtained from $M_{\rm SZ,500}$, {with the white cross showing the cluster centre}. The positions of the subtracted sources in the cluster region are highlighted with red crosses. The yellow polygon represents the area where the flux densities were measured.}
\end{figure*}

\begin{figure*}
\centering
{\includegraphics[width=0.45\textwidth]{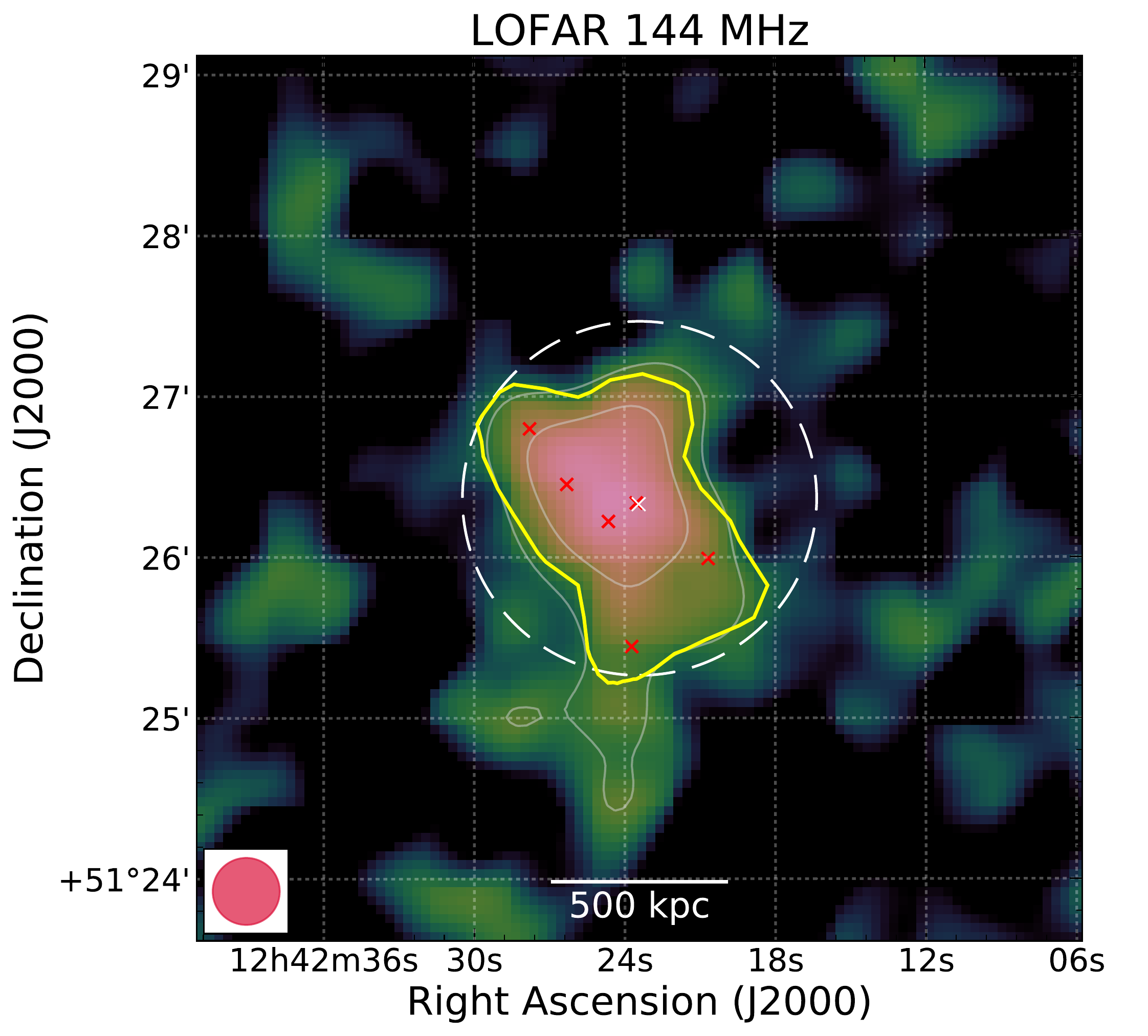}}
{\includegraphics[width=0.45\textwidth]{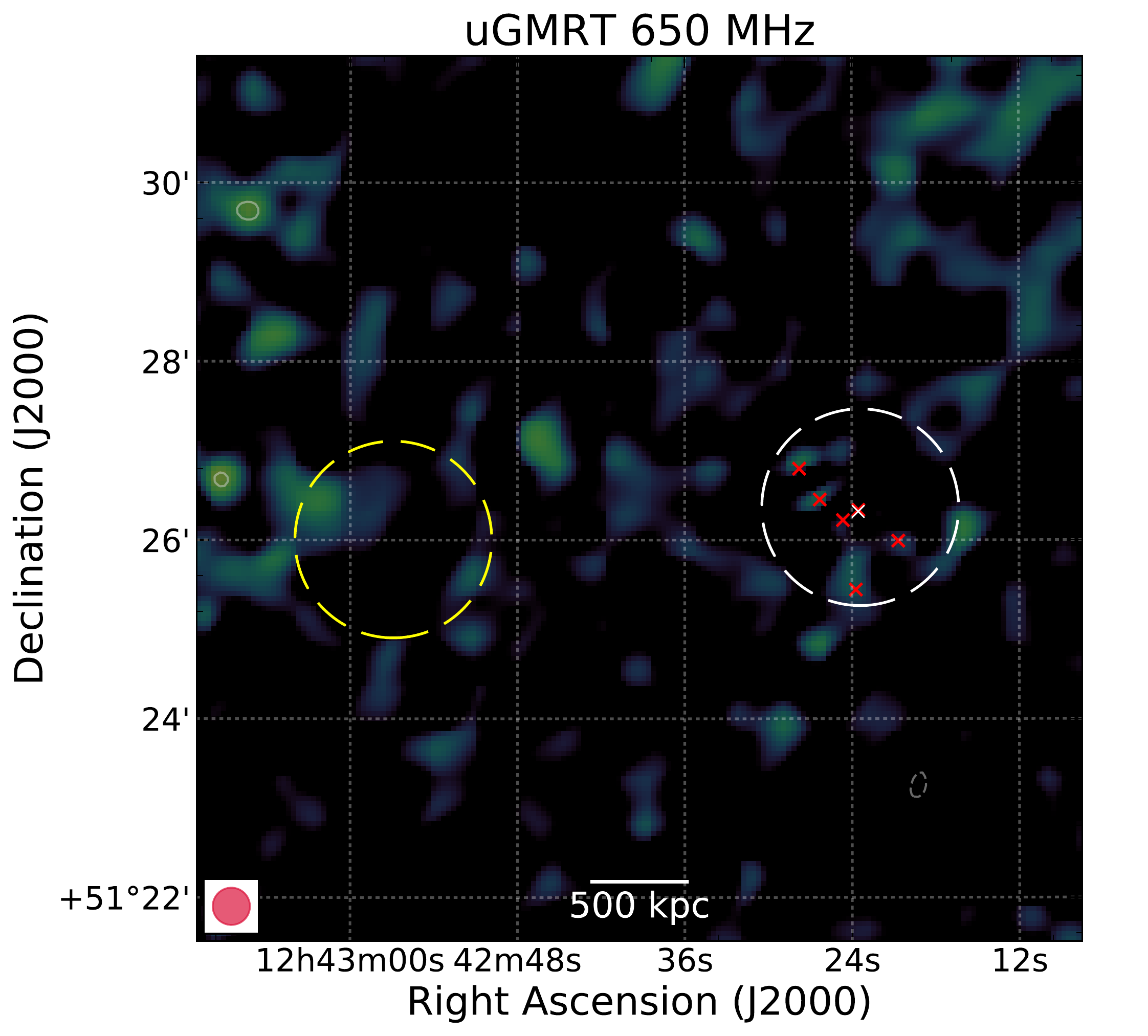}}
\caption{PSZ2\,G126.28+65.62. $25''$-resolution compact source-subtracted LOFAR and uGMRT images at 144 MHz (left) and 650 MHz(right). White-coloured radio contours at the same resolution are drawn at levels of $2.5\sigma_{\rm rms}\times[-1,1,2,4,8,16,32]$, with $\rm \sigma_{rms,144}=130~\mu Jy~beam^{-1}$ and $\rm \sigma_{rms,650}=80~\mu Jy~beam^{-1}$ the maps noise. The negative contour level is drawn with a dashed white line. The dashed white circle in each map shows the $R= 0.5 R_{\rm SZ,500}$ region obtained from $M_{\rm SZ,500}$, {with the white cross showing the cluster centre}, while the dashed yellow circle shows the position of the injected mock halo. The positions of the subtracted sources in the cluster region are highlighted with red crosses. The yellow polygon represents the area where the flux densities were measured.}
\end{figure*}

\begin{figure*}
\centering
{\includegraphics[width=0.45\textwidth]{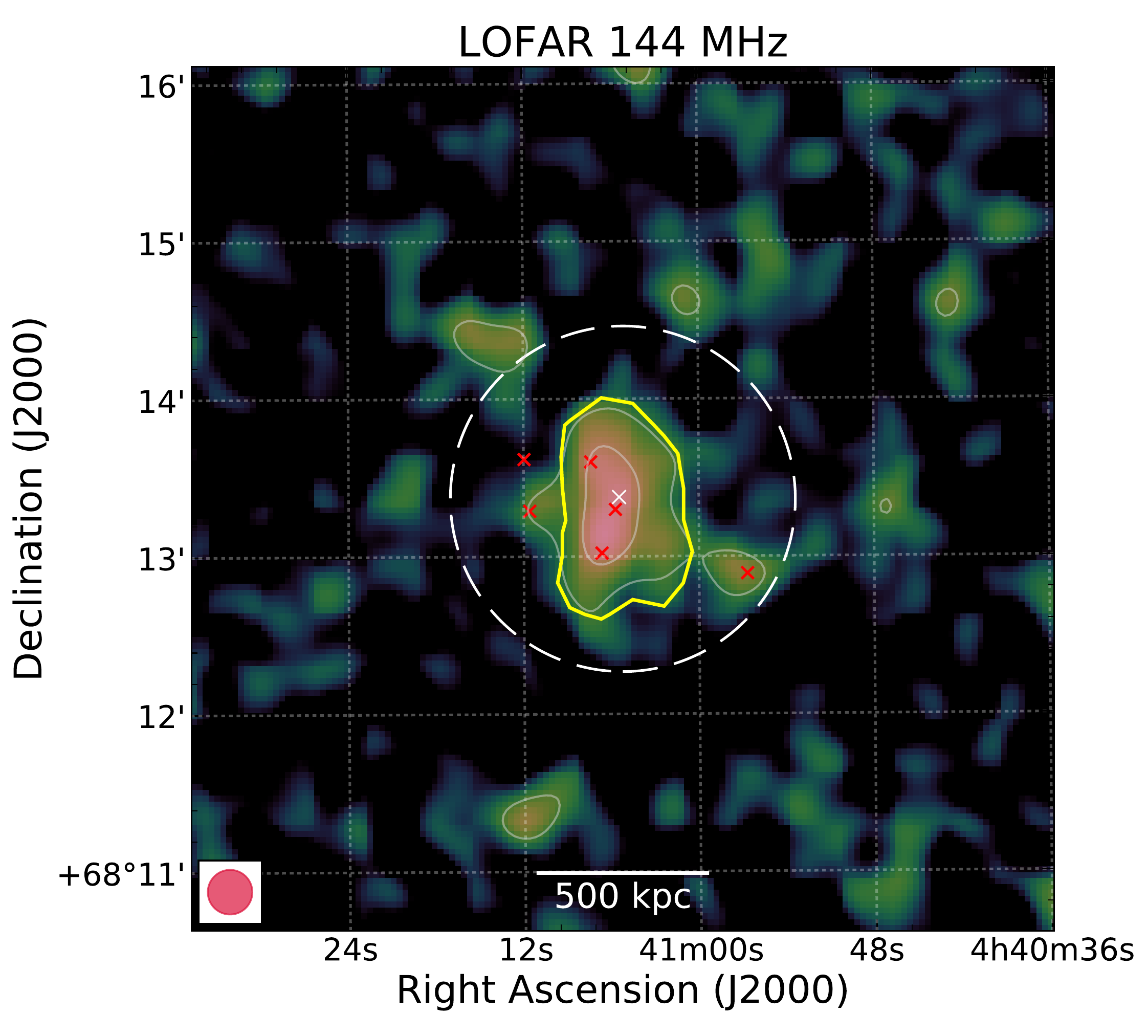}}
{\includegraphics[width=0.45\textwidth]{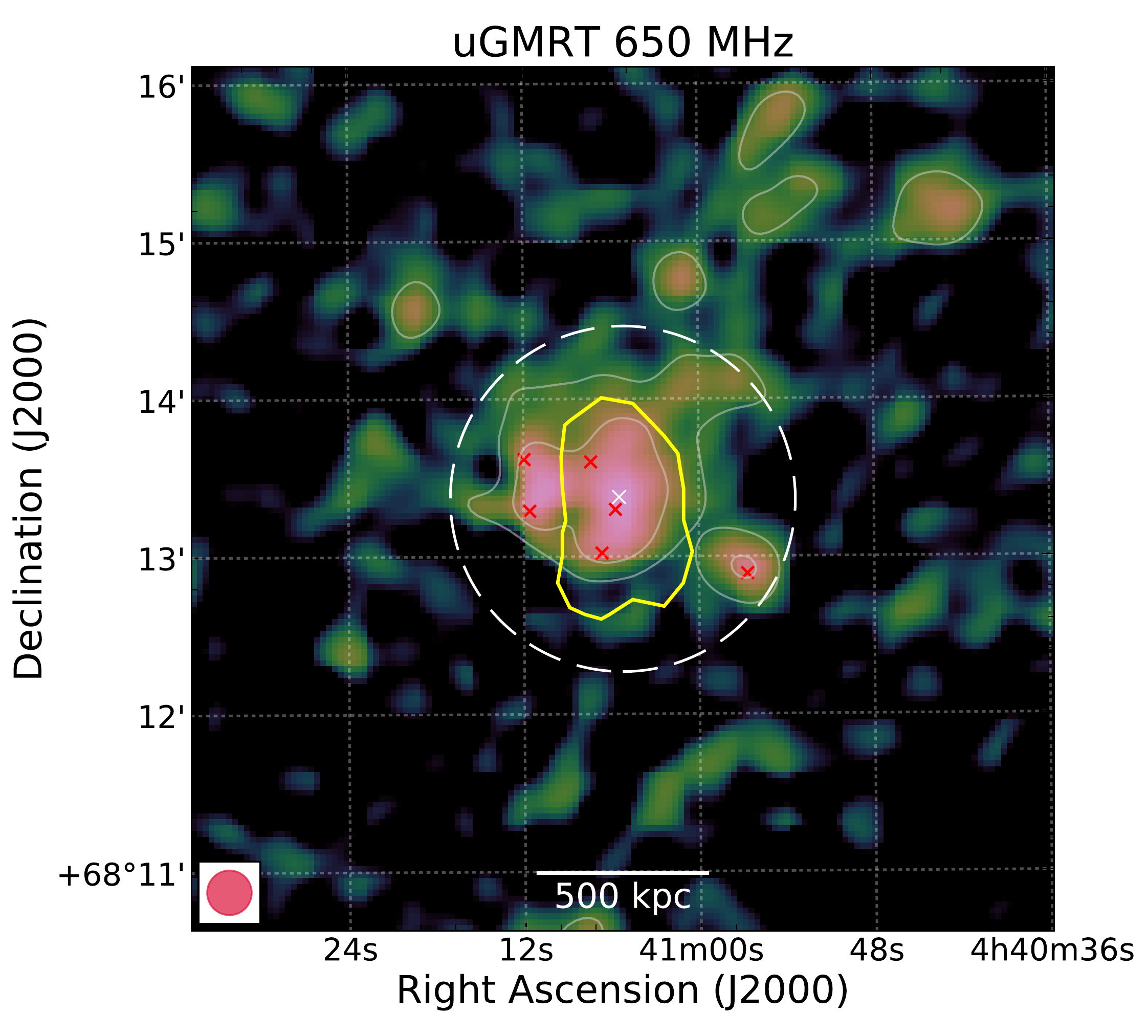}}
\caption{PSZ2\,G141.77+14.19. $17''$-resolution compact source-subtracted LOFAR and uGMRT images at 144 MHz (left) and 650 MHz(right). White-coloured radio contours at the same resolution are drawn at levels of $2.5\sigma_{\rm rms}\times[-1,1,2,4,8,16,32]$, with $\rm \sigma_{rms,144}=165~\mu Jy~beam^{-1}$ and $\rm \sigma_{rms,650}=20~\mu Jy~beam^{-1}$ the maps noise. The negative contour level is drawn with a dashed white line. The dashed white circle in each map shows the $R= 0.5 R_{\rm SZ,500}$ region obtained from $M_{\rm SZ,500}$, {with the white cross showing the cluster centre}. The positions of the subtracted sources in the cluster region are highlighted with red crosses. The yellow polygon represents the area where the flux densities were measured.}
\end{figure*}

\begin{figure*}
\centering

{\includegraphics[width=0.3\textwidth]{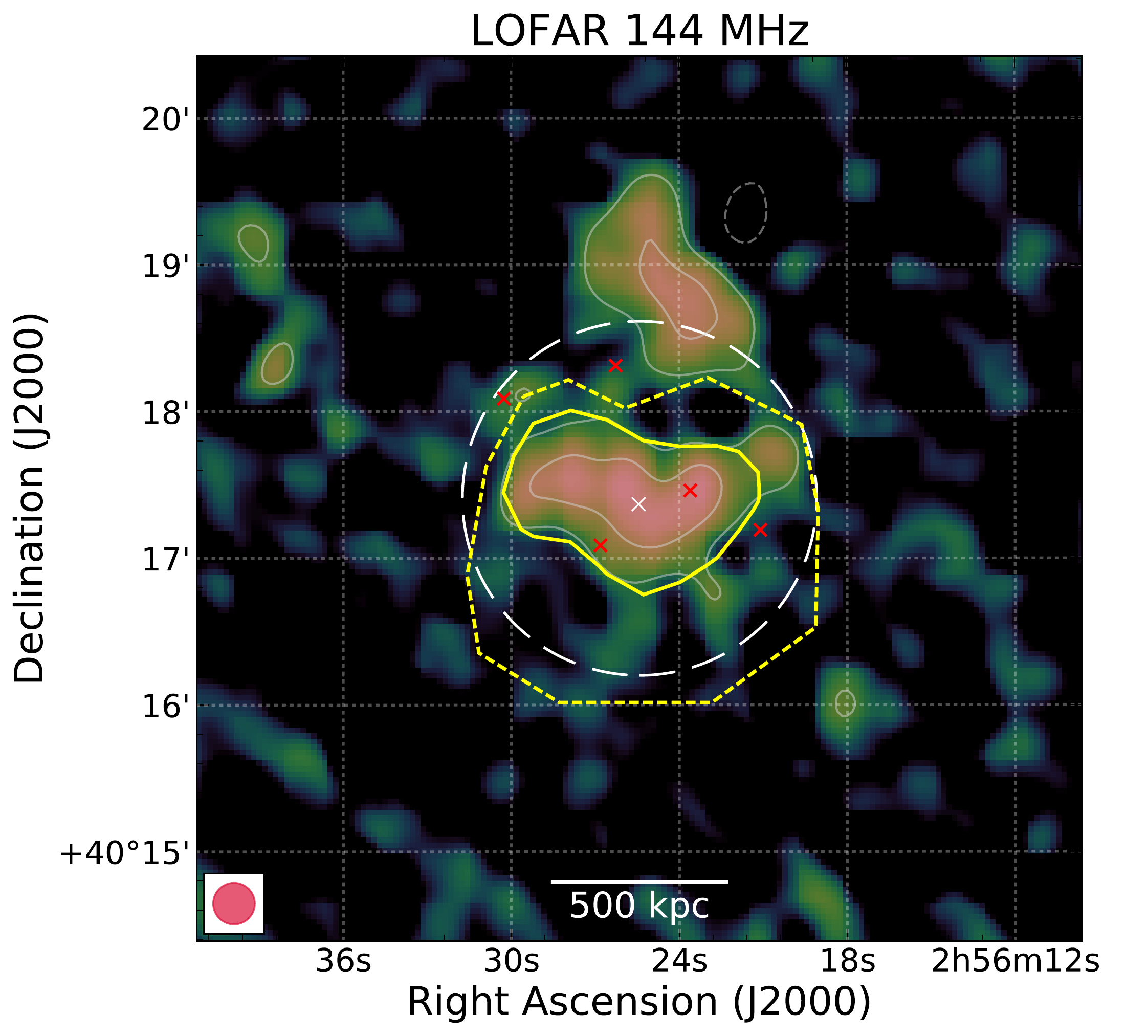}}
{\includegraphics[width=0.3\textwidth]{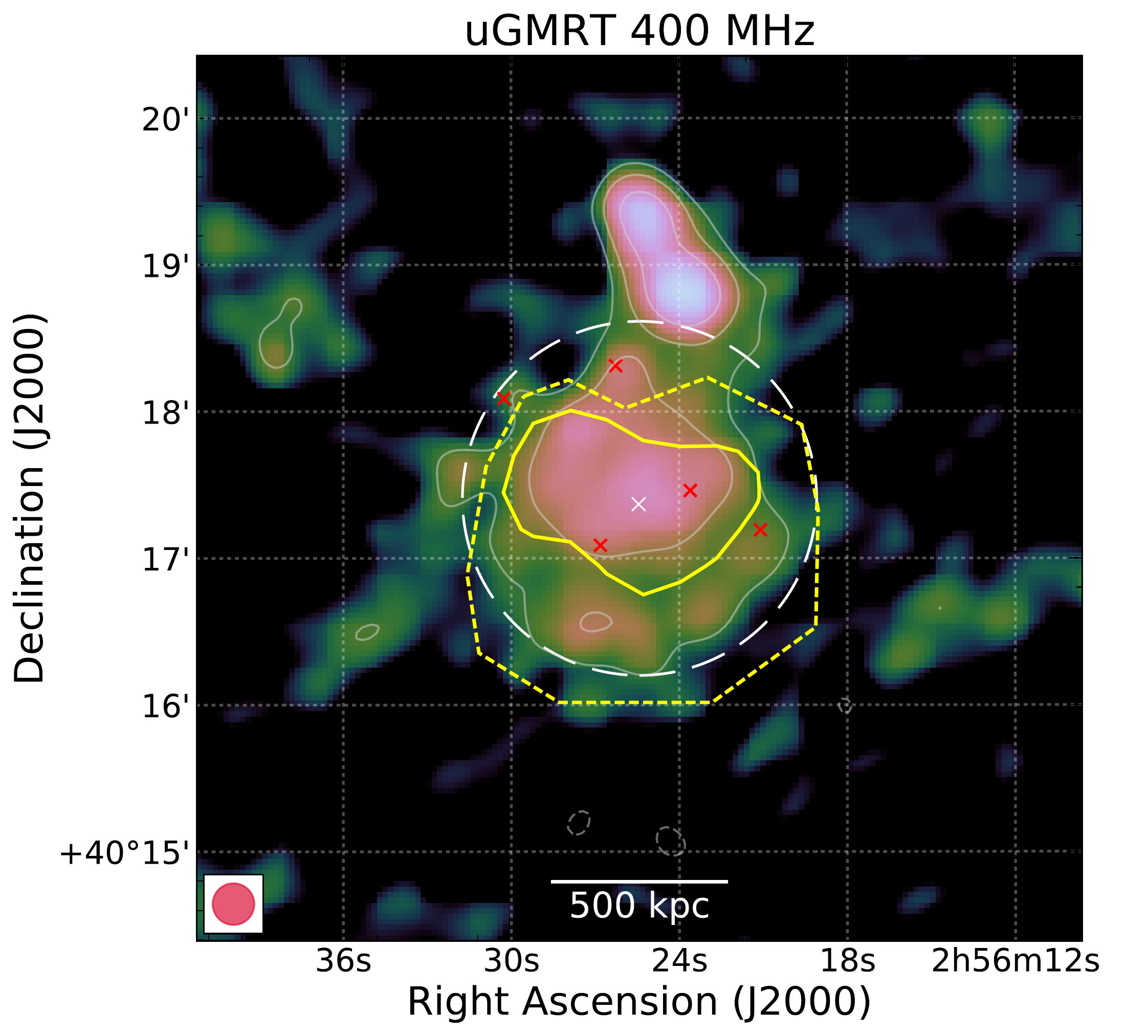}}
{\includegraphics[width=0.3\textwidth]{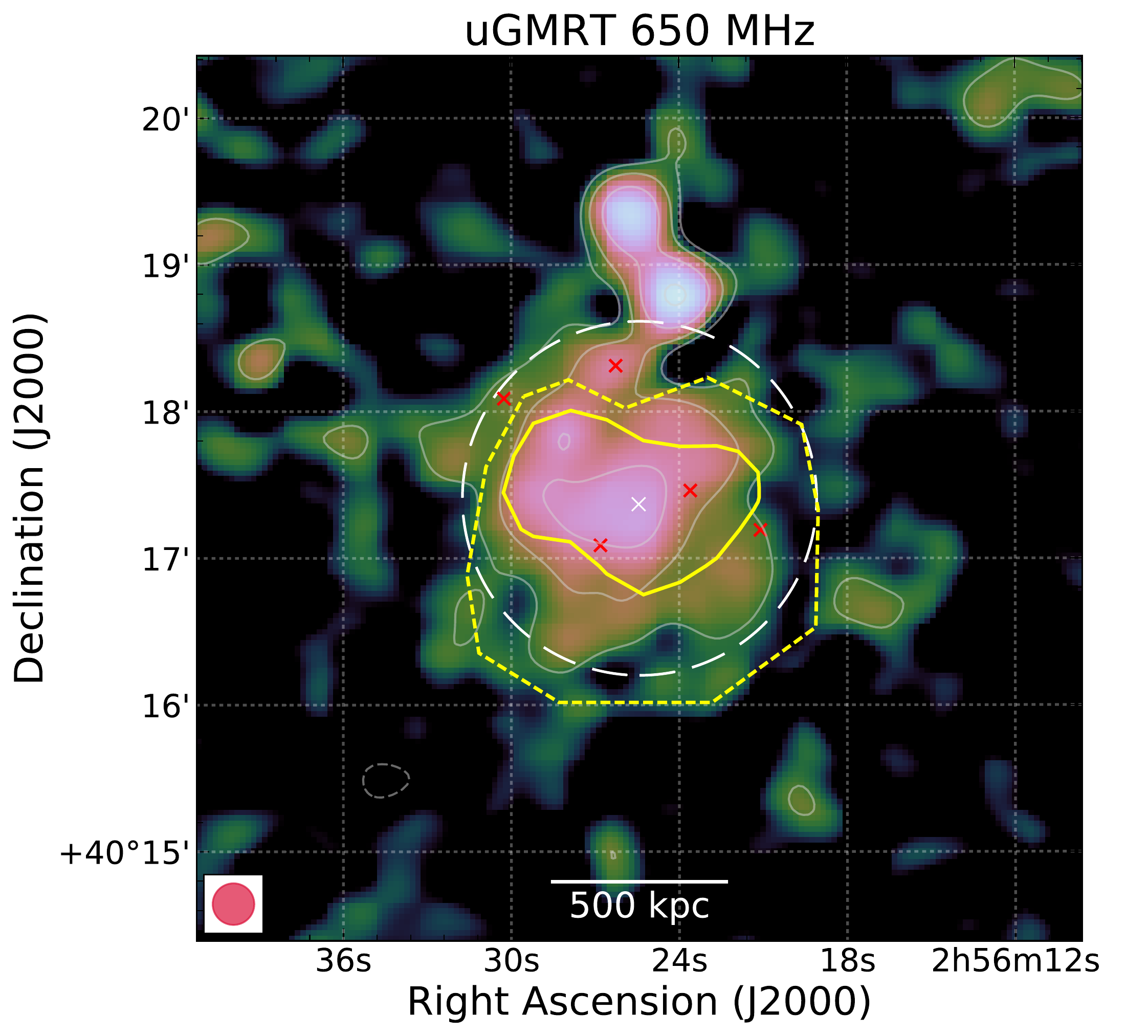}}
\caption{PLCK\,G147.3--16.6. $17''$-resolution compact source-subtracted LOFAR and uGMRT images at 144 MHz (left), 400 MHz (middle) and 650 MHz(right). White-coloured radio contours at the same resolution are drawn at levels of $2.5\sigma_{\rm rms}\times[-1,1,2,4,8,16,32]$, with $\rm \sigma_{rms,144}=300~\mu Jy~beam^{-1}$, $\rm \sigma_{rms,400}=61~\mu Jy~beam^{-1}$ and $\rm \sigma_{rms,650}=25~\mu Jy~beam^{-1}$ the maps noise. The negative contour level is drawn with a dashed white line. The dashed white circle in each map shows the $R= 0.5 R_{\rm SZ,500}$ region obtained from $M_{\rm SZ,500}$, {with the white cross showing the cluster centre}. The positions of the subtracted sources in the cluster region are highlighted with red crosses. The yellow solid and dashed polygons represent the area where the flux densities were measured.}
\end{figure*}

\begin{figure*}
\centering
{\includegraphics[width=0.45\textwidth]{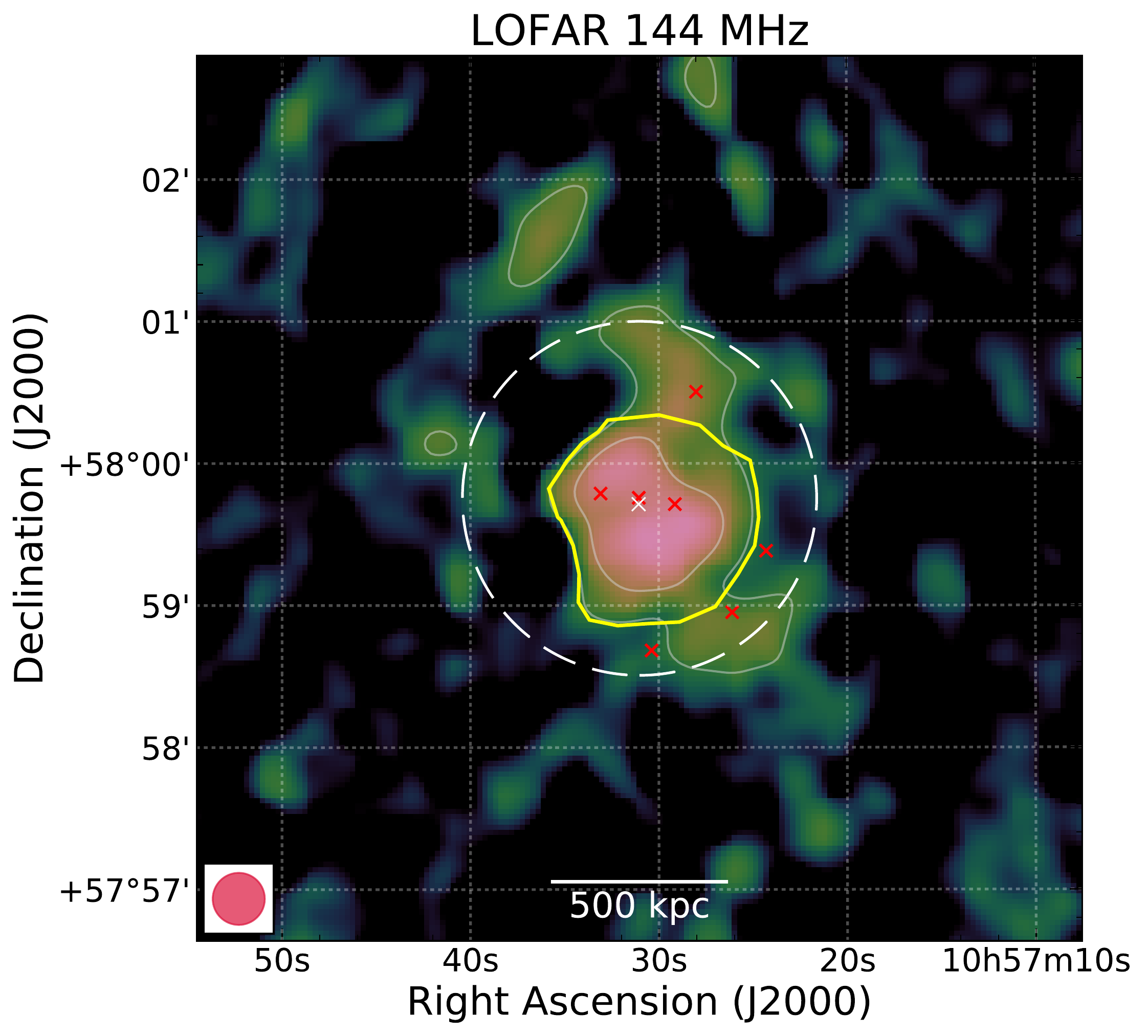}}
{\includegraphics[width=0.45\textwidth]{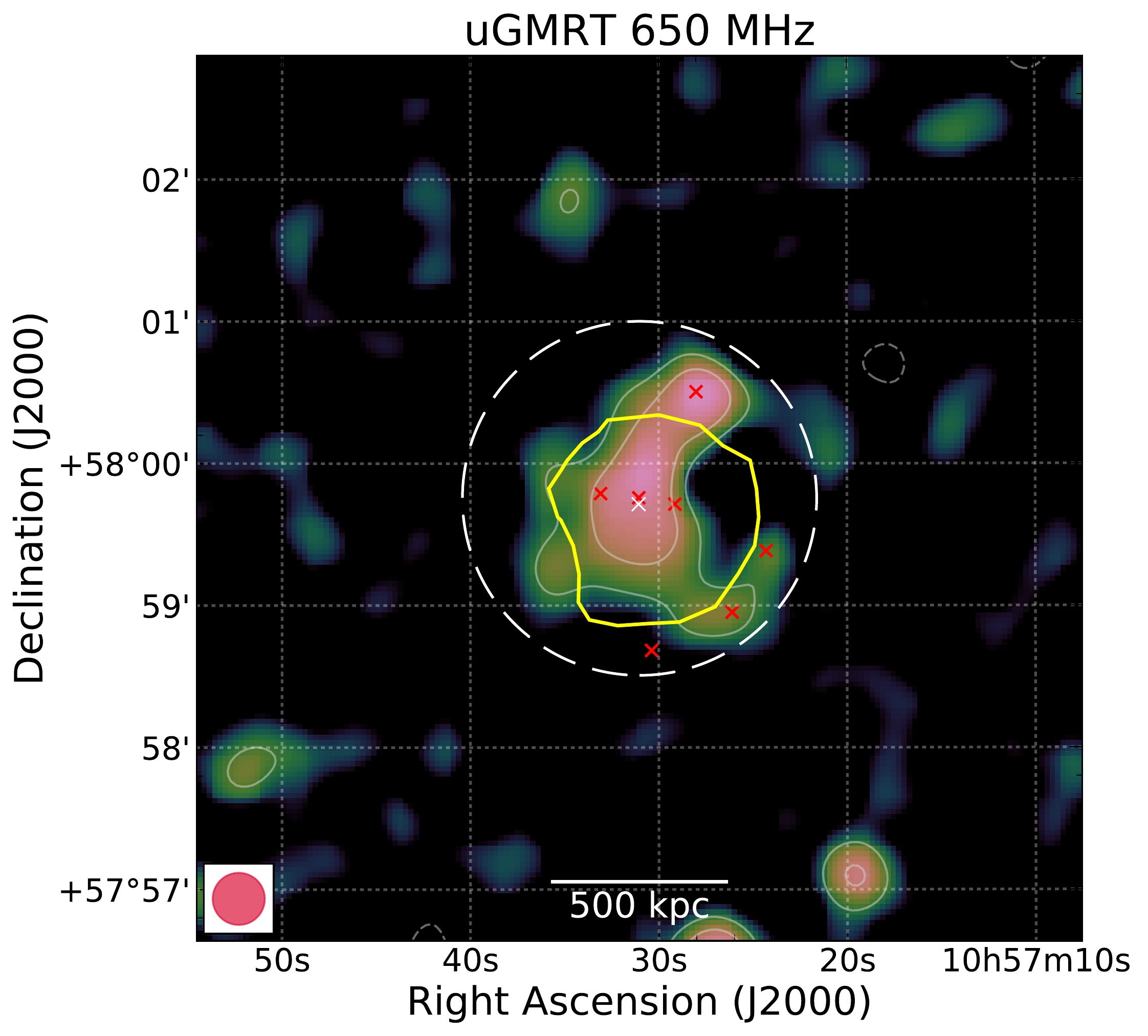}}
\caption{PSZ2\,G147.88+53.24. $22''$-resolution compact source-subtracted LOFAR and uGMRT images at 144 MHz (left) and 650 MHz(right). White-coloured radio contours at the same resolution are drawn at levels of $2.5\sigma_{\rm rms}\times[-1,1,2,4,8,16,32]$,with $\rm \sigma_{rms,144}=140~\mu Jy~beam^{-1}$ and $\rm \sigma_{rms,650}=35~\mu Jy~beam^{-1}$ the maps noise. The negative contour level is drawn with a dashed white line. The dashed white circle in each map shows the $R= 0.5 R_{\rm SZ,500}$ region obtained from $M_{\rm SZ,500}$, {with the white cross showing the cluster centre}. The positions of the subtracted sources in the cluster region are highlighted with red crosses. The yellow polygon represents the area where the flux densities were measured.}
\end{figure*}

\begin{figure*}
\centering
{\includegraphics[width=0.45\textwidth]{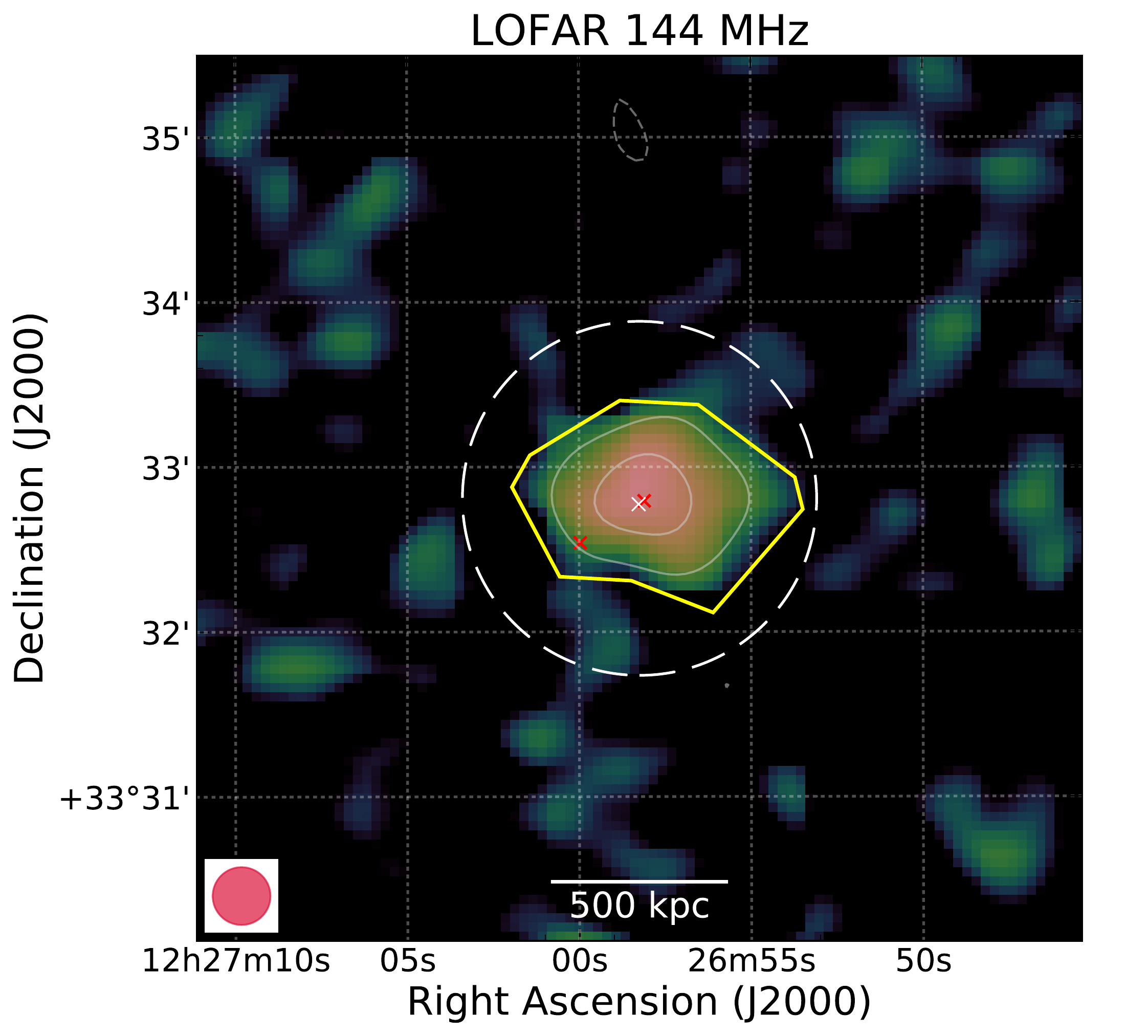}}
{\includegraphics[width=0.45\textwidth]{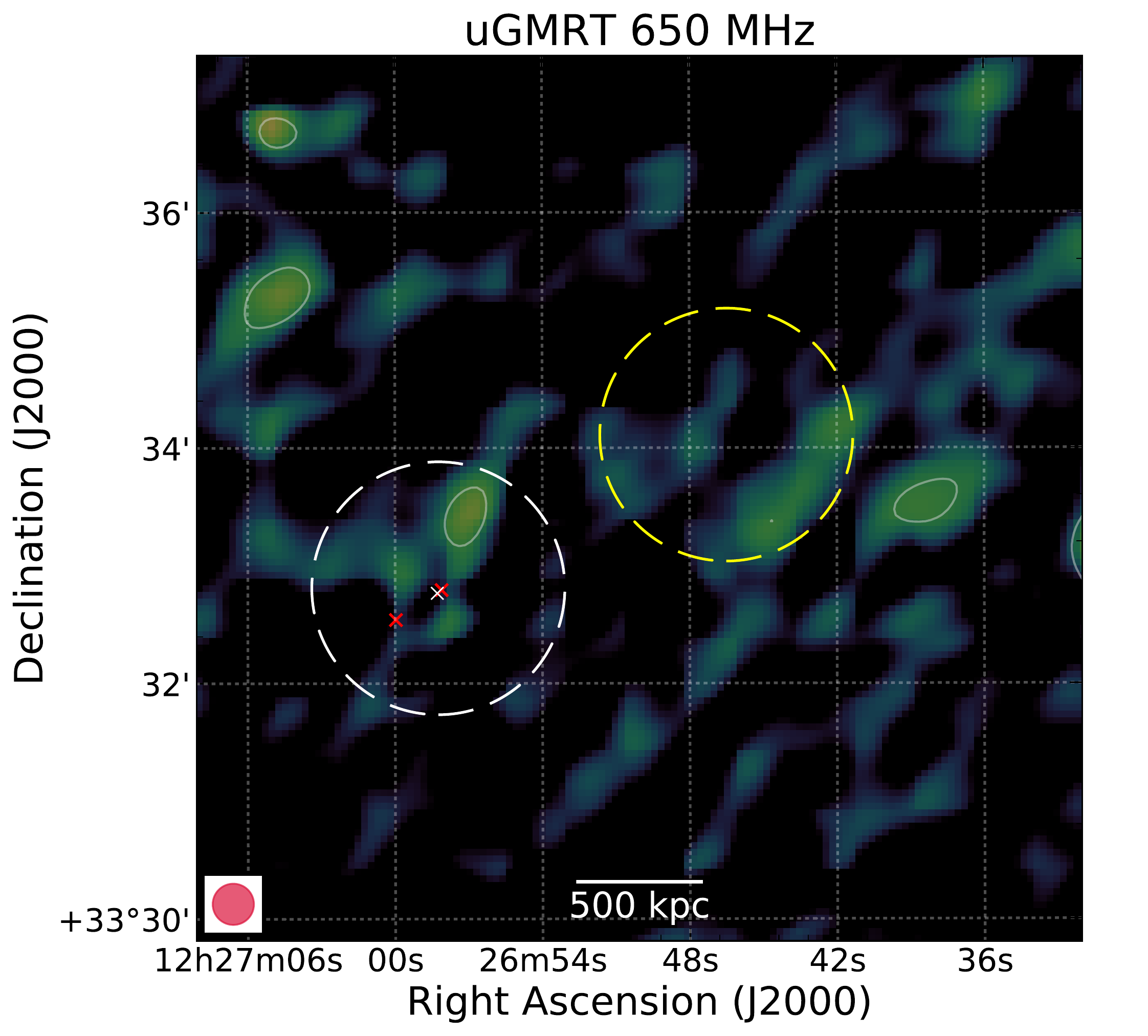}}
\caption{PSZ2\,G160.83+81.66. $22''$-resolution compact source-subtracted LOFAR and uGMRT images at 144 MHz (left) and 650 MHz(right). White-coloured radio contours at the same resolution are drawn at levels of $2.5\sigma_{\rm rms}\times[-1,1,2,4,8,16,32]$, with $\rm \sigma_{rms}$ the noise map (see Table \ref{tab:images_parameters}). The negative contour level is drawn with a dashed white line. The dashed white circle in each map shows the $R= 0.5 R_{\rm SZ,500}$ region obtained from $M_{\rm SZ,500}$, {with the white cross showing the cluster centre}, while the dashed yellow circle shows the position of the injected mock halo. The positions of the subtracted sources in the cluster region are highlighted with red crosses. The yellow polygon represents the area where the flux densities were measured.}
\end{figure*}

\section{Optical--radio overlays for PSZ2\,G091.83+26.11}\label{apx:optical_images}
Here we present the optical \emph{irg} image of PSZ2\,G091.83+26.11 taken from the PanSTARRS archive\footnote{\url{https://ps1images.stsci.edu/cgi-bin/ps1cutouts}} \citep[][]{panstarss16}. We overlay the radio contours of the uGMRT 650 MHz image after removing the contribution of the halo emission in order to investigate possible optical counterparts that generate the radio emission of the candidate radio relic.

\begin{figure*}
\centering
{\includegraphics[width=0.6\textwidth]{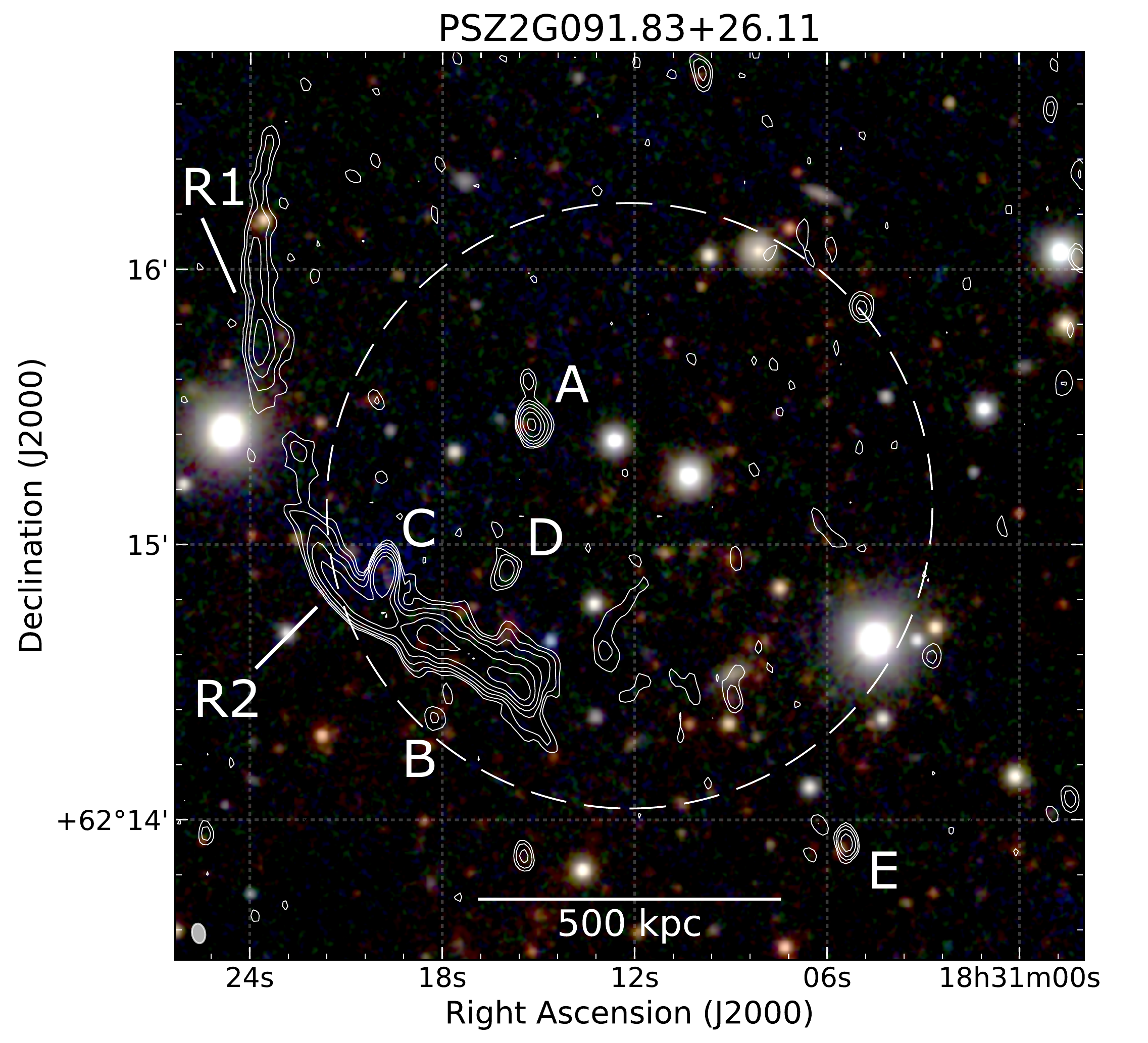}}
\caption{PanSTARRS optical \emph{irg} image of PSZ2\,G091.83+26.11. Radio contours at 650 MHz without the contribution of the radio halo (i.e. with the $uv$-cut at 500 kpc, see Sect. \ref{sec:images}) are overlaid in white. The white dashed circle represents the $R_{500}$ region. Labels follow Fig. \ref{fig:images_091}.}
\label{fig:apx_optical}
\end{figure*}

\bibliographystyle{aa}
\bibliography{biblio.bib}

\end{document}